\documentclass[amsmath]{aa}

\usepackage{txfonts}
\usepackage{float}
\usepackage{graphicx}
\usepackage{subcaption}
\usepackage{longtable}
\usepackage{natbib}
\usepackage{psfig}
\bibpunct{(}{)}{;}{a}{}{,}

\usepackage[colorinlistoftodos]{todonotes}
\usepackage[colorlinks=true, allcolors=blue]{hyperref}
\usepackage{siunitx}

\newcommand{\rms}{r{.}m{.}s{.}~}

\newcommand{\ie}{i{.}e{.}}

\newcommand{\fg}{Fig{.}~}
\newcommand{\fgs}{Figs{.}~}
\newcommand{\sct}{Sect{.}~}
\newcommand{\scts}{Sects{.}~}

\begin{document}

  \title{
  Aging of galaxies along the morphological sequence, marked by bulge growth and disk quenching \\
  }
   \author{Quilley, L.\thanks{email: louis.quilley@iap.fr} de Lapparent, V.\thanks{email: valerie.de\_lapparent@iap.fr}}

\titlerunning{Aging of galaxies along the morphological sequence}
\authorrunning{Quilley, de Lapparent}

   \institute{Institut d'Astrophysique de Paris, CNRS,
    Sorbonne Universit\'e, 98 bis boulevard Arago, 75014 Paris, France
}
  
   \date{Received 8 June 2022 / Accepted 3 August 2022}

  \abstract
{}
{We revisit the color bimodality of galaxies using the extensive EFIGI morphological classification of nearby galaxies.}
{The galaxy profiles from the Sloan Digital Sky Survey (SDSS) $gri$ images were decomposed as a bulge and a disk by controlled profile modeling with the Euclid SourceXtractor++ software. The spectral energy distributions from our resulting $gri$ SDSS photometry complemented with Galaxy Evolution Explorer (GALEX) $NUV$ photometry were fitted with the ZPEG software and PEGASE.2 templates in order to estimate the stellar masses and specific star formation rates (sSFR) of whole galaxies as well as their bulge and disk components.}
{The absolute $NUV-r$ color versus stellar mass diagram shows a continuous relationship between the present sSFR of galaxies and their stellar mass, which spans all morphological types of the Hubble sequence monotonously. Irregular galaxies to intermediate-type Sab spirals make up the ``Blue Cloud'' across 4 orders of magnitude in stellar mass but a narrow range of sSFR. This mass build-up of spiral galaxies requires major mergers, in agreement with their frequently perturbed isophotes. At high mass, the Blue Cloud leads to the ``Green Plain,'' dominated by S0a and Sa early-type spirals. It was formerly called the ``Green Valley,'' due to its low density, but we rename it because of its wide stretch and nearly flat density over $\sim2$ magnitudes in $NUV-r$ color (hence sSFR), despite a limited range of stellar mass (1 order of magnitude). The Green Plain links up the ``Red Sequence,'' containing all lenticular and elliptical galaxies with a 2 order of magnitude mass interval, and systematically higher masses for the ellipticals. We confirm that the Green Plain cannot be studied using $u-r$ optical colors because it is overlayed by the Red Sequence, hence $NUV$ data are necessary. Galaxies across the Green Plain undergo a marked growth by a factor 2 to 3 in their bulge-to-total mass ratio and a systematic profile change from pseudo to classical bulges, as well as a significant reddening due to star formation fading in their disks. The Green Plain is also characterized by a maximum stellar mass of $10^{11.7} M_\odot$ beyond which only elliptical galaxies exist, hence supporting the scenario of ellipticals partly forming by major mergers of massive disk galaxies.}
{The EFIGI attributes indicate that dynamical processes (spiral arms and isophote distortions) contribute to the scatter of the Main Sequence of star-forming galaxies (Blue Cloud), via the enhancement of star formation (flocculence, HII regions). The significant bulge growth across the Green Plain confirms that it is a transition region, and excludes a predominantly quick transit due to rapid quenching. The high frequency of bars for all spirals as well as the stronger spiral arms and flocculence in the knee of the Green Plain suggest that internal dynamics, likely triggered by flybys or (mainly minor) mergers, may be the key to the bulge growth of massive disk galaxies, which is a marker of the aging of galaxies from star forming to quiescence. The Hubble sequence can then be considered as an inverse sequence of galaxy physical evolution.}


  \maketitle
  

\section{Introduction                                \label{intro}}

The study of galaxy morphology started with Edwin Hubble classifying the different galaxies he observed according to their shapes and features \citep{1936rene.book.....H}. This process led to the Hubble sequence being established, which was then improved by De Vaucouleurs \citep{1959HDP....53..275D}. This system of visual classification has several drawbacks: it is a long and tedious work that requires prohibitive amounts of human time to be performed on large surveys; galaxies should be well resolved, hence at a limited distance depending on the telescope and instrument; it is based on the human eye, so the results are prone to errors, as well as systematic drifts across the sample.

To survey the diversity of galaxy population, establishing color-magnitudes diagrams have become a standard work-around approach, because it can be performed on all kinds of galaxies, even distant and poorly resolved ones, and it is a quantitative method that is significantly faster than visual classification. Using such diagrams, it was shown that early-type galaxies (ellipticals and lenticulars) are predominantly red, whereas late-type galaxies (spirals and irregulars) are blue \citep{2007ApJS..173..185G}. This dichotomy was first shown using the optical color $u-r$ (\citealt{2001AJ....122.1861S}, \citealt{2004ApJ...600..681B}, \citealt{2006MNRAS.373.1389C}), and it became even clearer when complementing the optical data with an ultraviolet (UV) band. Color-magnitude diagrams were proved to convey physical information about galaxies: for example, SED model-fitting showed that UV to optical color (in particular the $NUV-r$ color) is a tracer of star formation (\citealt{2007ApJS..173..267S}, \citealt{2007ApJS..173..357K}, \citealt{2007ApJS..173..619K}).

The star formation rate (SFR) of a galaxy is linked to its morphology \citep{1998ARA&A..36..189K}: elliptical and lenticular galaxies are generally quiescent and composed of old stars, whereas spiral galaxies show significant star formation in their disks, characterized by young stellar populations and HII regions delineating their spiral arms. The analysis of the UV-optical color magnitude diagram (\citealt{2007ApJS..173..342M}, \citealt{2007ApJS..173..293W}, \citealt{2007ApJS..173..267S}) shows a physical bimodality between the star-forming galaxies of the Blue Cloud, and the  so-called quiescent (or passive) galaxies of the Red Sequence, that is with very low or no star formation. The intermediate region between them was originally named the Green Valley because of its low number density of galaxies in the color-magnitude diagram.

Uncertainties remain as to how the individual Hubble types are located along these sequences, in particular for Green Valley galaxies. For example, the Galaxy Zoo \citep{2011MNRAS.410..166L} greatly expanded the number of classified galaxies, but the poor angular resolution of the majority of the galaxies (up to redshifts $\lesssim0.1$) only allows for a rough classification into ellipticals, spirals, or mergers. More recently, \cite{2017MNRAS.471.2687B} studied the star-forming state of galaxies with a visual morphological classification \citep{2010ApJS..186..427N}, and show the strong interdependence between SFR and morphology, being stronger than with the local density for both parameters. Studies based on bulge and disk decomposition of galaxies at redshifts from 0 to $\sim2.5$ 
show the role of bulge growth (\citealt{Lang_2014_bulge_growth_quenching_CANDELS}, \citealt{Bluck_2014_bulge_mass},
\citealt{2018MNRAS.476...12B}, \citealt{Bluck_2022_quenching_bulge_disk_ML}, \citealt{Dimauro_2022_bulge_growth}) and disk reddening (\citealt{2018MNRAS.476...12B}) in the transformation of galaxies from star-forming to quiescent.

In the present article, we analyze a sample of nearby, well-resolved galaxies extracted from the Sloan Digital Sky Survey images (SDSS), a subsample of which were thoroughly studied to create a visual morphological classification (``Extraction de Formes Idealis\'ees de Galaxies en Imagerie'' - EFIGI), and we examine them in terms of absolute colors and magnitudes. We complement the SDSS optical photometry with UV photometry obtained by the Galaxy Evolution Explorer (GALEX). In addition, we perform bulge and disk decomposition to further relate colors and the morphology.

In \sct \ref{data}, we present the data used for this study. In \sct \ref{methodology}, we detail the methodology used to perform model-fitting of the luminosity profiles using SourceXtractor++ \citep{2020ASPC..527..461B}, as well as of the spectral energy distribution (SED) using ZPEG \citep{2002A&A...386..446L}. We then analyze our results in \sct \ref{results}, by locating morphological types and their characteristics within the color-absolute magnitude diagram, as well as the color-stellar mass diagram. In \sct \ref{discussion}, we question the nature of the transition between star-forming and quiescent galaxies in view of additional EFIGI morphological attributes, and well as other published relevant analyses, and discuss the implications of our findings.

\section{Data                              \label{data}}

\subsection{MorCat}

The Morphological Catalogue (MorCat) is a complete catalog of galaxy images extracted from the Sloan Digital Sky Survey (SDSS) Data Release 8 (DR8) \citep{2011ApJS..193...29A} in its five bands $u$, $g$, $r$, $i$ and $z$, to an apparent magnitude limit of $g\le15.5$. To avoid solving the complex segmentation problem of identifying large galaxies (and all of them) directly from the images, we assume that all galaxies to that $g$ limit are included in HyperLeda \citep{2014A&A...570A..13M} to $B_T\le18.0$. After restricting this HyperLeda sample to a sky mask corresponding to the SDSS northern galactic cap, extracting the images of each galaxy from the SDSS image database, measuring their photometry by bulge and disk modeling using SExtractor \citep{1996A&AS..117..393B}, and discarding spurious sources (unresolved objects and image artifacts), a total of 20126 MorCat galaxies with $g\le15.5$ are obtained.

\subsection{EFIGI}

Within the intersection of their masks over the northern sky and the northern galactic cap (see Quilley \& de Lapparent, (\textit{in prep.}), EFIGI (``Extraction de Formes Idéalisées de Galaxies en Imagerie'') is a subsample of MorCat (EFIGI contains 657 galaxies outside this intersection, and located mostly within the southern galactic hemisphere of the northern sky). It contains 4458 galaxies with known morphological types and uncertainties from the RC3 Revised Hubble sequence (RC3-seq hereafter) \citep{1991rc3..book.....D}. 

A systematic visual classification process by a group of astronomers led to a new catalog with 16 morphological attributes: {\tt B/T} (as bulge over total flux ratio), spiral {\tt Arm Strength}, spiral {\tt Arm Curvature}, {\tt Bar Length}, {\tt Inclination-Elongation} (hereafter shortened to {\tt Incl-Elong}), {\tt Perturbation}, {\tt flocculence}, {\tt Hot Spots} and {\tt Visible Dust} are the 9 attributes used here; the other 7 attributes not directly used in the present analysis are {\tt Contamination}, {\tt Multiplicity}, spiral {\tt Arm Rotation}, {\tt Inner Ring}, {\tt Outer Ring}, {\tt Pseudo-ring}, and {\tt Dust Dispersion} \citep{2011A&A...532A..74B}. These attributes allow to better understand the nature of the Hubble sequence \citep{2011A&A...532A..75D}, and can be useful to understand some trends (or biases) in the results of analyses as a function of morphology.

The stage of each galaxy on a morphological sequence based on the RC3-seq, and which we call the EFIGI morphological sequence (EFIGI-seq hereafter), has also been measured for each EFIGI galaxy, with the following specificities: contrary to the RC3-seq, the EFIGI-seq only has one elliptical type, as the different elongation stages of elliptical galaxies in the RC3-seq are measured by the EFIGI {\tt Incl-Elong} attribute; the RC3-seq nonmagellanic irregulars (I0 type) are not considered as a separate type in the EFIGI-seq, but as galaxies of some other Hubble type which undergo some perturbation as measured by the corresponding attribute; in the EFIGI-seq, the dwarf elliptical (dE) and dwarf spheroidal (dSph) galaxies are in a separate class from the Ellipticals, named dE. The various types along the EFIGI-seq are listed in Table \ref{tab_types}. Table \ref{tab_types} also shows the mean difference of RC3 and EFIGI type for EFIGI galaxies grouped by EFIGI type: except for the difference due to different definitions (for types cE and dE), there is an overall agreement  between RC3-seq and EFIGI-seq, as shown by the rather small mean difference and moderate dispersion of these differences \citep[see also][\fg 19]{2011A&A...532A..74B}.

We note that the EFIGI sample was built with the goal of having a few hundreds of galaxies of each Hubble type, and mostly include galaxies with apparent diameter $\ge 1$ arcmin \citep{2011A&A...532A..75D}.
Therefore EFIGI is not magnitude limited and is not a representative sample of nearby galaxies, which MorCat is. Their largely overlapping masks in the sky allows to interpret MorCat results with EFIGI morphology information, and to generalize EFIGI results using MorCat.

\begin{table*}[ht]
\caption{Means and standard deviations of the difference between the RC3 and EFIGI Hubble types for EFIGI galaxies.}
\centering
\resizebox{\linewidth}{!}{%
\begin{tabular}{ c c c c c c c c c c c c c c c c c c c}
\hline\hline
EFIGI Hubble Type & cE & E & cD & S0$^+$ & S0 & S0$^-$ & S0a & Sa & Sab & Sb & Sbc & Sc & Scd & Sd & Sdm & Sm & Im & dE\\
\hline
RC3 number & -6 & -5 & -4 & -3 & -2 & -1 & 0 & 1 & 2 & 3 & 4 & 5 & 6 & 7 & 8 & 9 & 10 & 11 \\
\hline
Mean difference & 2.964 & 1.334 & 0.388 & -0.048 & 0.276 & -0.292 & -0.021 & 0.111 & 0.251 & 0.206 & -0.173 & -0.41 & -0.538 & -0.653 & -0.936 & -1.22 & -0.818 & -9.775 \\
\hline
Standard deviation & 3.96 & 1.267 & 1.703 & 1.386 & 1.552 & 1.733 & 1.645 & 1.349 & 1.303 & 1.664 & 1.604 & 1.46 & 1.468 & 1.403 & 1.989 & 2.306 & 1.85 & 5.38 \\
\hline
\end{tabular}}
\label{tab_types}
\end{table*}

\subsection{GALEX cross-match     \label{sec:galex}}

In order to complete the SDSS optical photometry with UV photometry, we perform a cross match between the EFIGI or MorCat catalog and the Galaxy Evolution Explorer (GALEX) catalog, which covers the whole sky, except for the plane of the Milky Way. To retrieve GALEX data, we used VizieR\footnote{\url{https://vizier.u-strasbg.fr/}} and performed a cross-match on celestial position, with a maximum distance of 5 arcsec, between two GALEX tables from the revised catalog of GALEX UV sources (GUVcat\textunderscore AIS DR6$+$7,  \citealt{2017ApJS..230...24B}): (i) the sample of the revised GALEX catalog of UV sources (GUVcat\textunderscore AIS) from DR6$+$7, providing magnitudes of all detected sources in the $FUV$ ($1350 - 1750 \AA $) and $NUV$ ($1750 - 2800 \AA $) bands; (ii) the list of galaxies larger than 1 arc minute included in the GUVcat footprint. Another cross match is then performed between the resulting GALEX table and the MorCat galaxies, based on the galaxy name list provided by HyperLeda \citep{2014A&A...570A..13M} for each MorCat object.

Among the 4458 EFIGI galaxies, 1848 have $NUV$ data, while only 1754 of the 16327 of MorCat galaxies outside of EFIGI do. In the $FUV$ band, we obtain magnitudes for 1301 EFIGI galaxies and for 1255 MorCat galaxies (outside of EFIGI). As we show below, the GALEX $NUV$ band is sufficient to estimate the star formation rate, whereas using instead the $FUV$ band would reduce the EFIGI and MorCat subsamples with UV photometry by 29.6\% and 29\% respectively. We therefore only make use of the $NUV$ data in the present analysis, and note EFIGI $\cap$ GALEX and MorCat $\cap$ GALEX to refer respectively to the EFIGI and MorCat subsamples with GALEX $NUV$ photometry. GALEX photometry is given only for the whole galaxies, it is not separated into bulge and disk components. While SDSS photometry is subject to splitting issues (see \sct 5.3 and \fg 24 in \cite{2011A&A...532A..74B}), hence the need to remeasure magnitudes ourselves (see \sct\ref{methodo_srx}), we keep the magnitudes provided by the pipeline in GALEX. Indeed, \cite{2007ApJS..173..659B} explain that the shredding issue only applies to a few objects in GALEX AIS catalog, so it cannot jeopardize our results.

\begin{figure*}
\includegraphics[width=\textwidth]{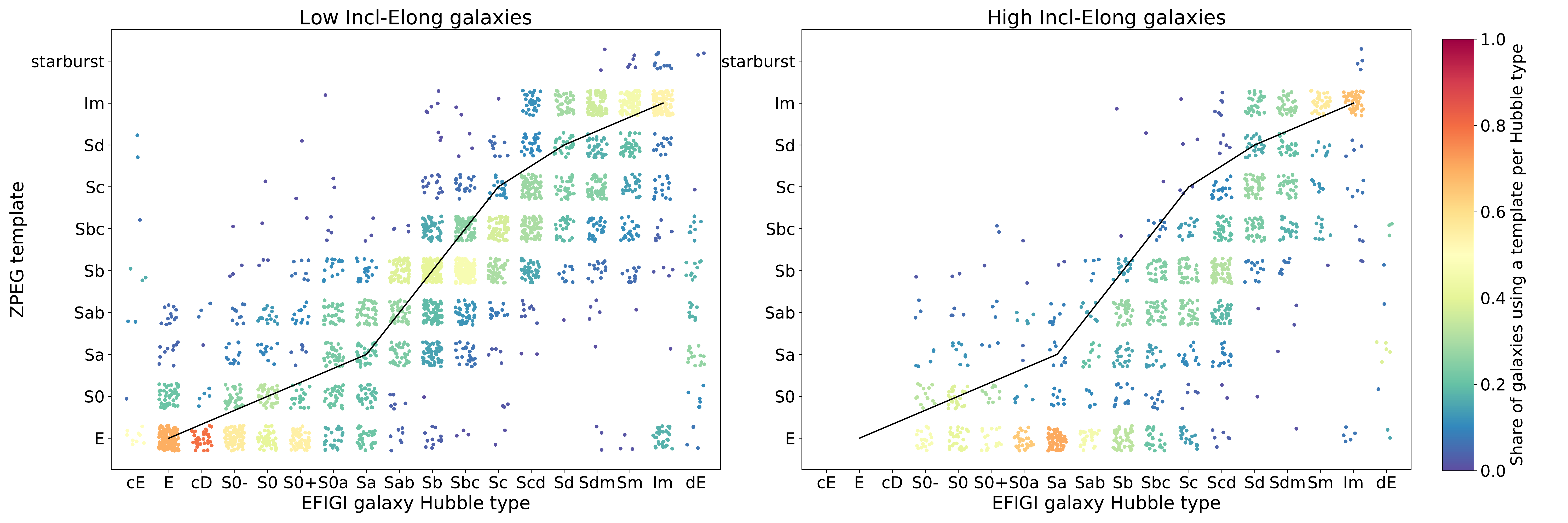}
\caption{Comparison of the EFIGI Hubble type to the spectral type defined by the SED template found by ZPEG to match the measured apparent magnitudes and redshift. The black solid line is the type identity line. The color represents the fraction of galaxies of a specific Hubble type that is best fitted by a given spectral template (sum of vertical fraction is 1 for each Hubble type). The left and right panels are restricted to EFIGI galaxies with {\tt Incl-Elong} $\leq 2$ and {\tt Incl-Elong} $\geq 3$ respectively, showing a dispersion between the morphological and spectral type as well as a systematic shift toward later spectral types for a given morphological type, due to disk inclination and internal dust.}
\label{compare_types_by_Incl}
\end{figure*}

\section{Methodology                                \label{methodology}}

\subsection{Galaxy profile-fitting with SourceXtractor++    \label{methodo_srx}}

SExtractor bulge and disk photometry is performed for all MorCat and EFIGI galaxies, and displays on synthetic images of galaxies with similar properties, a high accuracy in the modeled total apparent magnitudes. However the EFIGI morphological attributes allow one to detect systematic biases due to degeneracies between the bulge and disk components, and which affect the separate bulge and disk photometry. 

To obtain more reliable bulge and disk magnitudes for EFIGI galaxies, we use the SourceXtractor++ software \citep{2020ASPC..527..461B} on EFIGI galaxies only, which is a followup to SExtractor \citep{1996A&AS..117..393B} developed in the context of the Euclid space mission \citep{2021arXiv210801201S}. The great advantage of this new software is the possibility to perform simultaneous multiband model-fitting for as many S\'ersic (or exponential) profiles as desired for each object, and with a control on all the model parameters: one can put priors on the S\'ersic index, the effective radius or the aspect ratio ($=b/a$ with $a,b$ the major and minor axis of the profile respectively), and one can also control the relative values of a given aforementioned parameter between the bulge and disk components, or across filters. In particular, we make sure that values of both the bulge and disk effective radii are similar (within some adjustable margin) between filters (details on the adopted priors are provided in Quilley \& de Lapparent (\textit{in prep.}).  Because many SDSS images of the EFIGI galaxies in the $u$ and $z$ bands are noisier than in the higher signal-to-noise $g$, $r$, and $i$ bands, and therefore degrade the multiband bulge and disk model-fitting when included, we limit here these fits with SourceXtractor++ to the 3 bands $g$, $r$ and $i$ (and, as already mentioned, do not include the $NUV$ band in this decomposition).

We use and compare two different methods to model the galaxy profiles as sums of disks and bulges with SourceXtractor++. On the one hand, we have a prior-less two-component model as the sum of a S\'ersic profile (intended for the bulge) and an exponential profile (intended for the disk). On the other hand, we develop a two-component model with priors defined by automatically zooming into the galaxy disk in order to model specifically the bulge with a S\'ersic profile in the $g$, $r$ and $i$ bands separately, while treating the disk as a background component. The level of zoom is a function of the minor and major axis of the segmented area for the galaxy, and in any case smaller than $1/2$ of the smallest of the two. We then use the derived median bulge parameters over the $g$, $r$, and $i$ bands as bulge priors for the two-component bulge and disk model. These two-component models are performed simultaneously in the $g$, $r$ and $i$ bands, with Gaussian priors on ratios or differences of profile parameters between bands. More information on the SourceXtractor++ configuration will be given in Quilley \& de Lapparent (\textit{in prep.})
For consistency between the EFIGI and MorCat results derived  below, we calculate that the distribution of difference between SourceXtractor++ and SExtractor apparent magnitudes of EFIGI galaxies per apparent magnitude intervals ($[8,12]$, $[12,14]$, $[14,16]$ and $[16,18]$) in the $g$, $r$, and $i$ bands, peaks between $0.02$ and $0.06$ when more than 100 galaxies per interval, and with an \rms\ dispersion varying between 0.1 and 0.2 for all 3 bands for the 3 most luminous intervals, and reaching 0.4 for $g$ in the magnitude interval $[16,18]$ (this bin is too poorly populated in the $r$ and $i$ bands, with less than 100 galaxies). These differences may result from the use of an adaptive mesh size for estimating the sky background in the SourceXtractor++ analysis (5 times the isophotal diameter up to 2048 pixels), whereas the mesh size is fixed to 1024 pixels in the SExtractor analysis (except for heavily contaminated galaxies in both analyses).

The PSF is calculated using a sufficient number of stars around each galaxy using PSFEx, and is provided to SourceXtractor++ with each image in order to be convolved with the model before computing the distance to the data image, that is used to perform the model-fitting. Because the EFIGI galaxies are typically larger than 1 arcmin, the PSF has little impact except in the central parts of steep profiles, and for texture (HII regions, dust) on scales on the order of the seeing ($\sim1.3$ arcsec) of the SDSS images.

\begin{figure*}
\includegraphics[width=\linewidth]{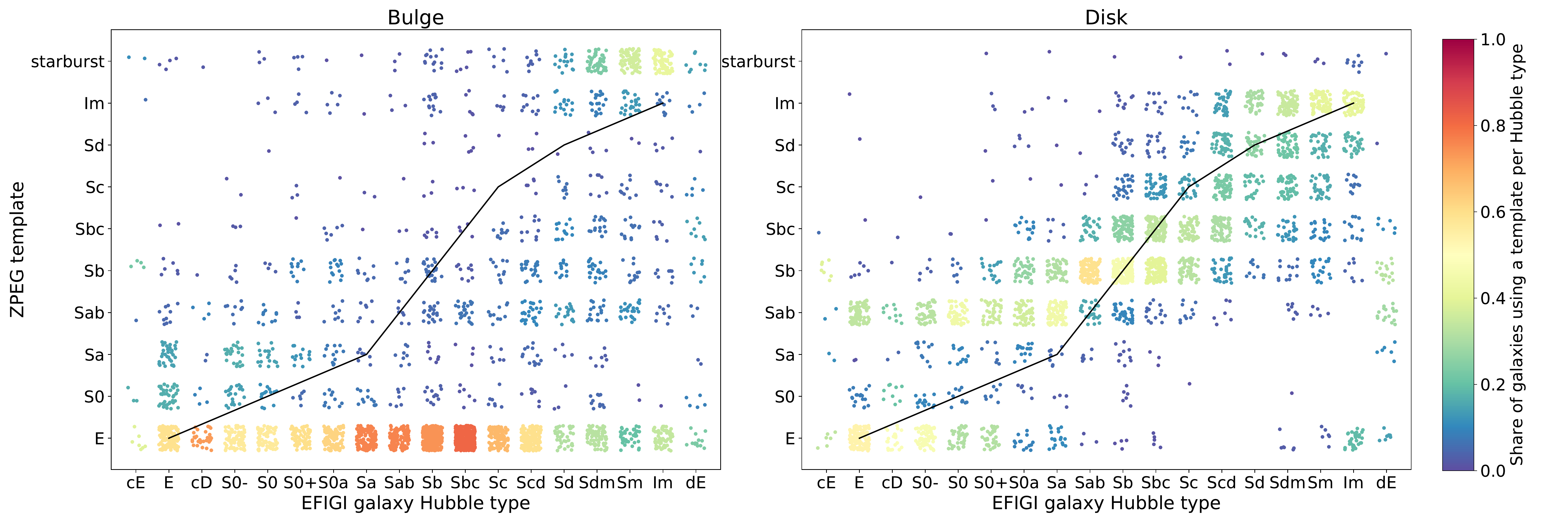}
\caption{Same as \fg\ref{compare_types_by_Incl} for the bulges (left) and the disks (right) of EFIGI galaxies with {\tt Incl-Elong} attribute values of 0, 1 or 2. For the bulges SED model-fitting, the E template is largely favored to fit the bulge magnitudes of all Hubble types having a bulge component, from E to Sc; while the later types Sd to Im are shared between the E and the starburst template. For the disks, there is a wider distribution around the identity line, as for whole galaxies in \fg\ref{compare_types_by_Incl}.}
\label{comparing_EFIGI_and_ZPEG_bulge_disk}
\end{figure*}

\subsection{SED model-fitting with ZPEG    \label{zpeg}}

\subsubsection{Absolute magnitudes, stellar masses and star formation rates \label{sec:absmag}}

To obtain the absolute (rest-frame) magnitudes and colors of EFIGI and MorCat galaxies, we use ZPEG \citep{2002A&A...386..446L}. This software receives as inputs the apparent magnitudes in the $NUV$, $g$, $r$, $i$ bands for EFIGI galaxies measured by SourceXtractor++ (with $NUV$ only when available from GALEX) and in the $NUV$, $u$, $g$, $r$, $i$ and $z$ bands for MorCat galaxies measured by SExtractor, as well as the HyperLeda redshifts corrected for Virgocentric infall  (see \sct 2.2 of \citealt{2011A&A...532A..75D}). ZPEG fits to these apparent magnitudes the SEDs of families of templates from the PEGASE.2 library \citep{1999astro.ph.12179F}, including all scenarios adjusted to the major galaxy types; they are mainly characterized by specific functions for the evolution of the star formation rate with time: E, S0, Sa, Sab, Sb, Sbc, Sc, Sd, Im, starburst. There is an age constraint on these templates with all types from E to Sd, Im and starburst having a minimum age of 11 Gyr, 9 Gyr and 0 Gyr respectively. ZPEG offers the option for a variable internal extinction; we however deactivated this option because it leads to larger discrepancies between the spectral and morphological types (see below). Through this SED model-fitting, ZPEG yields the age of the scenario corresponding to the best-fit template, as well as several galaxy parameters including the mean stellar age, the stellar mass $M_\ast$ and the star formation rate SFR, from which we derive the specific star formation rate $\mathrm{sSFR}= \mathrm{SFR}/M_\ast$.

To test the robustness of this SED model-fitting, \fg\ref{compare_types_by_Incl} compares the Hubble types as classified in EFIGI to the spectral type of the best fit template determined by ZPEG, which we note ``spectral type'' hereafter. Because the presence of dust and the inclination of a galaxy disk are two factors that lead to its reddening, we split EFIGI galaxies at an {\tt Incl-Elong} attribute value of $2$, corresponding to an inclination of 70° for disks and an elongation of 0.7 for disk-less galaxies (see \citealt{2011A&A...532A..74B}). The left panel of \fg\ref{compare_types_by_Incl} is restricted to galaxies with {\tt Incl-Elong} $\leq 2$ and show that face-on and galaxies with intermediate {\tt Incl-Elong} have a spectral type approximately ``aligned'' with their morphological type with some scatter and a systematic offset toward earlier spectral types for a given morphological type, likely to be due to internal reddening. For edge-on and nearly edge-on galaxies corresponding to {\tt Incl-Elong} $\geq 3$, shown in the right panel, the stronger systematic shift toward earlier spectral types illustrates the additional effect of disk inclination, which also increases the reddening due to dust.

\subsubsection{SED of bulges and disks \label{sed_bulge_disk}}

To evaluate whether the multiband bulge and disk apparent magnitudes obtained by the SourceXtractor++ bulge and disk profile-fitting can be used to perform ZPEG SED model-fitting to derive separate parameters for these components, one must evaluate whether the PEGASE.2 SED templates make sense for a bulge and a disk across all types in the Hubble morphological sequence. The following arguments favor this unconventional use of the PEGASE.2 templates :

\begin{enumerate}
    \item For elliptical galaxies, the bulge component dominates the galaxy profile with $B/T$ close to 1, the disk component is essentially used in the fits to correct some irregularities in the observed profile compared to the S\'ersic profile, or to account for a disk component that was not visible from SDSS images during the visual classification (these objects may be of type S0$^-$).
    \item For lenticular galaxies, the SourceXtractor++ photometry shows that the bulge and disk colors are similar (Quilley \& de Lapparent, \textit{in prep.}), so the S0 PEGASE.2 template should be appropriate for both components as long as it is appropriate for the whole galaxy.
    \item For early-type spirals, \cite{2006MNRAS.371....2A} showed that the light profile of bulges are similar in terms of color and profile slope (S\'ersic index) to those of elliptical galaxies, which justifies the use of the E template for these central components.
    \item For the early and intermediate type spirals (Sa to Scd), we assume that the presence of a bulge simply shifts the disk template to a later type compared to that for the whole galaxy, and we use the a posteriori quality of the SED fits to validate this argument (see below).
    \item For very late-type spirals (Sd, Sdm, Sm) and irregulars, with very weak or no bulge, SourceXtractor++ yields very low values values of $B/T$ $\leq 0.01$, so the disk component constitutes the whole galaxy and can be modeled by the Sd or Im scenarios. In the cases where a strong HII region is modeled by a bulge component, it is expected to be best fit with the Im or starburst template.
\end{enumerate}

\begin{figure}
\includegraphics[width=\columnwidth]{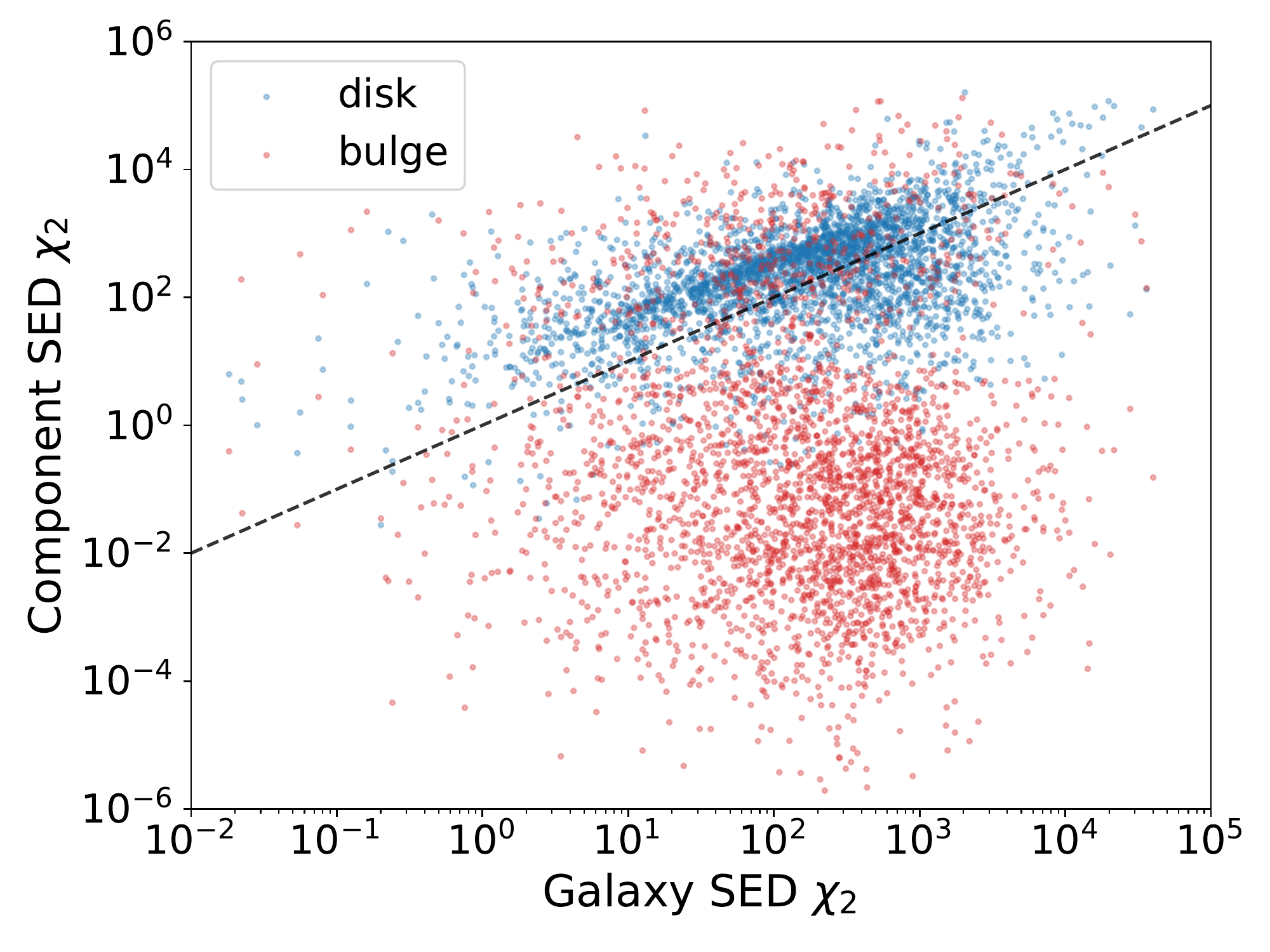}
\caption{Distribution of $\chi^2$ of ZPEG best-fit PEGASE.2 template with the EFIGI $NUVgri$ photometry (or only $gri$ when $NUV$ is not available) for whole galaxies compared with the $\chi^2$ obtained with the $gri$ only photometry of the bulge and disk components extracted with SourceXtractor++ by bulge and disk modeling. The 1 to 3 order of magnitude lower $\chi^2$ for the bulges compared those for the whole galaxies, and the factor of 2-3 higher for disks validate the use of templates designed for whole galaxies.}
\label{zpeg_chi2}
\end{figure}

We therefore apply the ZPEG SED model-fitting on the bulge and disk apparent magnitudes in the $g$, $r$, and $i$ bands derived from the SourceXtractor++ bulge and disk modeling with the same PEGASE.2 templates as for the entire galaxy, which allows us to obtain estimates of absolute magnitudes and stellar masses for the bulge and disk components separately. \fg\ref{comparing_EFIGI_and_ZPEG_bulge_disk} shows that bulges are better fitted by E spectral templates down to spiral Hubble types Sd, but the E template competes with the Im or starburst spectral templates for Sdm, Sm and Im Hubble types, in agreement with the fact that these very late types have very weak or no bulge, hence a star-forming region is sometimes modeled as such. In contrast the disk of S0 and Sa Hubble types are best fitted by E or Sab spectral types, and those of Sab to Im Hubble types by Sb or later spectral types. The 1-type mismatch between the Hubble types and PEGASE.2 templates may call for a readjustment of the PEGASE.2 scenarios, as the EFIGI catalog was not available when the scenarios were designed. \fg\ref{comparing_EFIGI_and_ZPEG_bulge_disk} is further discussed in \sct \ref{disk_reddening}.

\fg\ref{zpeg_chi2} shows the distribution of $\chi^2$ derived by ZPEG for the best-fit template to each EFIGI galaxy compared to the $\chi^2$ for the best-fit templates to the bulge and disk components separately. First, it is noteworthy that the $\chi^2$ of the majority of bulges (in red), that are best-fit by the E template (as shown in the left panel of \fg\ref{comparing_EFIGI_and_ZPEG_bulge_disk}), is $\sim 2-5$ orders of magnitude lower than the $\chi^2$ for the whole galaxy-template match, reinforcing the fact that the PEGASE.2 elliptical template is an appropriate description of any real bulge of all morphological types in EFIGI (only galaxies with a very small, hence unreal bulge, enclosing a few percents of the galaxy light have a ratio of bulge to total SED $\chi^2$ in the range 0.1 to 100 - see \sct \ref{disk_reddening}). The blue cloud of points in \fg\ref{zpeg_chi2} shows that the $\chi^2$ for the disk is predominantly higher than for the whole galaxy, by a factor 3 to 4, but there are only 3\% of galaxies for which the disk to galaxy $\chi^2$ ratio is $\gtrsim10$, and 22\% of galaxies for which the ratio is $\lesssim 0.1$. Using templates outside of their intended range of application explain these slightly poorer fits. Note also a tail of 5\% of galaxies for which the disk to galaxy $\chi^2$ ratio is $\lesssim 0.01$, indicating that for these galaxies there is an important gain in the bulge and disk decomposition in terms of SED fitting. Altogether, these results lend joint credibility to our bulge and disk decompositions using SourceXtractor++, and to their SED fitting using ZPEG and PEGASE.2 templates. 

\section{Results                                \label{results}}

\subsection{How galaxy fluxes and colors change with morphology        \label{res_color_morpho}}

\begin{figure*}
\includegraphics[width=\columnwidth]{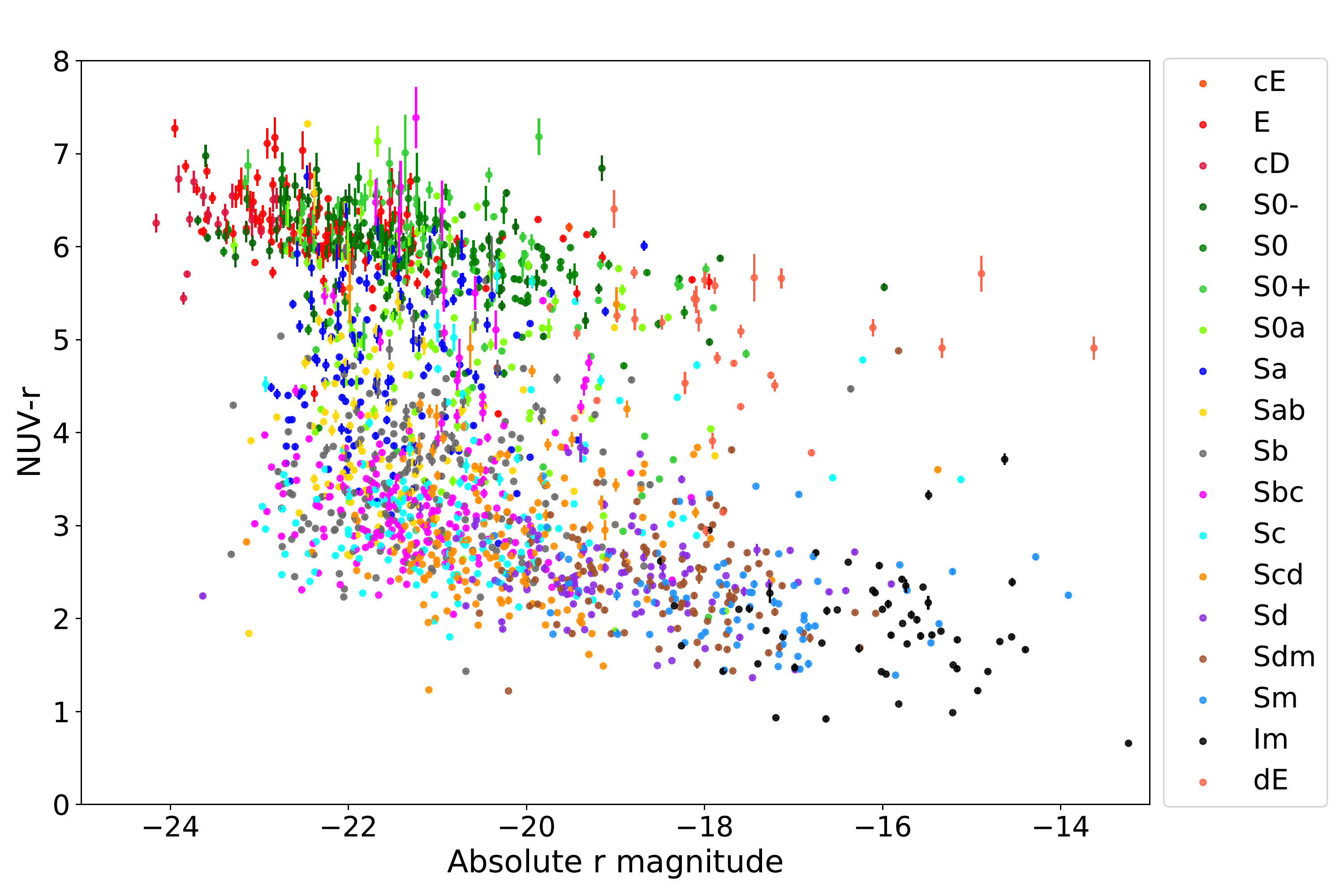}
\includegraphics[width=\columnwidth]{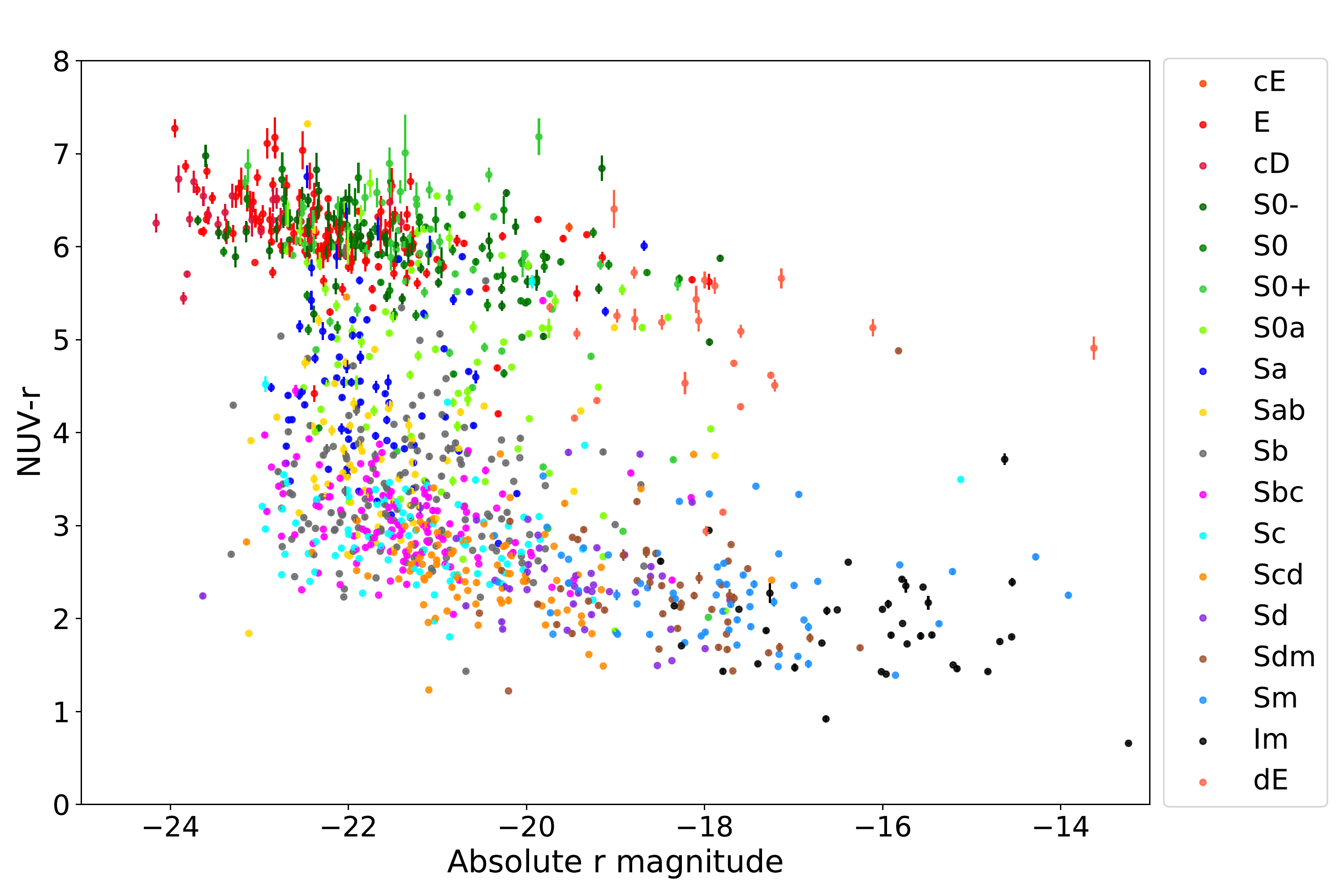}
\caption{Color-magnitude diagram defined as $NUV-r$ absolute color versus $r$ absolute magnitude for the 1848 EFIGI $\cap$ GALEX galaxies at all inclinations (\textbf{left panel}), and with {\tt Incl-Elong} $\leq2$ (\textbf{right panel}). The color of the points indicates their Hubble Type as classified in the EFIGI morphological catalog. The distributions of galaxies in both graphs exhibit the bimodality between the ellipticals and lenticulars in the upper region (the Red Sequence), and spirals of type Sab and later in the lower region (the Blue Cloud). A lower density region (the Green Valley) connects both structures, with galaxies of types S0a and Sa.}
\label{NUV-r-face-on}
\end{figure*}

\begin{figure}
\includegraphics[width=\columnwidth]{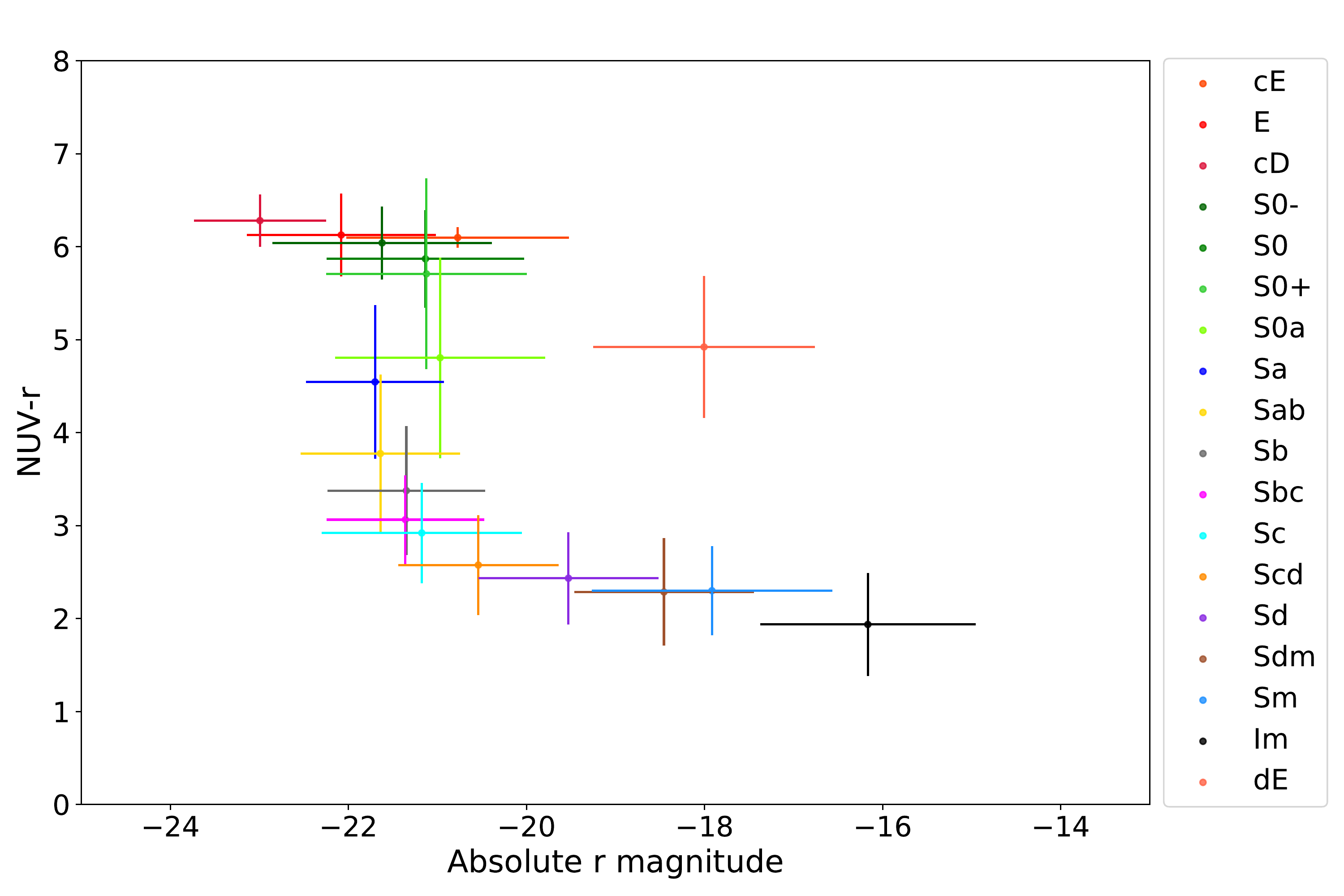}
\caption{Mean absolute color $NUV-r$ versus mean $r$ absolute magnitude for each Hubble type for EFIGI $\cap$ GALEX galaxies with {\tt Incl-Elong} $\leq 2$; the error bars represent the \rms deviation within each type. The Hubble sequence draws an ``S'' shape that could be parameterized monotonously by the mean Hubble type, with the exception of dE at $(-18,4.9)$.}
\label{mean-NUV-r-face-on}
\end{figure}

In the left panel of \fg\ref{NUV-r-face-on}, we plot the $NUV-r$ absolute color versus the absolute $r$ magnitude (denoted $M_r$ hereafter) for the 1848 EFIGI galaxies with GALEX $NUV$ photometry. Galaxies are grouped in three main regions: on one hand, the Red Sequence around $NUV-r = 6$ and the Blue Cloud containing galaxies bluer than $NUV-r = 4$ and extending toward very faint, and very blue galaxies; both regions define the well-known galaxy population bimodality, with specific characteristics like the greater extent of the Blue Cloud compared to the Red Sequence; on the other hand, one can observe between the two high density regions a low density one, originally named Green Valley, and first studied by \citet{2007ApJS..173..342M}, \citet{2007ApJS..173..293W} and \citet{2007ApJS..173..267S}. We note that the EFIGI Green Valley (that we define in \sct \ref{result-fading} as $NUV-r\in[3.77, 5.71]$) extends further up than the $[4, 5]$ interval quoted in the review of this region by \cite{2014SerAJ.189....1S}: this author finds a Red Sequence at lower, hence bluer values of $NUV-r$ than in our analysis, with an approximate shift of $\Delta_{NUV-r}\simeq0.75$, whereas there is no shift in our respective Blue Clouds. This prevents a simple interpretation in terms of difference between our respective samples, or the SDSS photometry and ours (using SourceXtractor++). We point out that the errors in the absolute magnitudes and colors are plotted on all graphs of this study: they are mostly very small and indistinguishable from the points except for objects with a low $NUV$ flux, hence predominantly in the Red Sequence.

The left panel of \fg\ref{NUV-r-face-on} suggests that each Hubble type populates a specific part of the color-magnitude plane. Because disk inclination tends to redden galaxy colors \citep{2007MNRAS.379.1022D}, we show in the right panel of \fg\ref{NUV-r-face-on} the same graph restricted to EFIGI galaxies with {\tt Incl-Elong} $\leq 2 $: this corresponds to face-on or intermediate inclination galaxies for disks, that is $\leq 70^\circ$, and elongation $\leq0.7$ for disk-less galaxies; this removes highly inclined disks, but keeps all E galaxies as their values of {\tt Incl-Elong} are between 0 and 2 \citep{2011A&A...532A..74B}. This leads to a ``cleaner'' sample in which galaxies of each morphological type are less dispersed in $NUV-r$: for example, Sc galaxies shown as cyan symbols in \fg\ref{NUV-r-face-on} mostly have $NUV-r\lesssim3.7$ in the right panel of \fg\ref{NUV-r-face-on}, whereas they spread above $NUV-r=5$ in the left panel. We therefore choose to focus on galaxies with {\tt Incl-Elong} $\leq 2$ hereafter.

The right panel of \fg\ref{NUV-r-face-on} better shows a remarkable coherence of the morphological sequence along the color-magnitude diagram: types occupy specific ranges of color and magnitudes, with these ranges being contiguous and overlapping. This can be better seen using the mean colors and magnitude of each morphological type with {\tt Incl-Elong} $\leq2$, plotted in \fg\ref{mean-NUV-r-face-on}. All ellipticals and S0$^-$, S0 and S0$^+$ lenticulars populate the Red Sequence.  Note also the presence of dwarf ellipticals (dE) at absolute magnitude $M_r\sim-18$, as faint as Sm galaxies, whereas they have $NUV-r\sim 5$, as red as S0a galaxies.
The Green Valley is dominated by galaxies of the intermediate type between lenticulars and spirals, that is S0a, as well as of the earliest spiral type, Sa, with fractions of 57.5\% and 72.4\% resp{.} of these types in the $NUV-r$ interval $[3.77, 5.71]$ (see right panel of \fg\ref{NUV-r-face-on} and \fg\ref{mean-NUV-r-face-on}). The S0a galaxies have a morphology similar to lenticulars in terms of bulge-to-total ratio and old stellar population, with some additional evidence for tenuous and blue portions of spiral arms, hence star formation, within the disk. Moving along the color-magnitude diagram, \fgs \ref{NUV-r-face-on} and \ref{mean-NUV-r-face-on} show that the Blue Cloud is populated by immediately later spiral types, that is from Sab all the way to Sm, followed by the Magellanic Irregulars (Im; referred to as irregulars hereafter), each type being bluer and fainter as one advances along the Hubble sequence.

\begin{figure*}
\includegraphics[width=\columnwidth]{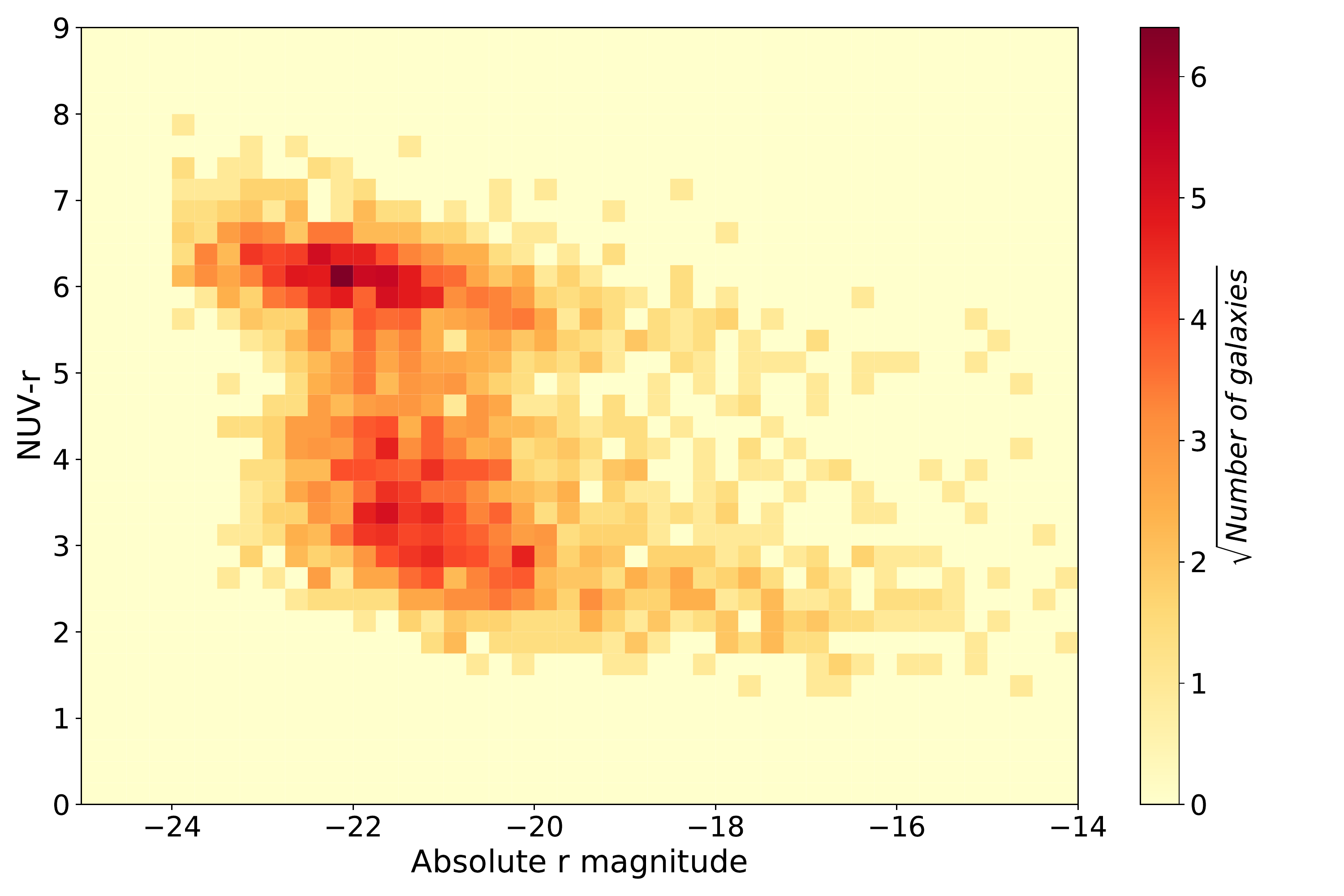}
\includegraphics[width=\columnwidth]{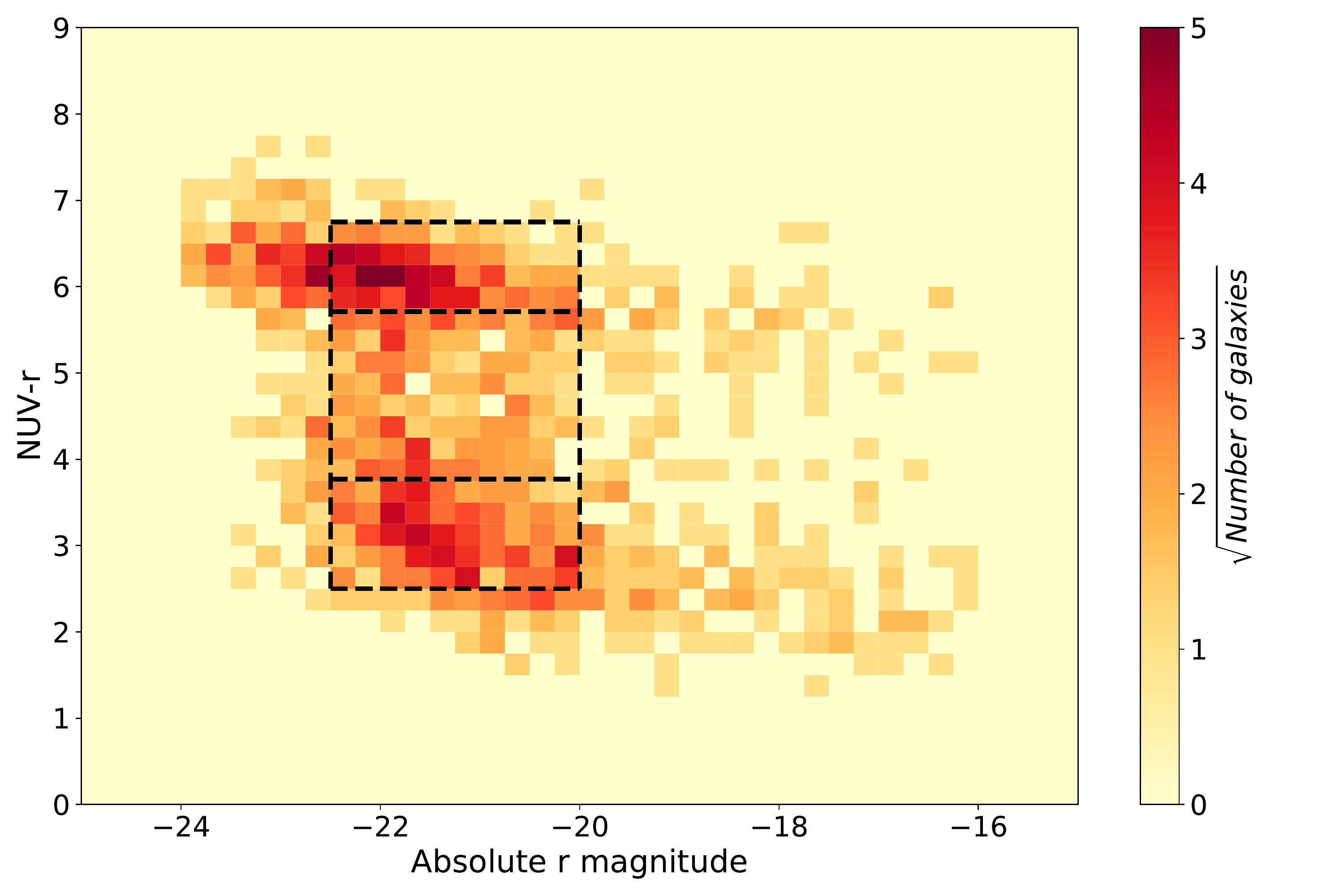}
\caption{MorCat color-magnitude diagrams in square cells of color and magnitude of length 0.25. To retrieve a representative picture of nearby galaxies, galaxies were randomly drawn from the MorCat $\cap$ GALEX sample in 1 magnitude intervals of $M_r$, in order to match the $M_r$ distribution of MorCat. \textbf{Left panel} shows the distribution of all MorCat galaxies, whereas the \textbf{right panel} is restricted to galaxies with an aspect ratio $\geq0.61$, in order to exclude highly inclined and edge-on disks. Both graphs display the bimodal distribution of galaxy population with the two density peaks of the Red Sequence and the Blue Cloud, separated by the Green Valley or Green Valley. The dashed lines in the right panel delimit the regions of the Red Sequence, the Green Valley, and the Blue Cloud that are respectively used in \sct \ref{result-fading} to illustrate the stretching of the $NUV-r$ that generates the Green Valley, and lead us to rename it Green Plain.}
\label{density_NUV}
\end{figure*}

Altogether, the Hubble sequence displays an ``S'' shape in the color-magnitude diagram. Both panels of \fg\ref{NUV-r-face-on} show that moving back up the Hubble Sequence starting from the irregulars, the ``S'' shape results from a strong increase in $M_r$ over 8 magnitudes across the Blue Cloud, with a progressive but weak reddening from $NUV-r=2$ to $3.5$. Then there is an abrupt and vertical change of $NUV-r$ color between 4 and 5.5 across the Green Valley.  The third and final trend is again a strong luminosity increase by 4 magnitudes along the Red Sequence, when going from S0 to ellipticals, correlated to a weak reddening. Moreover, lenticulars have redder colors (in the interval [5.5, 6.5]) than early spiral types (Sa-Sab), but they are spread over the same magnitude range with $M_r \in [-23.0, -20.0]$, spanning the Green Valley; on the other hand, ellipticals have slightly redder colors in the interval [5.75, 7] and significantly brighter magnitudes ($M_r \in [-24, -21]$) compared to lenticulars and early spirals. There is, however, a lower luminosity tail in lenticular galaxies than in the Sa type dominating the Green Valley. Indeed, the mean magnitude of S0a, S$0^+$, S$0^-$ galaxies are 0.73, 0.58 and 0.57 magnitude fainter respectively than for Sa galaxies. \cite{1990ApJ...348...57V} already noticed the lower-luminosity of S0 galaxies and inferred that only a fraction of lenticular galaxies, the most luminous ones, could be seen as an intermediate state between E and Sa galaxies, and that the S0 class was likely to group objects with different evolution histories. See \citet{Barway_2007_S0_length_correlations}, \citet{Barway_2009_S0_relations}, and \citet{Barway_2013_S0_colors} for further characterization of faint and bright S0 galaxies.

Importantly, \fg\ref{mean-NUV-r-face-on} shows that the S shape of the color-magnitude diagram could be parameterized with the mean Hubble type: displacements along the diagram can be defined as a monotonous function of the Hubble type. This smooth and continuous transition between morphological properties and types of galaxies was already shown by \cite{2007AJ....134.1508V}, but the use of only optical colors prevents this analysis from detecting the Green Valley (see \sct\ref{NUV_detect_GV}). 
Hereafter, we refer to the \textbf{knee} of the Green Valley as the region of the graph where the most luminous and bluest Sab, Sb, Sbc, Sc, and Scd galaxies reside, that is for $M_r\in[-23,-21]$ and $NUV-r\in[2.3,3.5]$, as it is the pivotal region of the S shape color-magnitude sequence between variations dominated by a change in magnitude (Blue Cloud) and by a change in color (Green Valley; see also \sct \ref{result-fading}).\\

Moreover, the equality we find between the Red Sequence and the early-type galaxies of the Hubble Sequence (E+S0) contradicts the statement made by \cite{2014SerAJ.189....1S} that the Red Sequence can not be equated to early-type galaxies because it includes almost all of Sa and Sb galaxies. However, he refers to the results of \cite{2007ApJS..173..185G}, in which the considered Sa-Sb galaxies, from the GALEX Ultraviolet Atlas of Nearby Galaxies, do have intermediate $FUV-K$ colors, in agreement with the results obtained here for EFIGI sample. \cite{2006MNRAS.373.1389C} also show that Green Valley galaxies are predominantly of Sa and Sab types, using $U-B$ colors of RC3 galaxies, however within the limitations of using only optical colors (see \sct\ref{NUV_detect_GV}). \cite{2018MNRAS.476...12B} also describes the Green Valley population as dominated by early-type spirals.

\subsubsection{From EFIGI to MorCat: The representative galaxy density \label{from_efigi_to_morcat}}

\begin{figure}
\includegraphics[width=\columnwidth]{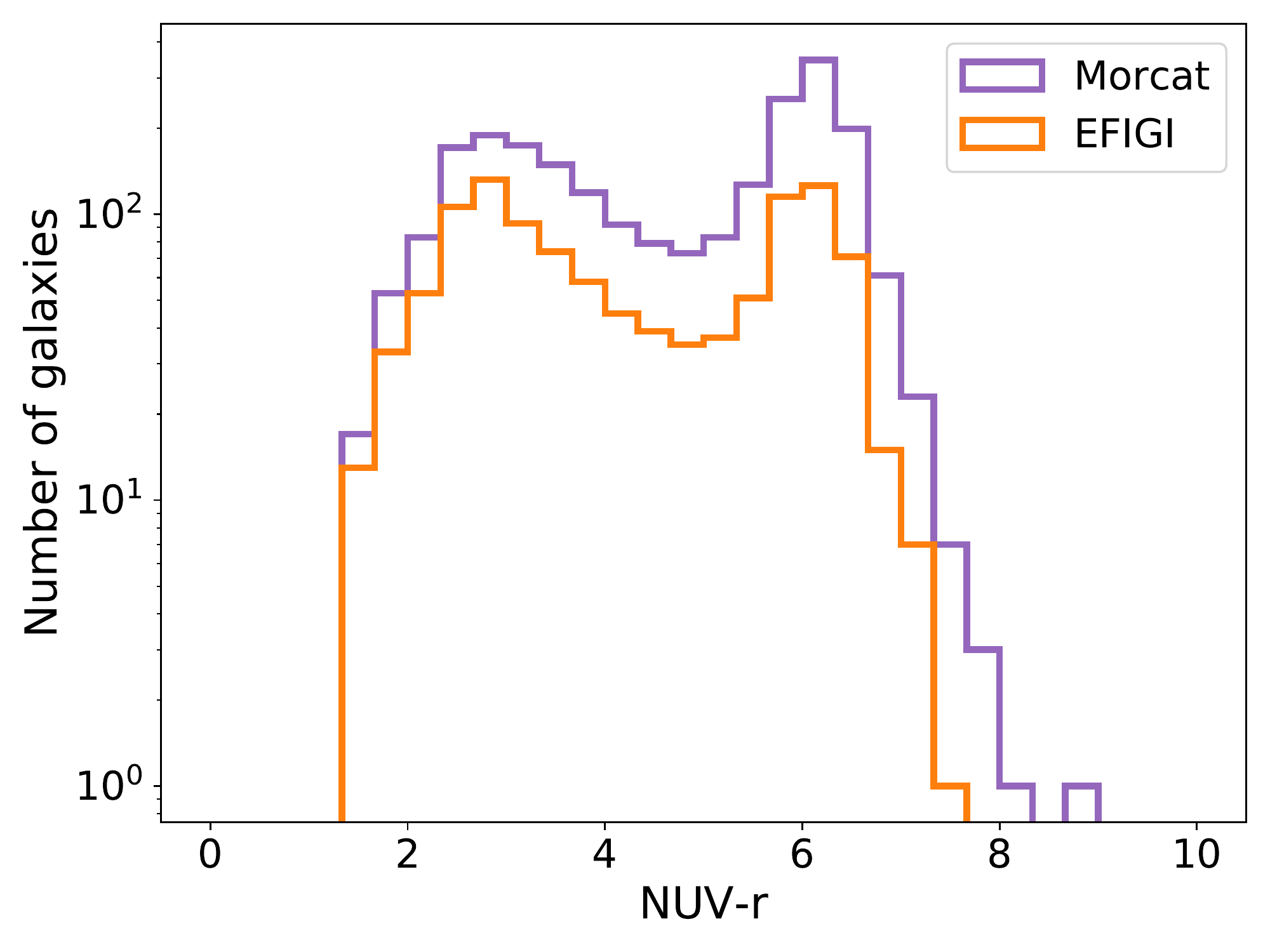}
\caption{$NUV-r$ color distribution for all MorCat $\cap$ GALEX galaxies with aspect ratio $\geq0.61$ (purple), and for all EFIGI $\cap$ GALEX with {\tt Incl-Elong} $\leq2$ (orange). One can see that the galaxy bimodality is present in both samples with two peaks of density at $NUV-r \sim 2.5$ for the Blue Cloud, and $\sim 6.0$ for the Red sequence, while the Green Valley corresponds to the under-density for $NUV-r \in 3.7, 5.7]$.}
\label{NUV-r_morcat_efigi}
\end{figure}

In order to evaluate how the relative galaxy densities seen in the Red Sequence, Green Valley and Blue Cloud of \fg\ref{mean-NUV-r-face-on} are affected by the type selection biases of the EFIGI morphological sample, we use the magnitude-limited MorCat sample (see \sct \ref{data}). One must correct for the fact that only 3149 of the $20,126$ MorCat galaxies have a $NUV$ magnitude in GALEX, and that the fraction of galaxies with $NUV$ data varies with absolute $r$ magnitude ($M_r$). We thus compute the distribution of $M_r$ for both MorCat and MorCat $\cap$ GALEX and for each bin of $M_r$ (of width 1, between -24 and -14) we randomly draw galaxies from MorCat $\cap$ GALEX to create a subsample with a similar $M_r$ distribution as the full MorCat sample (based on a constant fraction of galaxies with $NUV$ equal to that in the bin where it is the lowest). The resulting $NUV-r$ distribution is shown in the left panel of \fg\ref{density_NUV}. The three major features of the EFIGI S shape color-magnitude diagram (\fg\ref{NUV-r-face-on}) remain present when using the magnitude-limited MorCat sample, with similar characteristics: the Red Sequence has higher density than the Blue Cloud, and they are separated by an under-dense region, the Green Valley. 

As the dispersion in the $NUV-r$ color for the EFIGI data can be reduced using the {\tt Incl-Elong} attribute (see \fg\ref{NUV-r-face-on}), we use the aspect ratio of the isophotal profiles of MorCat galaxies measured by SExtractor to identify highly inclined disk galaxies. To this end, we perform a polynomial fit of the aspect ratio as a function of the {\tt Incl-Elong} attribute for EFIGI galaxies. We obtain that  {\tt Incl-Elong} $\le2$ can be approximated by an aspect ratio $\ge$ 0.61, and show in the right panel of \fg\ref{density_NUV} the resulting color-magnitude distribution for the corresponding MorCat subsample. The lower density of the Green Valley compared to the two dense regions appears accentuated.

\fg\ref{NUV-r_morcat_efigi} shows the $NUV-r$ histograms for the right panels of \fgs\ref{NUV-r-face-on} and \ref{density_NUV} (restricted to weakly elongated or weakly inclined galaxies): the peaks of the Red Sequence and Blue Cloud are located at identical $NUV-r$ colors in both EFIGI and MorCat, hence are not sensitive to the EFIGI type selection effects. If one considers only bins with more than $\sim$ 10 galaxies, they are represented in similar proportions for MorCat and EFIGI for $NUV-r \in$ $[3.5, 6.0]$. However, the bluest EFIGI galaxies, with $NUV-r \lesssim 3.5$ (that is spiral types later than Sb) are over-represented compared to MorCat. This is due to the fact that EFIGI was created so as to represent densely every morphological type, including the rarer late spirals and irregulars.

\subsubsection{Abrupt reddening across the Green Plain
\label{result-fading}}

In the right panel of \fg\ref{density_NUV}, we delineate as dashed lines regions of the Red Sequence, Green Valley, and Blue Cloud in the common $M_r\in[-22.5,-20]$ interval: comparison of their $NUV-r$ extent and galaxy densities shows that the Green Valley is characterized by a significant width and low and nearly flat density over $\sim2$ magnitudes in $NUV-r$. We therefore rename it the ``Green Plain'' hereafter, and further characterize this region and justify this name change.

Let us scrutinize the mean colors of morphological types across the Green Plain in  \fg\ref{mean-NUV-r-face-on}. In the approximate center of the Green Plain lie the S0a and Sa types (from top to bottom; order also used hereafter), separated by only $0.26$ in $NUV-r$. Immediately above and below are types S0$^+$ and Sab, respectively separated by $0.90$ and $0.78$ from the S0$^+$ and Sab respectively. These wide steps cause the large color extent of the Green Plain. Then the next color steps to the outside types shrink to $0.16$ and $0.40$ for types S0 and Sb, respectively; the average of these steps is $0.28$, comparable to the step between the central types. If one scales down both stretched intervals S0$^+$-S0a and Sa-Sab to $0.28$, the full $NUV-r=[3.77-5.71]$ interval from the mean Sab to the mean S0$^+$ color shrinks from $1.94$ down to $0.84$, that is by a factor of $2.36$.

The mean colors $5.71$ and $3.77$ of types S0$^+$ and Sab, respectively, are those used to split regions of the Red Sequence ($NUV-r\in[5.71, 6.75]$), Green Plain ($NUV-r\in[3.77, 5.71]$) and Blue Cloud ($NUV-r\in[2.5, 3.77]$), using the dashed lines in the right panel of \fg\ref{density_NUV}. The number of galaxies within these color intervals and the common $M_r\in[-22.5, -20]$ interval are 377, 174 and 317 respectively. As such, the density of galaxies per color-magnitude cells is smaller in the Green Plain than in either the Blue Cloud or the Red Sequence by a factor of 1.8 or 2.2 respectively. Squeezing of the $NUV-r$ interval by the $2.36$ factor proposed above would then increase the number density of MorCat Green Plain galaxies to levels similar or higher than in the Red Sequence or the Blue Cloud. With this higher density of galaxies in the color-magnitude graph, this region would therefore not appear anymore as a Plain, but as a ``bridge'' connecting the Blue Cloud and the Red Sequence, with a region of similar galaxy density as both regions.

\begin{figure*}
\includegraphics[width=\columnwidth]{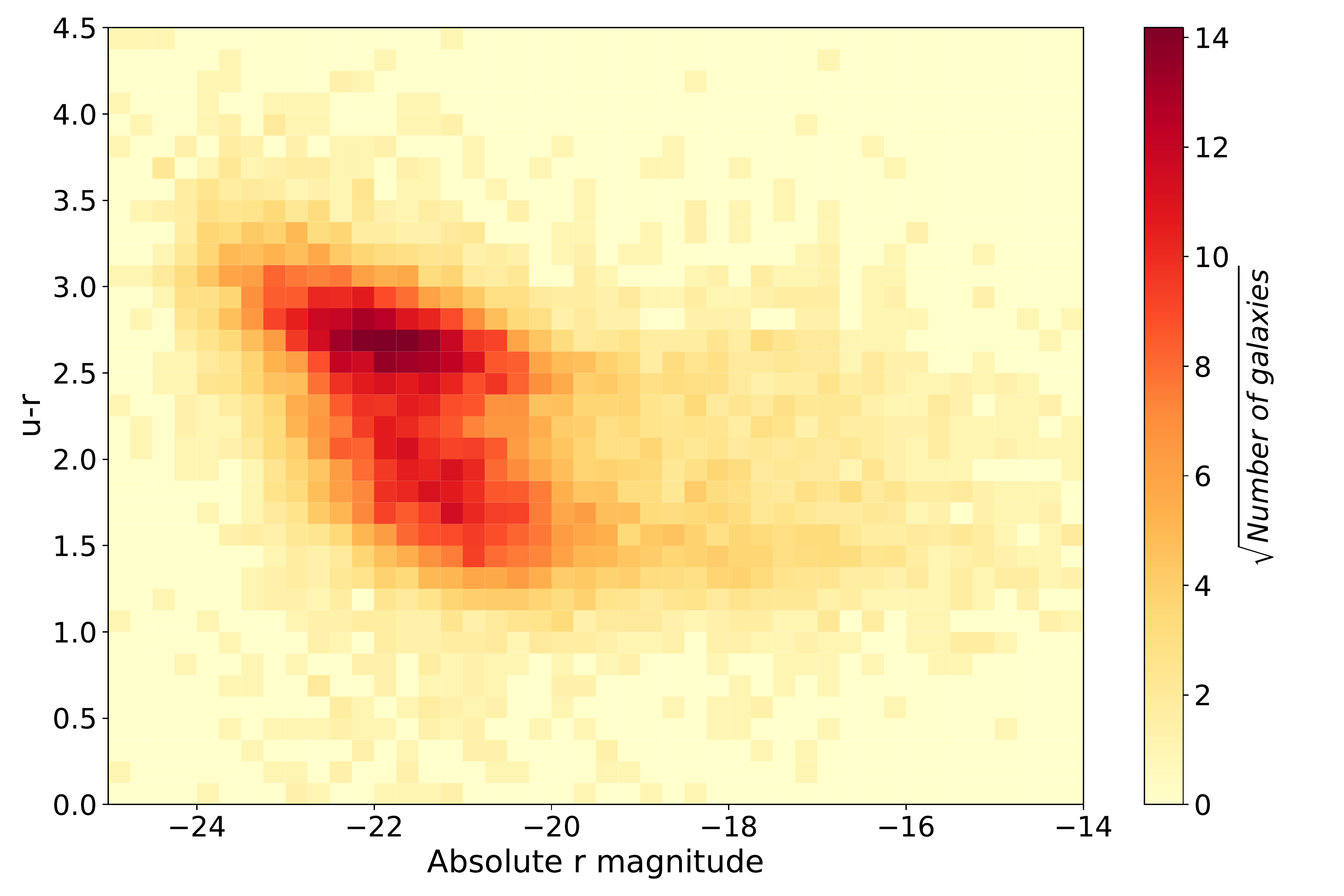}
\includegraphics[width=\columnwidth]{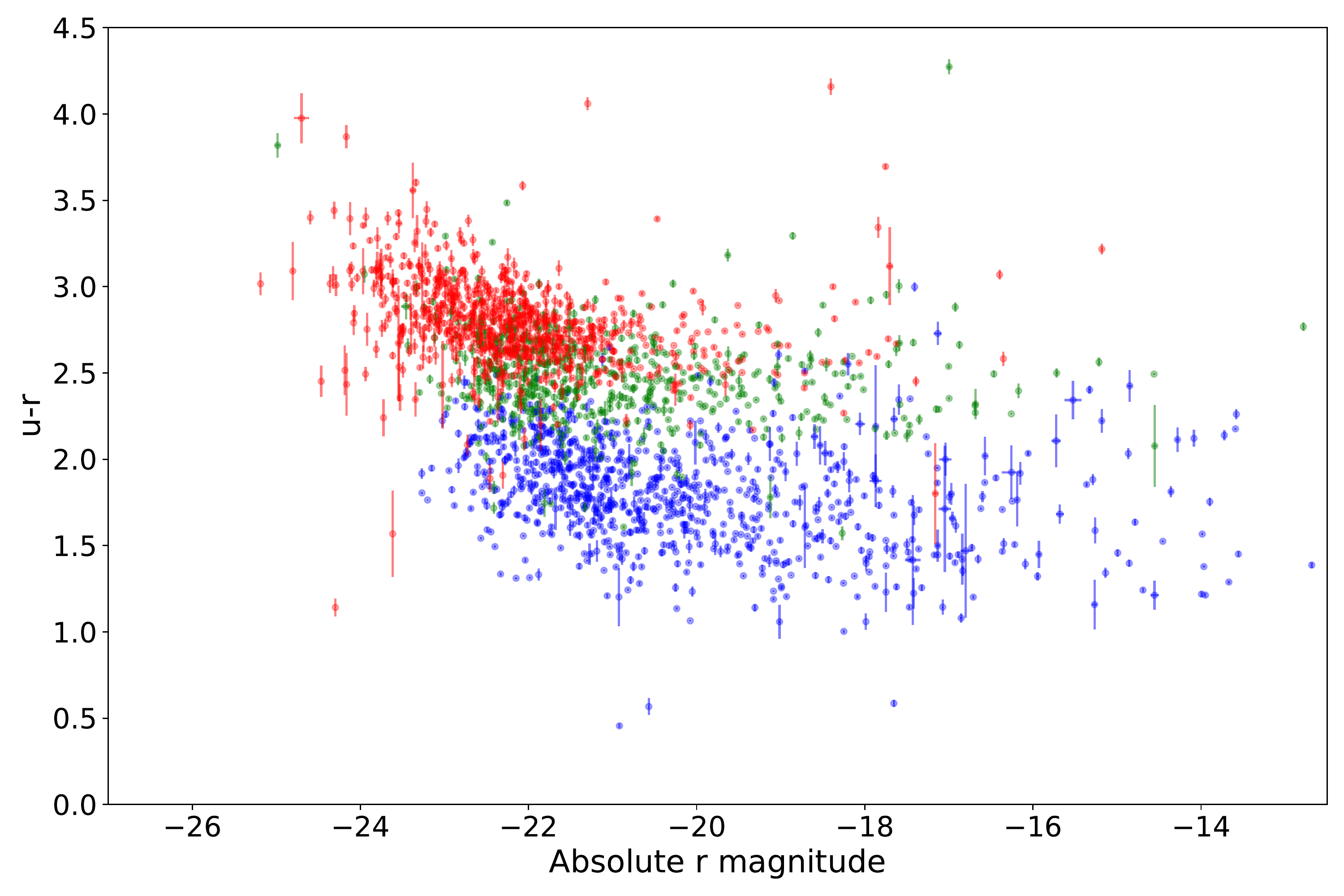}
\caption{Color-magnitude diagram for all MorCat galaxies with aspect ratio $\geq0.61$ using the absolute optical color $u-r$ and the absolute magnitude in the $r$ band, from the SExtractor photometry. \textbf{Left:} number of galaxies in cells of both color and magnitude. Each cell has a width of 0.25 and a height of 0.125 dex. \textbf{Right:} Individual galaxies are plotted in blue for the Blue Cloud ($NUV-r < 3.77$), green for the Green Plain ($NUV-r \in [3.77, 5.71]$) and red in the Red Sequence ($NUV-r > 5.71$). The dichotomy between the Red Sequence and the Blue Cloud is still present but the Green Plain seen in $NUV-r$ (\fg\ref{NUV-r-face-on}) has disappeared, being superimposed on the Red Sequence.}
\label{u-r_density}
\end{figure*}

\begin{figure*}
\centering
    \includegraphics[width=\columnwidth]{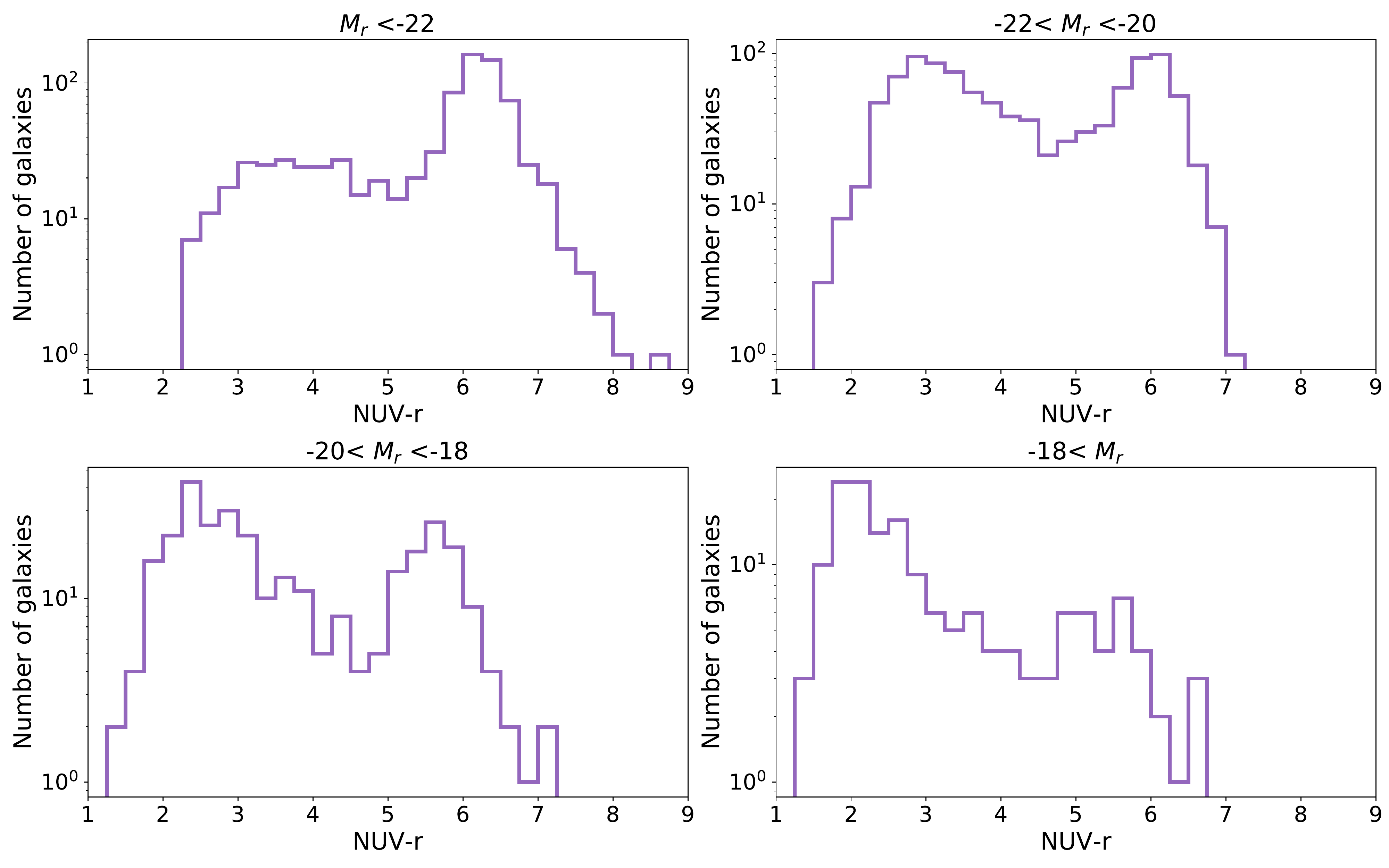}
    \includegraphics[width=\columnwidth]{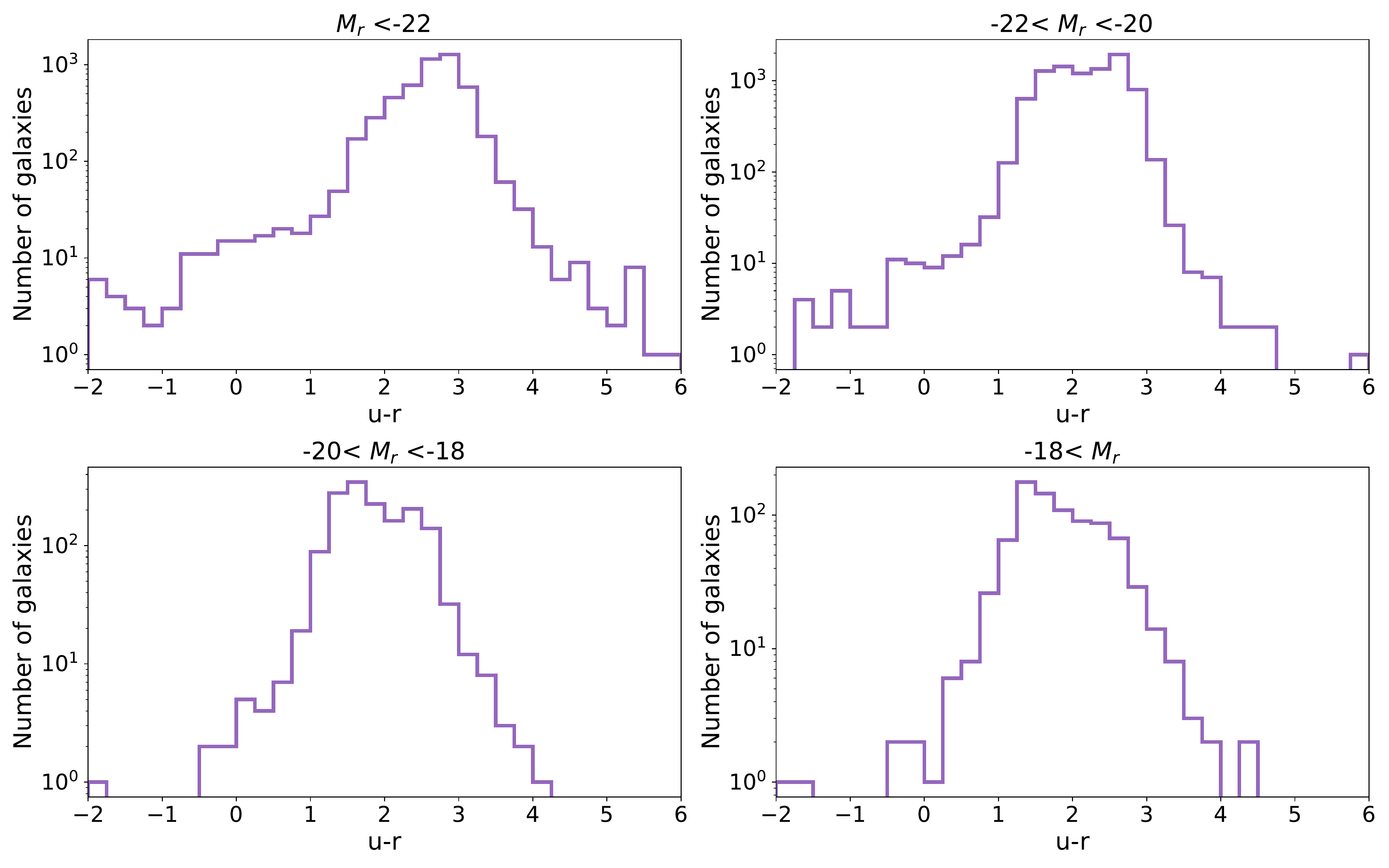}
    \caption{Comparison of the $NUV-r$ and $u-r$ absolute color distributions for all MorCat $\cap$ GALEX galaxies with aspect ratio $\geq0.61$, in bins of absolute r band magnitude, using SExtractor photometry data. \textbf{Left:} The distribution of $NUV-r$ colors shows the bimodality caused by the presence of the Green Plain for all absolute magnitude intervals. \textbf{Right:} The Green Plain, hence the bimodality disappear in the distributions of $u-r$ colors for all $r$ intervals.}
    \label{color_distrib_per_mag}
\end{figure*}

\begin{figure}
\includegraphics[width=\columnwidth]{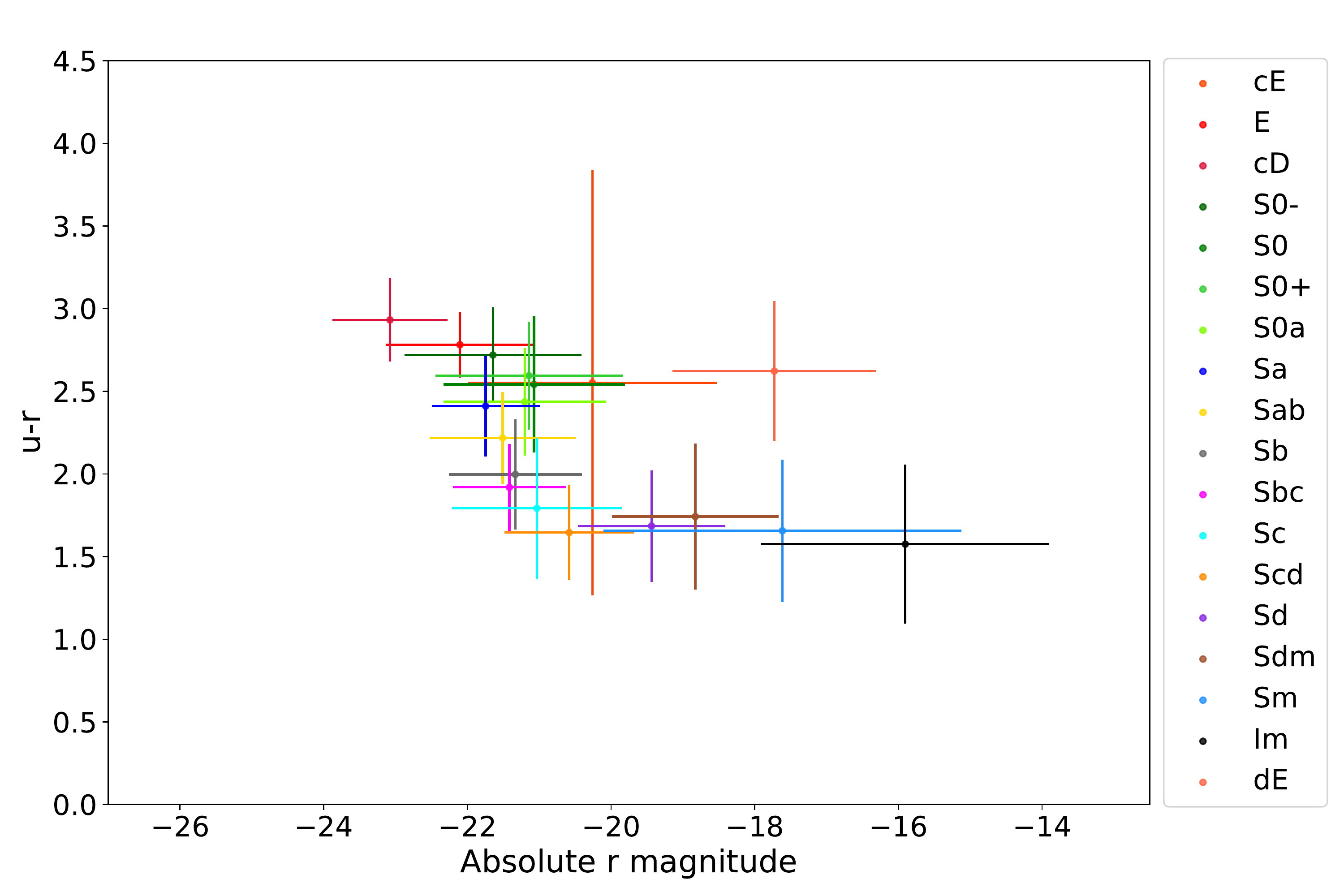}
\caption{Mean position of each Hubble Type in the color-magnitude diagram, using the absolute $u-r$ color for all EFIGI galaxies with {\tt Incl-Elong} $\leq 2$. 
One sees a similar ``S'' sequence as in $NUV-r$ color (see \fg\ref{mean-NUV-r-face-on}), but without the additional spacing of S0$^+$-S0a and Sa-Sab types that generates the Green Plain. The central point with very large error bars correspond to the rare cE types, whereas dE are located at $(-17.8,2.6)$, away from the ``S'' sequence.}
\label{u-r_mean_Ht}
\end{figure}

This rough calculation illustrates that the Green Plain is caused by a strong reddening rather than by a decrease in the number density of the corresponding morphological types. We show in turn in \sct\ref{results_ssfr_m} that this reddening results from a strong fading in the specific star formation rate of the corresponding Hubble types. The larger $NUV-r$ color range seen in \fg\ref{mean-NUV-r-face-on} for the Green Plain compared to the other 2 sequences could also be contributed to by a larger dispersion in color for the Green Plain morphological types that for other types. At last, the fact that S0a, and even more so, Sa and Sab galaxies may have a strong dust content visible in the optical, as shown in EFIGI \citep{2011A&A...532A..75D} could introduce an additional dispersion in the color to that due only to the star formation rate. The EFIGI classification shows a high dust content for Sb types \citep{2011A&A...532A..75D}, which may also explain the stronger dispersion in color of the morphological types at the bright end of the Blue Cloud.

\subsubsection{Ultraviolet band for detecting the Green Plain    \label{NUV_detect_GV}}

Optical-optical colors are often used to study the color-magnitude bimodality of galaxies \citep{2001AJ....122.1861S,2004ApJ...600..681B,2014MNRAS.440..889S}, but \cite{2014SerAJ.189....1S} warns that the Green Plain can only be seen using an UV to optical color. For direct comparison of both approaches with a single sample, we examine the SDSS $u-r$ colors for MorCat, that are available for all galaxies, and compare with the colors based on the $NUV$ data from GALEX, that are available for only $\sim1/7$ of MorCat. 

The left panel of \fg\ref{u-r_density} shows that for MorCat galaxies with aspect ratio $\geq0.61$, there is only one dense cloud of points extending from $u-r \sim 3$ to $u-r \sim 2.5$, with a diffuse tail of bluer galaxies down to $u-r \sim 1.5$. Both features extend to faint galaxies down to $M_r = -18$. If a red over-density (with $u-r \sim 3$ to 2.5) remains visible, the bimodality with an excess of blue galaxies is not as clear. Indeed, the right panel of \fg\ref{u-r_density} shows the color magnitude diagram in $u-r$, with the points color-coded according to their $NUV-r$ color and location within the Red Sequence (red), Green Plain (green), and Blue Cloud (blue), using the limits used in \sct \ref{result-fading}. One can see that in $u-r$, Green Plain galaxies have a comparable range of $u-r$ colors to Red Sequence galaxies, leaving no room in the $u-r$ versus $r$ color-magnitude diagram for a transition region between the locus of quiescent and star-forming galaxies. This is due to the fact that the blue optical bands are only weakly sensitive to star formation, therefore the emission from old stars may dominate the $u-r$ color, in particular in the case of a significant bulge. Indeed the total $NUV-r$ range of the full color-magnitude diagram is $\sim5$ magnitudes (\fg\ref{NUV-r-face-on}) compared to $\sim1.5$ magnitude in $u-r$ (\fg\ref{u-r_density}). Moreover, the significant bulges in the Green Plain also weight the optical colors toward being redder (see \sct \ref{bulge}). Therefore, we confirm that complementing optical data with a UV band is a requirement to study the Green Plain.

We then compare in \fg\ref{color_distrib_per_mag} the $NUV-r$ (left panel) and $u-r$ histograms (right panel) of MorCat galaxies with aspect ratio $\geq0.61$as a function of absolute magnitude interval. Left panel of \fg\ref{color_distrib_per_mag} shows two peaks in the $NUV-r$ distribution, one around $NUV-r \sim 6$ and another below $NUV-r = 3$, corresponding respectively to the Red Sequence and the Blue Cloud identified in \fg\ref{density_NUV}. The $NUV-r$ ``red'' peak appears at all $M_r$ magnitudes, whereas the  ``blue'' peak appears at $M_r>-20$. The peaks of both features are shifted in $M_r$ for different intervals of $M_r$ because the two sequences are inclined in the $NUV-r$ versus $r$ color-magnitude diagram (\fg\ref{density_NUV}). In contrast, the right panel of \fg\ref{color_distrib_per_mag} shows only one peak in the $u-r$ histograms, near $u-r=2.8$ for both bright magnitude intervals, and around $u-r=1.5$ for both fainter magnitude intervals (these peak values are in agreement with the results of \citealt{2001AJ....122.1861S} and \citealt{2004ApJ...600..681B}).

The disappearance of the Green Plain in $u-r$ color is also illustrated in \fg\ref{u-r_mean_Ht}, which shows the mean position of Hubble types in the $u-r$ versus $r$ absolute color-magnitude diagram. A similar S shape sequence as in $NUV-r$ is present (\fg\ref{NUV-r-face-on}), spanning all morphological types. But the additional color spacings between morphological types S0$^+$-S0a and Sa-Sab in \fg\ref{mean-NUV-r-face-on}, that contribute to generate the Green Plain in $NUV-r$ are not present in $u-r$ (see also \sct \ref{result-fading}).

\subsection{Specific star formation rates (sSFR) and stellar masses \label{results_ssfr_m}}

\begin{figure}
\includegraphics[width=\columnwidth]{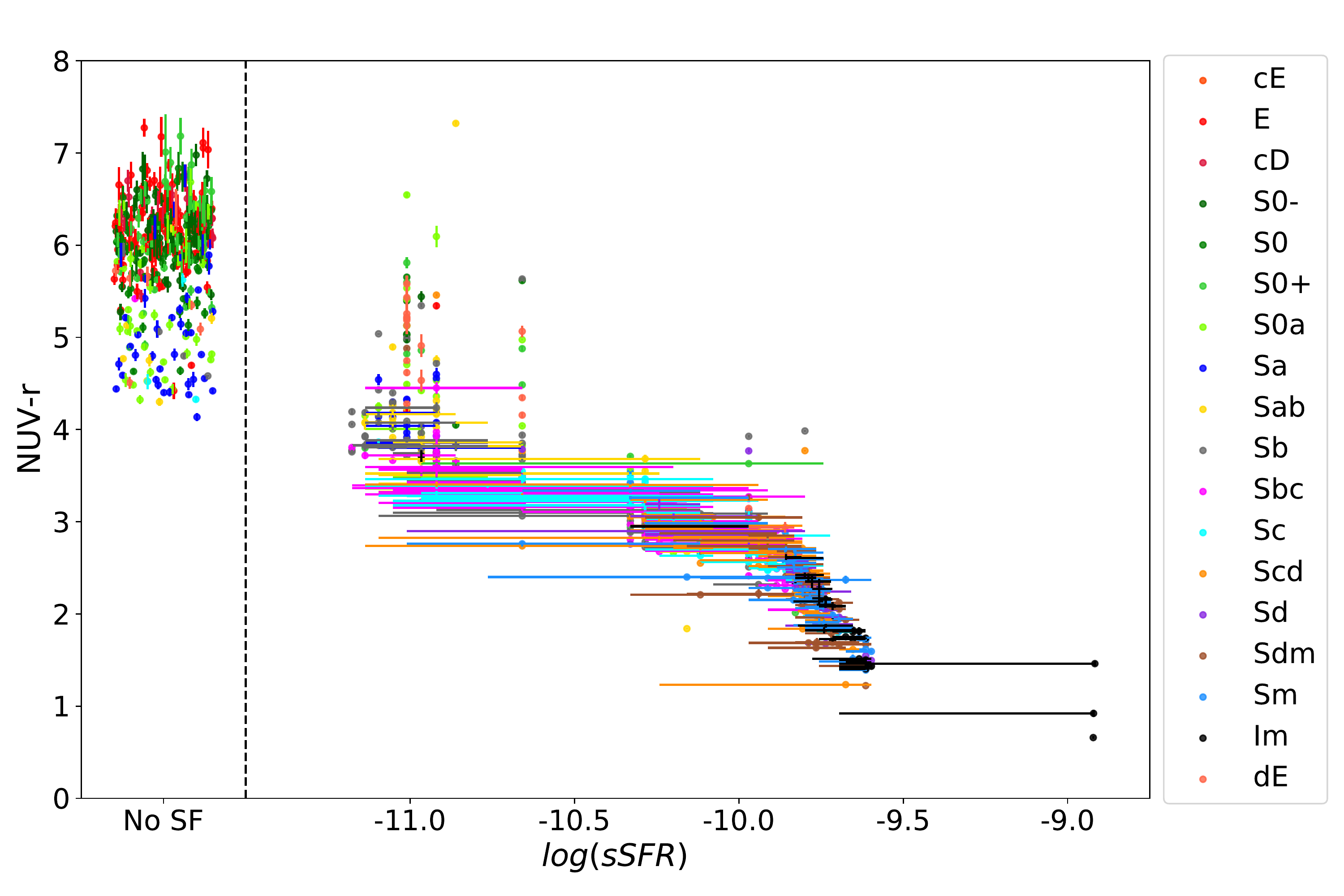}
\caption{Relation between the $NUV-r$ color and the sSFR inferred by SED model-fitting for EFIGI $\cap$ GALEX galaxies with {\tt Incl-Elong} $\leq 2$, showing the correlation between these two parameters, with the bluest galaxies being the most star forming, while red sequence galaxies mostly show no star formation. Intermediate colors are prone to larger uncertainties.}
\label{ssfr_color}
\end{figure}

\begin{figure}
\includegraphics[width=\columnwidth]{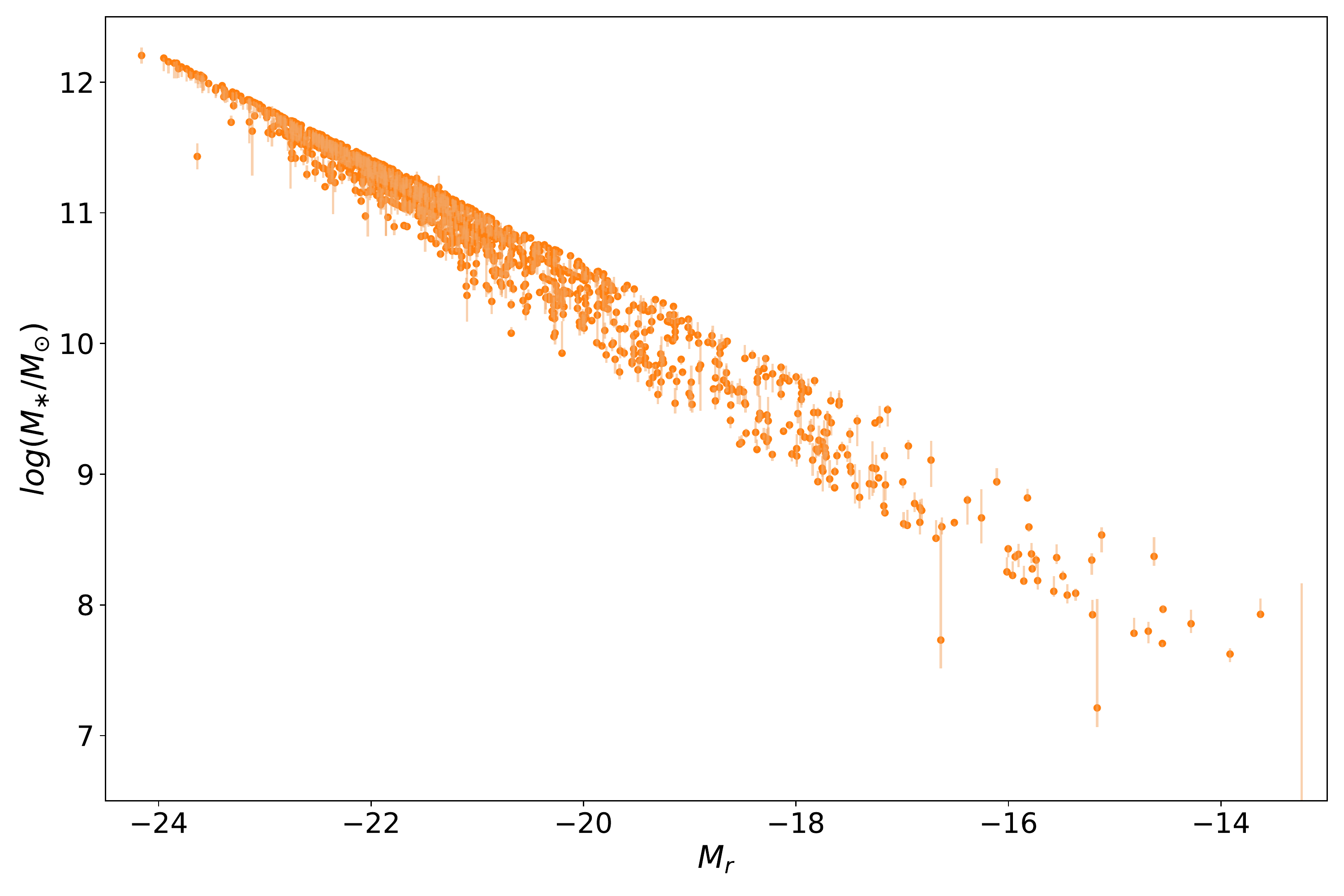}
\caption{Relation between the absolute magnitude in the r band $M_r$ and the stellar mass $M_\ast$ inferred by SED model-fitting for all EFIGI $\cap$ GALEX galaxies with {\tt Incl-Elong} $\leq 2$, showing a clear anticorrelation between these two parameters.}
\label{mass_mag}
\end{figure}

\begin{figure*}
    \includegraphics[width=\columnwidth]{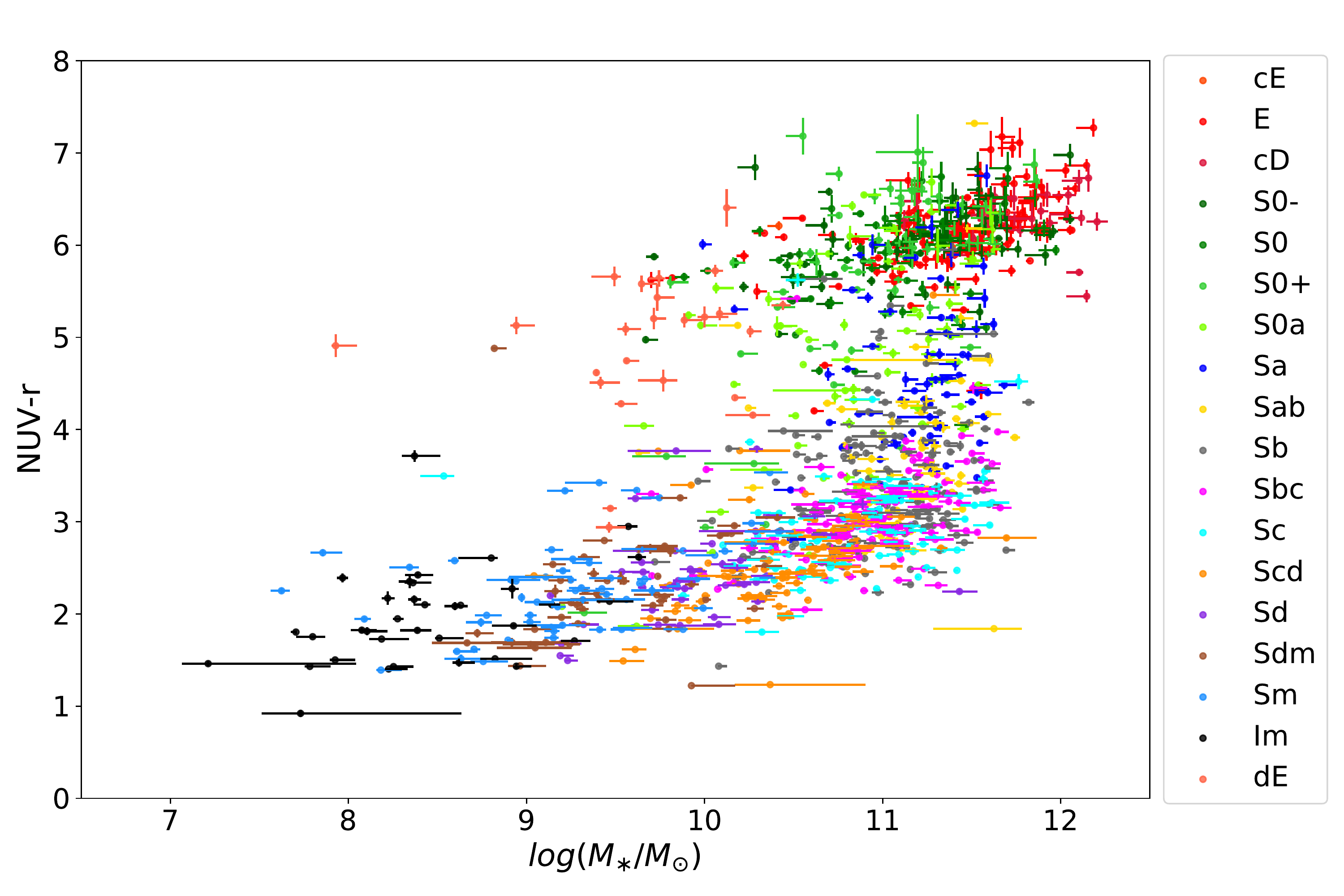}
    \includegraphics[width=\columnwidth]{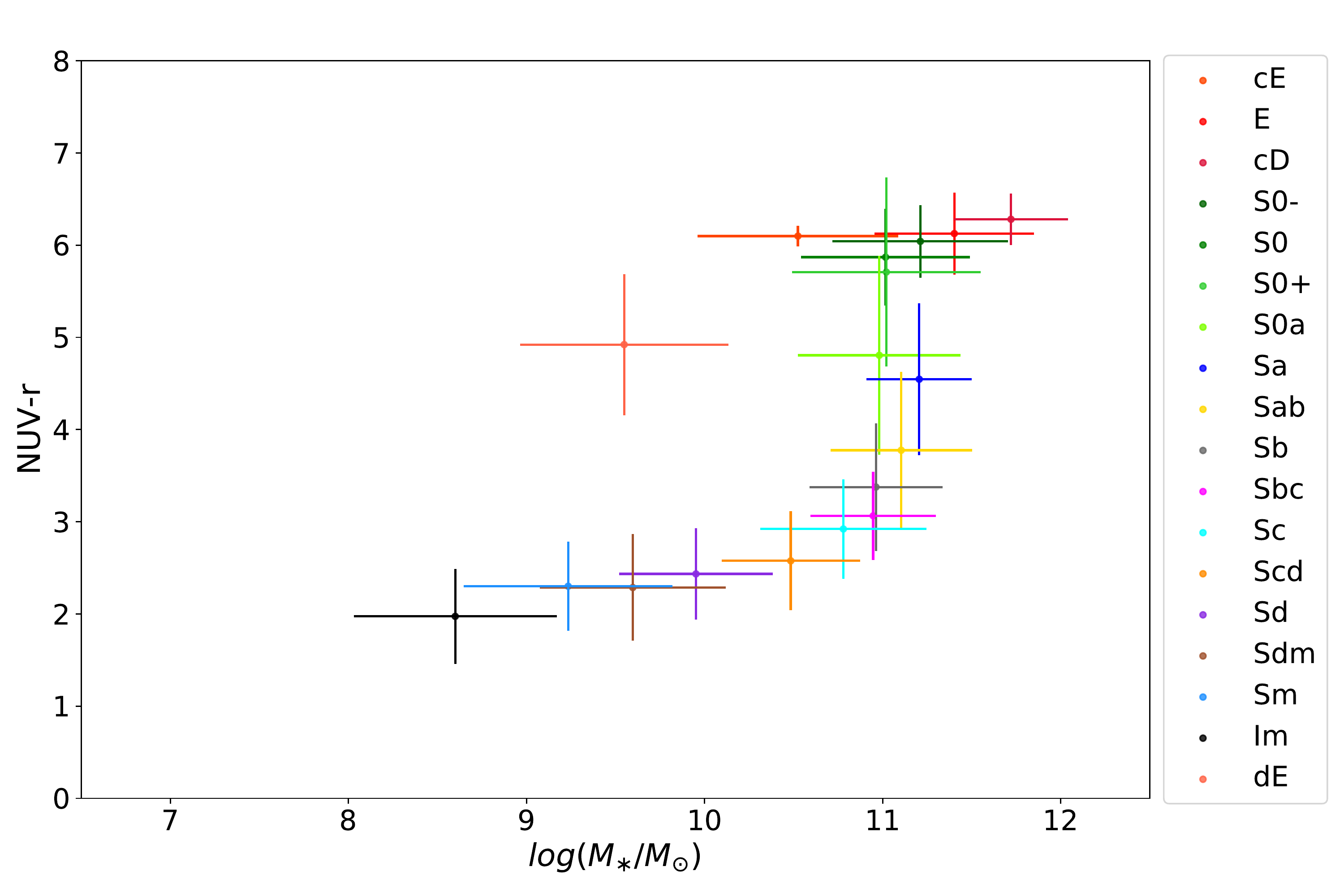}
    \caption{Color-mass diagram. \textbf{Left:} Color-mass diagram for all EFIGI $\cap$ GALEX galaxies with {\tt Incl-Elong} $\leq 2$. The color of the points indicates their Hubble Type as classified in the EFIGI morphological catalog. The distribution of galaxies exhibits the bimodality between the ellipticals and lenticulars in the upper region (the Red Sequence), and spirals of type Sab and later in the lower region (the Blue Cloud). A lower density region (the Green Plain) connects both structures, with galaxies of types S0a and Sa. \textbf{Right:} Mean absolute $NUV-r$ color versus mean stellar mass for each Hubble type for EFIGI $\cap$ GALEX galaxies with {\tt Incl-Elong} $\leq 2$; the error bars represent the \rms deviation within each type. The Hubble sequence draws an ``S'' shape that could be parameterized monotonously by the mean Hubble type, with the exception of dE at $(9.6,4.9)$.}
    \label{color_mass}
\end{figure*}

The interest of the $NUV-r$ color is that it can be used as a proxy for the specific star formation rate (sSFR) of galaxies \citep{2007ApJS..173..267S}. Indeed we show in \fg\ref{ssfr_color} the correlation between the sSFR inferred from SED model-fitting with ZPEG (see \sct \ref{zpeg}) and the $NUV-r$ color.  Indeed, blue galaxies with $NUV-r < 3$ are the most star-forming, with $\log(\mathrm{sSFR}) > -10$, whereas most red galaxies with $NUV-r > 5$ show no star formation (sSFR = 0), they are completely quiescent. Between these two extrema, there is a correlation between color and sSFR but it undergoes large uncertainties. This is partly due to the fact that there are only 10 ZPEG scenarios (see \fg\ref{zpeg}, which lead to a finite number of templates in terms of types and ages at each plausible redshift, therefore a limited number of values can be taken by the sSFR. The discreteness in the scenarios can be seen on the plot as vertical alignments of points. In the color range [3, 4] which corresponds to the reddest part of the Blue Cloud, the uncertainty in the sSFR is large because of the large step between the discrete values of sSFR for the subsequent Sab and Sb templates. Moreover, there is in the top-left of the plot an overwhelming majority of red galaxies that are completely quiescent while galaxies in the Green Plain ($NUV-r \in [3.75,5.75]$) have either no star formation at all, or the lowest values available $\log(\mathrm{sSFR[yr^{-1}]}) \sim -11.0$; these values of sSFR also include some galaxies from both the bluest part of the Red Sequence and the reddest part of the Blue Cloud.

Another parameter inferred by the SED model-fitting which is useful to study galaxy evolution is the stellar mass. \fg\ref{mass_mag} shows that the stellar mass of EFIGI galaxies is strongly correlated to the absolute magnitude in the $r$ band (we remind that in the ZPEG fits used in the present section, $gri$ SourcExtractor++ photometry is used for all objects, complemented by the available $NUV$ GALEX photometry for $1/2.4$ of the objects, see \scts \ref{sec:galex} and \ref{sec:absmag}.)
Consequently, all the diagrams shown in the previous section have their equivalent in the sSFR vs $M_\ast$ plane. For clarity, and because our derived values of sSFR are discrete and related to the limited number of PEGASE.2 scenarios used in ZPEG, we continue to use $NUV-r$ color in the rest of the article. We however replace absolute $r$ magnitude with the galaxy stellar mass $M_\ast$.

\fg\ref{color_mass} shows in the left panel all EFIGI galaxies with {\tt Incl-Elong} $\leq 2$ in the color-mass diagram, whereas the right panel show the mean stellar mass and color per morphological type: both panels are almost mirror images of the right panel of \fg\ref{NUV-r-face-on} and of \fg\ref{mean-NUV-r-face-on} respectively.

One noticeable difference is the larger error bars along the x-axis representing stellar mass (in the left panel), because masses inferred from the SED model-fitting are subject to larger uncertainties than those for the absolute magnitudes from which they were derived. The same observations that were made about the specific locations of each Hubble type, and how the Hubble sequence traces an S shape in \fg\ref{NUV-r-face-on} can be made from \fg\ref{color_mass}. There is also a correlation between the reddening of a galaxy and its mass growth in both the Blue Cloud and the Red Sequence, with an almost 4 orders of magnitude increase in stellar mass from the irregulars to the early spirals. In both sequences, mass growth along them translates into a morphological change between subsequent Hubble types, which is not the case in the Green Plain where mass growth at a fixed color happens at constant morphological type. In the Red Sequence, the stellar mass shift between lenticulars and ellipticals persists in their mass distributions, with mean masses located at $\log(M_\ast/M_\odot)\sim 11-11.2$ and $\log(M_\ast/M_\odot)\sim 11.4-11.7$ respectively (see also \fg\ref{ell_vs_len_mass} below). 

Moreover, \fg\ref{color_mass} shows that the Green Plain is a low-density region of the color-mass diagram with a limited mass range of $\log(M_\ast/M_\odot) \in [10.5, 11.7]$ across the same wide color interval of $NUV-r \in [3.75,5.75]$. Therefore, in order to study the quenching of galaxies from star-forming to quiescence, one must consider the full $NUV-r$ interval corresponding to the mass range $\log(M_\ast/M_\odot) \in [10.5, 11.7]$, that is $NUV-r=6.8$ to $2.3$: it represents as much as 83\% of the full $NUV-r$ range (1.5 to 6.8) spanned by all galaxy types (see also \sct \ref{result-fading}).

\begin{figure}
\includegraphics[width=\columnwidth]{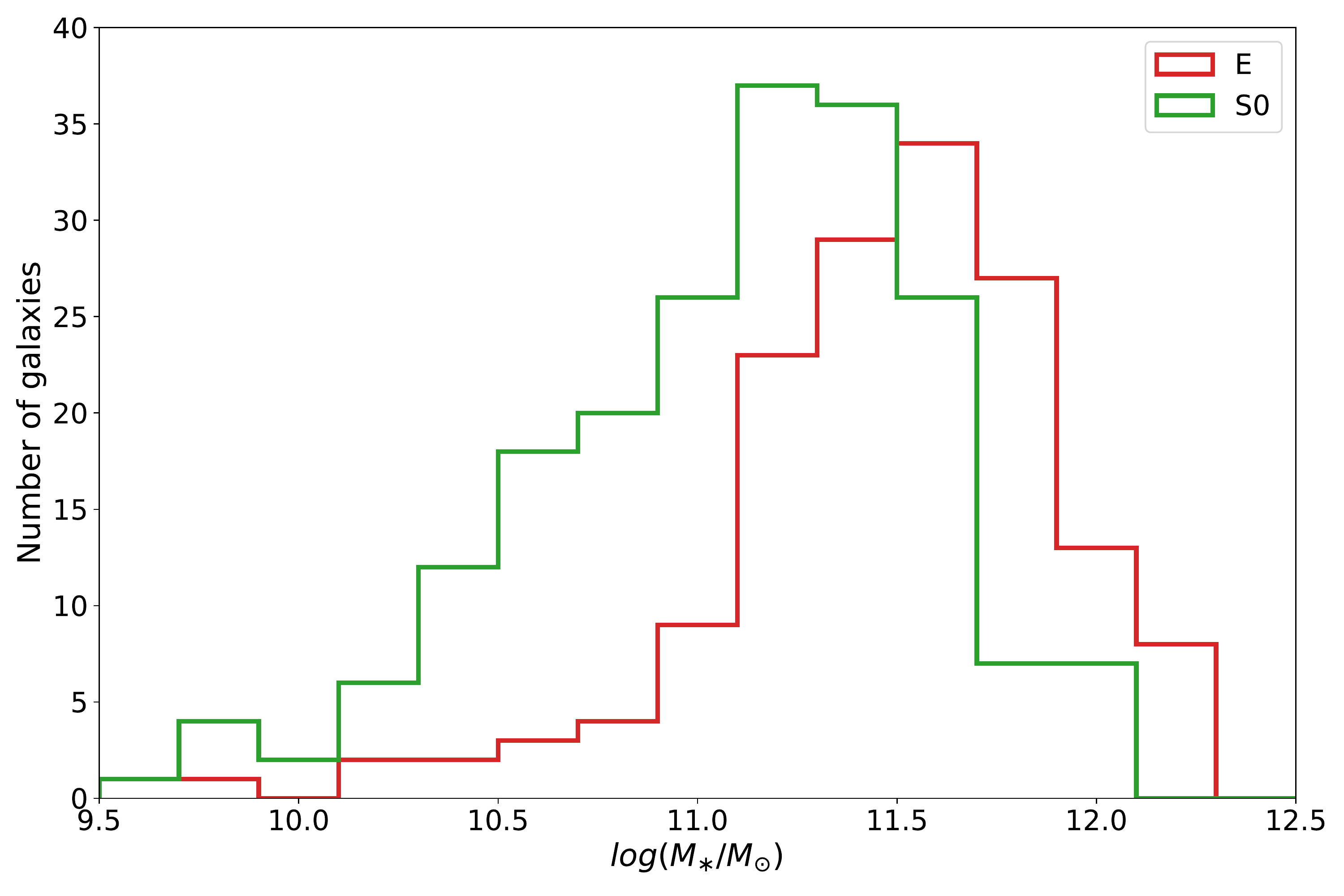}
\caption{Stellar mass distributions for EFIGI ellipticals (E and cD galaxies noted as E) and lenticulars (S0, S0$^+$ and S0$^-$ galaxies noted as S0) with {\tt Incl-Elong} $\leq 2$ in the Red Sequence. Ellipticals dominate the high-mass end of the Red Sequence, where $\log(M_\ast/M_\odot) > 11.5$, and lenticulars dominate the lower-mass part.}
\label{ell_vs_len_mass}
\end{figure}

Lastly, \fg\ref{color_mass} interestingly shows a common mass limit at $\log(M_\ast/M_\odot) = 11.7$ for all types from Sa to Sc, suggesting that there exists a critical mass limit over which spiral galaxies cannot exist. This limit is also visible in absolute $r$ magnitude in \fg\ref{NUV-r-face-on}. Therefore, if, through a major merger event a galaxy more massive than this limit is formed, it must end up in the Red Sequence and more often as an elliptical because the part of the Red Sequence with $\log(M_\ast/M_\odot) \ge 11.7$ contains $75\%$ E and $25\%$ S0, as shown in \fg\ref{ell_vs_len_mass} presenting the distribution of stellar masses for both morphological types. This graph also shows that the $\log(M_\ast/M_\odot) = 11.7$ mass limit corresponds to the peak of the elliptical distribution, which also dominates the high-mass part of the Red Sequence. Using the magnitude-limited MorCat $\cap$ GALEX would probably yield different relative distributions of S0 and E types than with EFIGI $\cap$ GALEX, but the trend itself cannot be a selection effect (see Quilley \& de Lapparent, \textit{in prep.}). This mass limit and the presence of mostly ellipticals beyond it, therefore supports a scenario in which mergers of massive spiral galaxies yields elliptical galaxies (see further discussion in \sct \ref{discussion-masslimit}).

\subsection{Morphological changes around the Green Plain        \label{change_GV}}

\subsubsection{Bulge growth  \label{bulge}}

\begin{figure}
\includegraphics[width=\columnwidth]{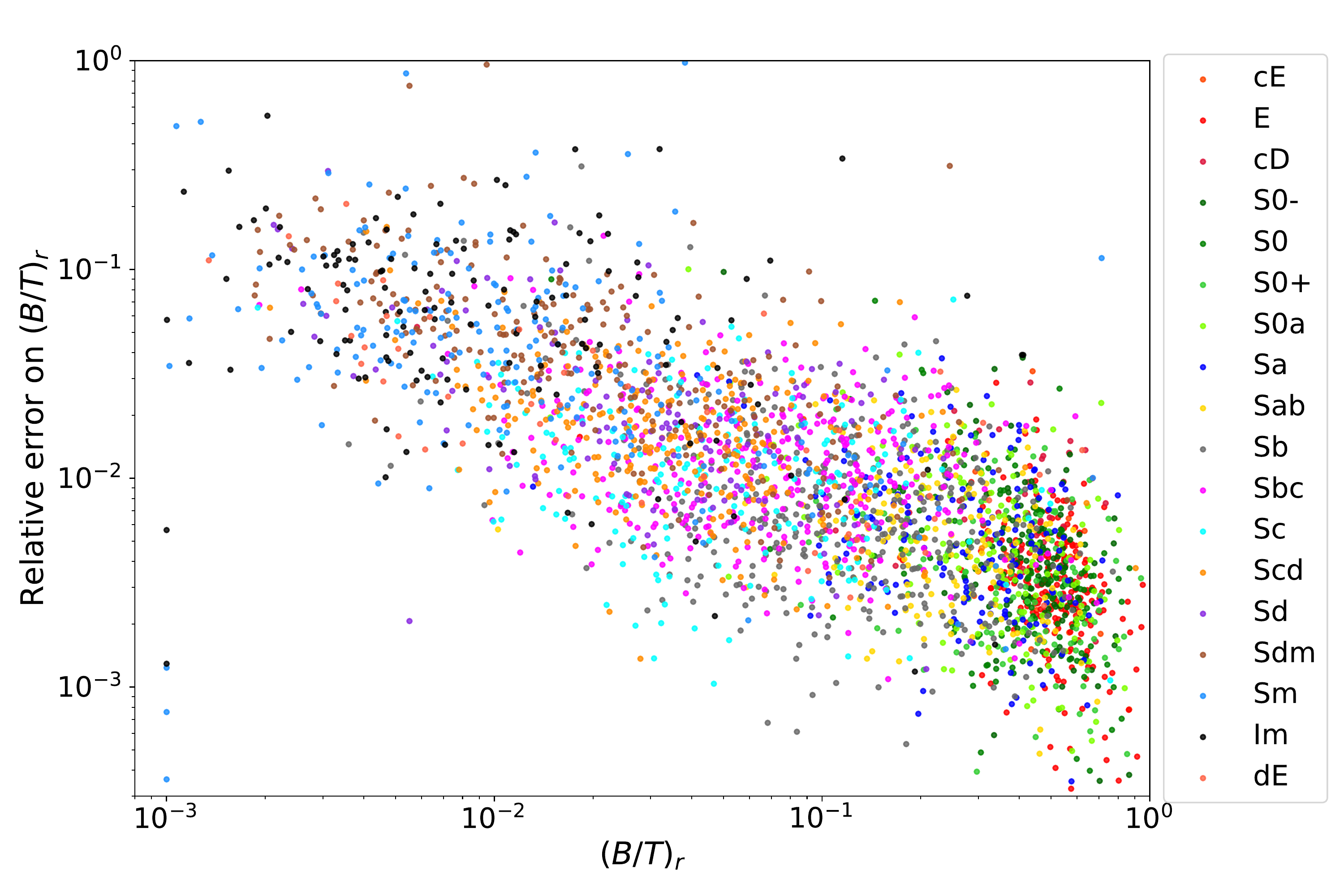}
\caption{Relative error on the bulge-to-total ratio in the $r$ band $({B/T}_{err})/{(B/T)}$ as a function of $B/T$ for EFIGI galaxies with {\tt Incl-Elong} $\leq 2$, the color of the points indicating the galaxy Hubble type. There is a trend of lower $B/T$ (that correspond to later-type galaxies) showing higher relative errors. All errors are small enough to not alter the results of the present analysis.}
\label{err_BT}
\end{figure}

\begin{figure*}
\includegraphics[width=\columnwidth]{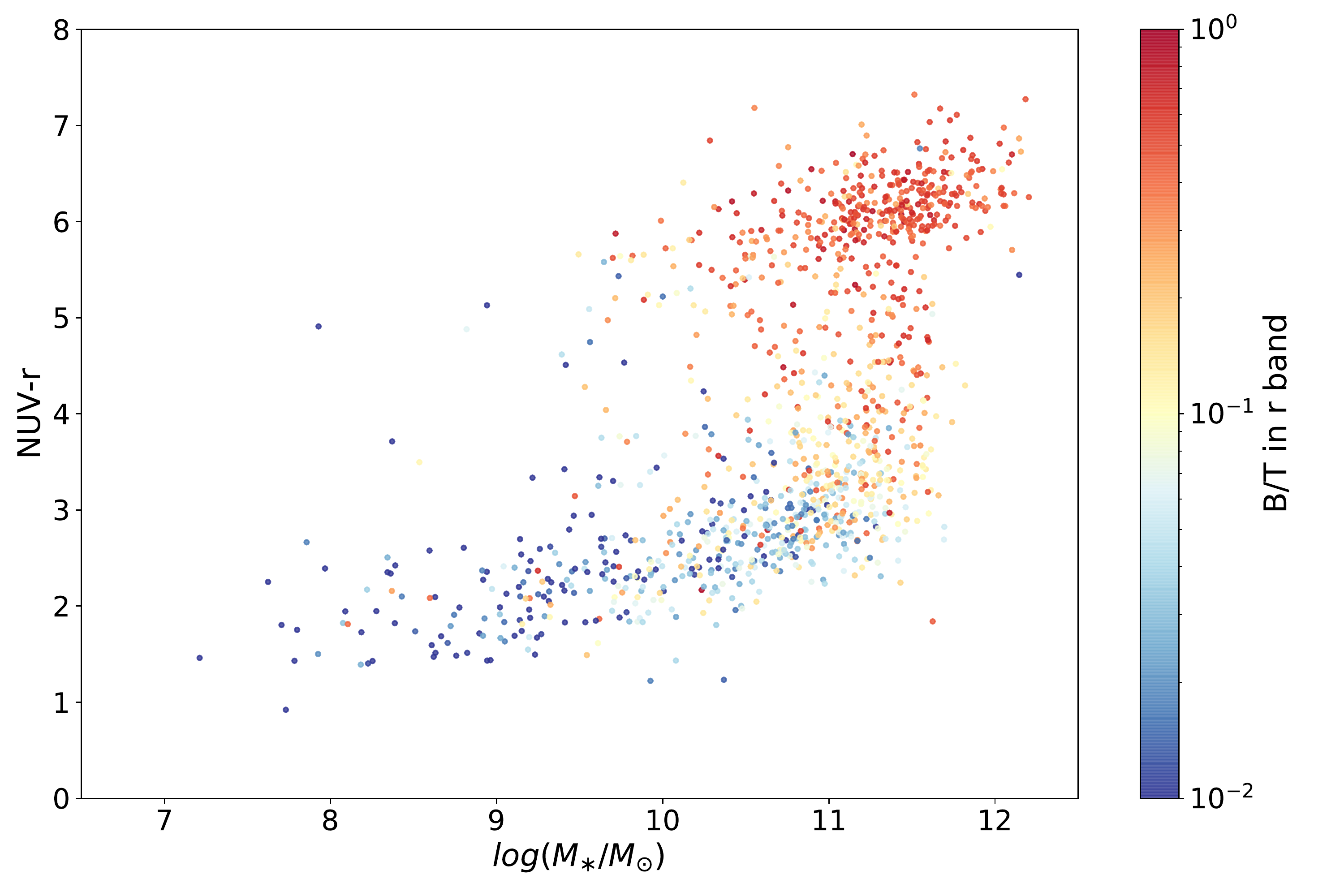}
\includegraphics[width=\columnwidth]{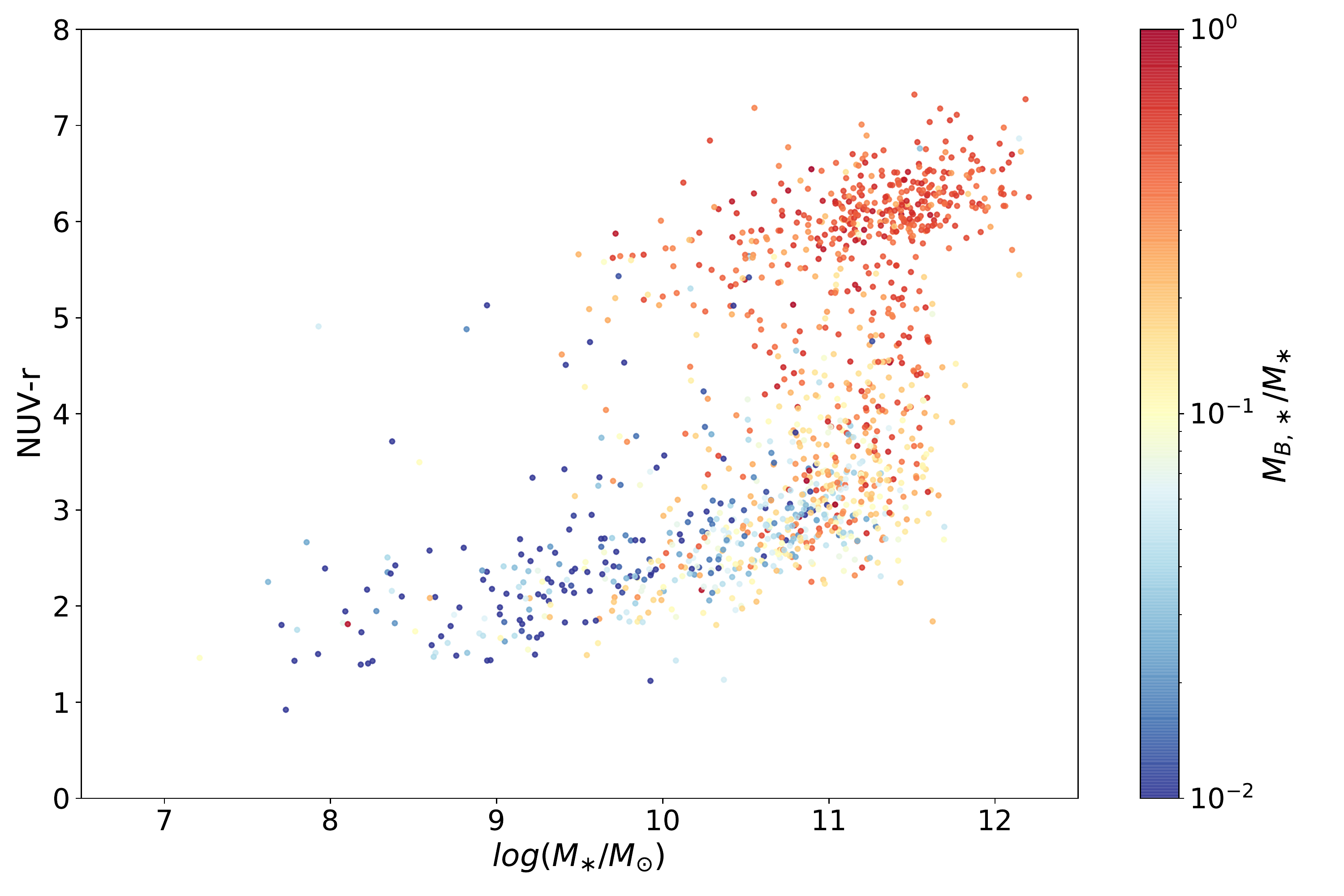}
\caption{Color-mass diagram for all EFIGI $\cap$ GALEX galaxies with {\tt Incl-Elong} $\leq 2$, in which the color of the points represents in the \textbf{left panel}, the bulge over total luminosity ratio ($B/T$) in the $r$ band, and in the \textbf{right panel}, the bulge over total stellar mass ratio. Both graphs show the same trend of an increase from the tail of the Blue Cloud at the lowest stellar masses, to the Red Sequence.} 
\label{BT}
\end{figure*}

\begin{figure*}
\includegraphics[width=\columnwidth]{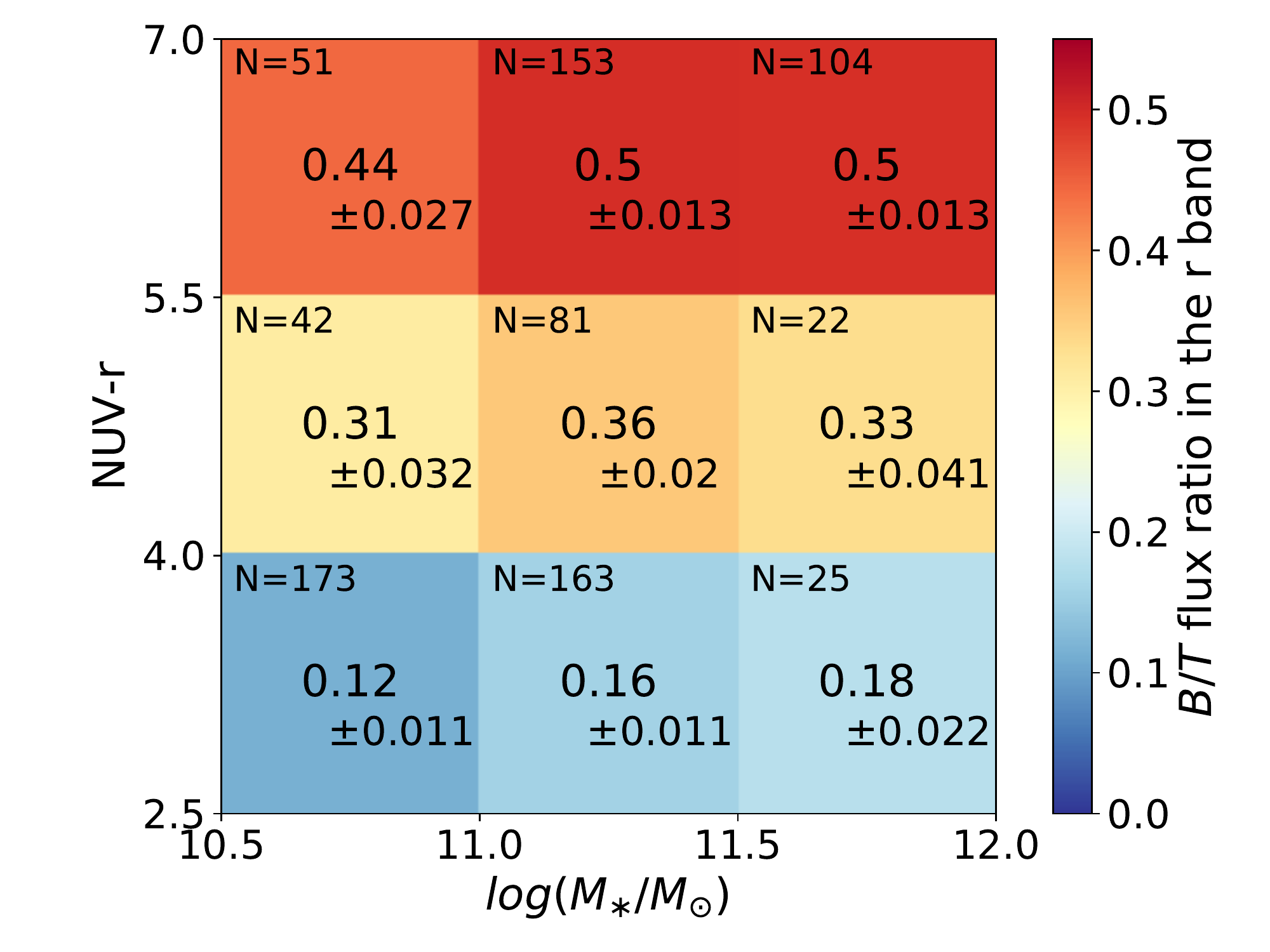}
\includegraphics[width=\columnwidth]{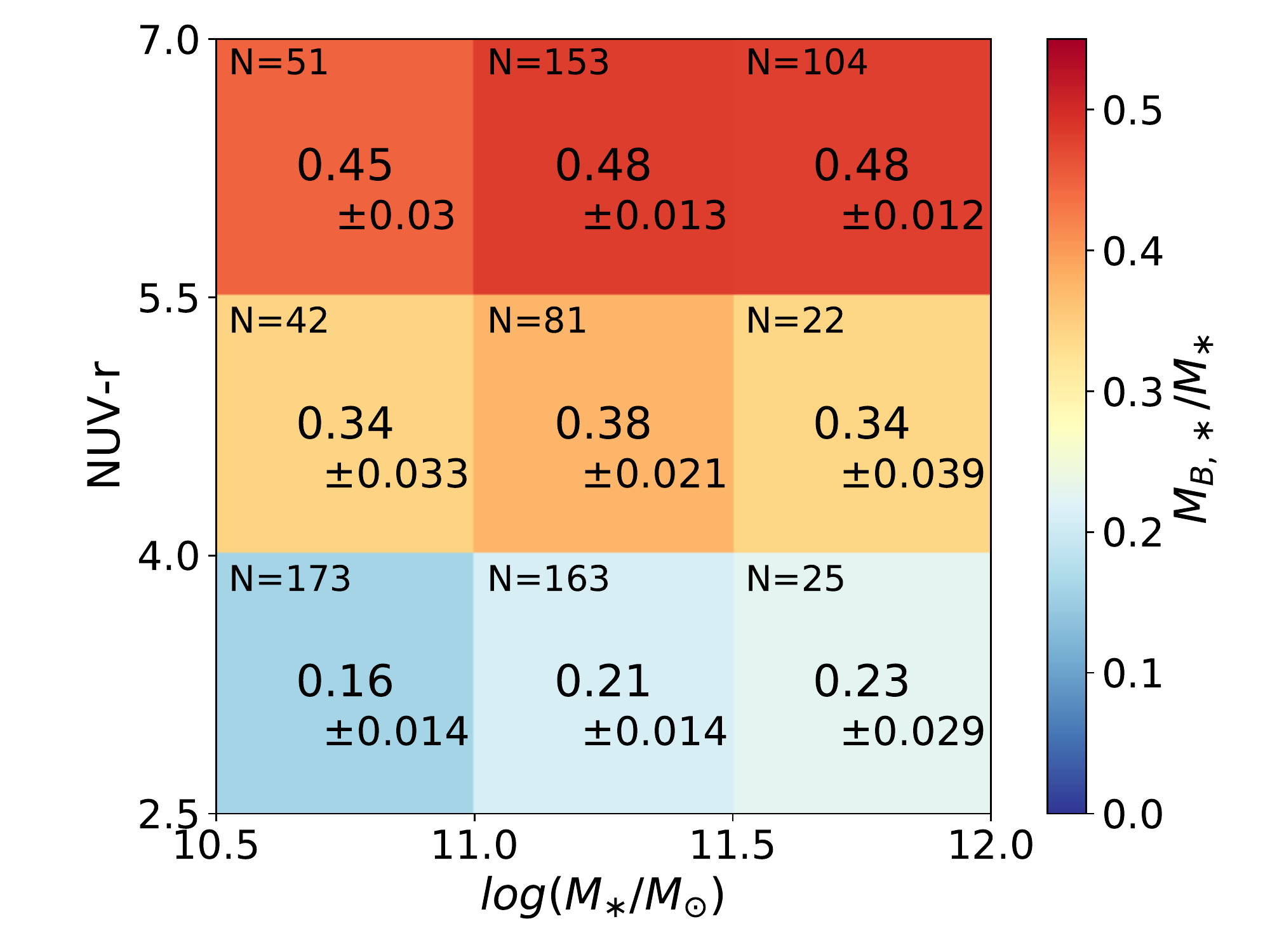}
\caption{Mean value of the bulge-to-total ratio for EFIGI $\cap$ GALEX galaxies with {\tt Incl-Elong} $\leq 2$ in color-mass cells of 0.5 dex in total stellar mass and 1.5 in $NUV-r$ color, and derived from \fg\ref{BT}. Only the bright part of the Blue Cloud, the Green Plain and the Red Sequence are shown. The associated error on the mean, and the number of galaxies in each cell are indicated. \textbf{Left:} Luminosity ratio in the r band, \textbf{Right:} Stellar mass ratio. There is a significant increase in the bulge-to-total ratio in both luminosity and mass with the $NUV-r$ color across the Green Plain, by a factor of $\sim1.5-2.6$ and $\sim2.1-3.6$ from the Blue Cloud to the Green Plain and the Red Sequence respectively. There is no effect with the galaxy stellar mass across the Green Plain, but a moderate one across the Red Sequence and sampled Blue Cloud.}
\label{GV_bulge_ratio}
\end{figure*}

To further characterize the morphology of EFIGI galaxies using SourceXtractor++ parameters, we examine the derived bulge-to-total luminosity ratio $B/T$, as this parameter is key for determining of the morphological types. Indeed, early-type spirals are defined as having a larger $B/T$ than their late-type counterparts, and ellipticals are pure bulge. \fg\ref{err_BT} shows that the relative errors on $B/T$ obtained through our SourceXtractor++ bulge and disk decomposition in the $r$ band (see \ref{methodo_srx} describe a large range from $\sim 10^{-1}$ to $\sim 10^{-3}$, and depends strongly on the value of the $B/T$ ratio itself. Obviously, higher $B/T$ have smaller relative errors, with ellipticals and lenticulars having errors mostly on the order of 0.1\%, while for late-type spirals with $B/T<0.1$, it increases to around 10\%. As we show below, the range of uncertainties in $B/T$ plotted in \fg\ref{err_BT} does not impair the subsequent analysis. 

To examine $B/T$ variation along the Hubble sequence, we show in the left panel of \fg\ref{BT} the color-magnitude diagram in which the points are color-coded with the value of $B/T$ in the $r$ band. We find the highest values of $B/T$ in the Red Sequence because it is populated by E galaxies that are almost pure-bulge, and lenticulars that have a dominant bulge. From early-type to late-type spirals, one can observe the expected decrease in the $B/T$, as this flux fraction is, with the pitch angle of the spiral arms, one of the criteria to classify spiral galaxies along the Hubble sequence \citep{1959HDP....53..275D}.

\begin{figure*}
\includegraphics[width=1.06\columnwidth]{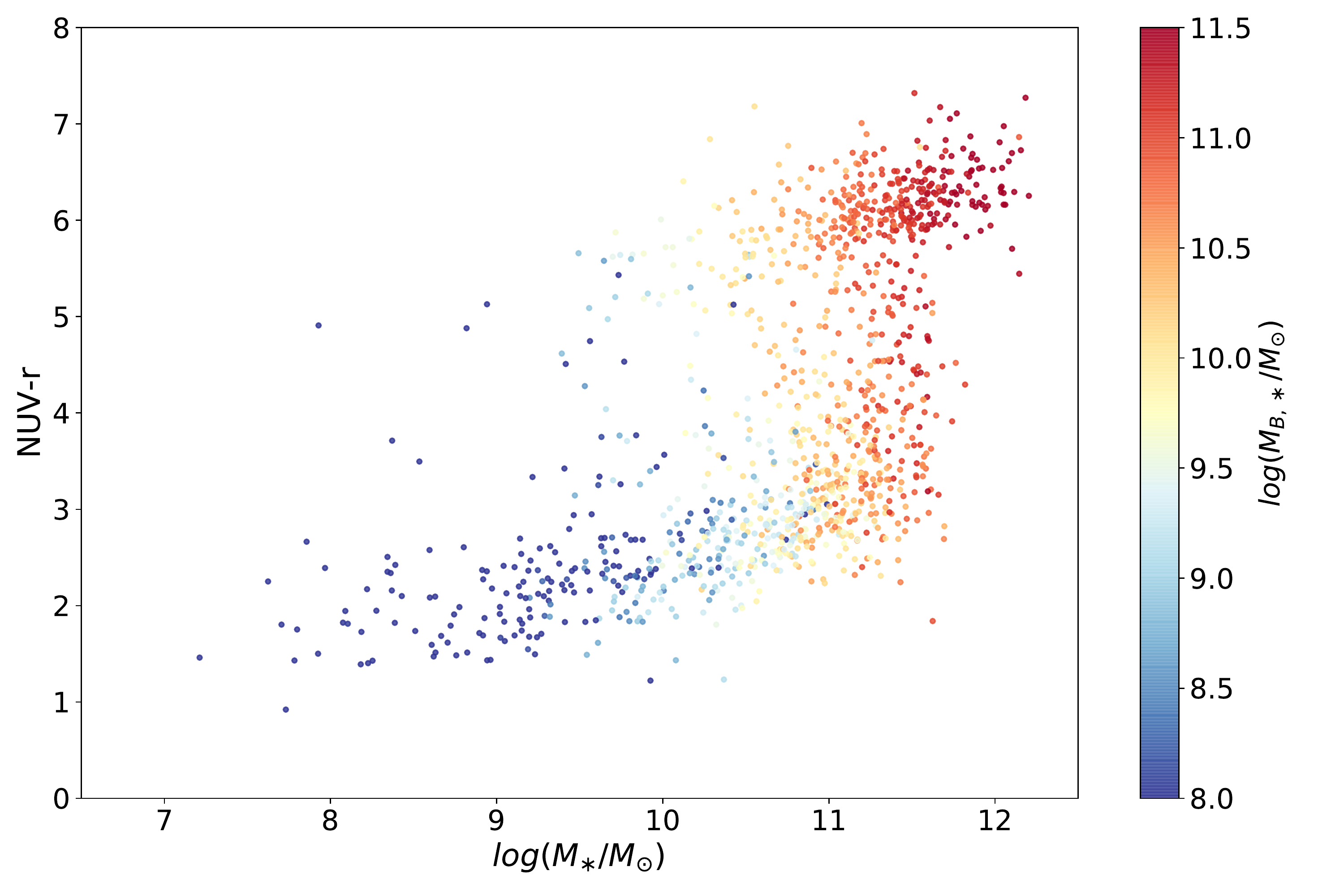}
\includegraphics[width=0.94\columnwidth]{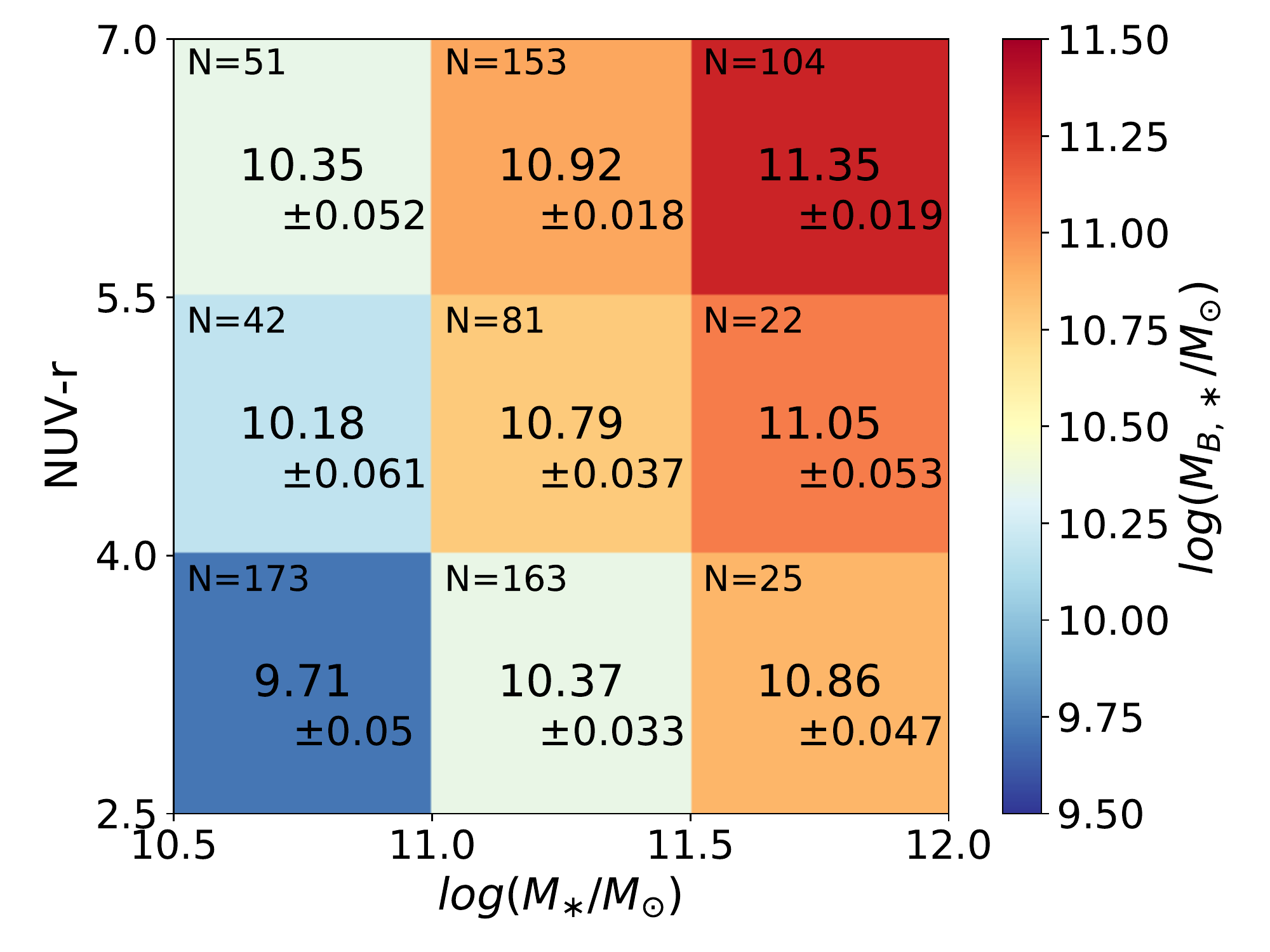}
\caption{Bulge stellar mass evolution in the galaxy color-mass diagram. \textbf{Left:} Color-mass diagram for all EFIGI $\cap$ GALEX galaxies with {\tt Incl-Elong} $\leq 2$, in which the color of the points represent the bulge stellar mass inferred from bulge and disk SED model-fitting. \textbf{Right:} Mean value of the bulge stellar mass in color-mass cells of 0.5 dex in total stellar mass and 1.5 in $NUV-r$ color. Only the bright part of the Blue Cloud, the Green Plain and the Red Sequence are shown. The associated error on the mean, and the number of galaxies in each cell are indicated. The bulge stellar mass increases with both the total stellar mass and the $NUV-r$ color of the galaxy.}
\label{bulge_mass}
\end{figure*}

In our SourceXtractor++ bulge and disk decompositions (\sct \ref{methodo_srx}), the S\'ersic component aims at adjusting a central concentration within the disk. For bulgeless galaxies (irregulars and very-late type spirals: Sd and later types), the sum of the S\'ersic model and the exponential profile may be inappropriately used to model the whole galaxy in variable proportions, but with comparable effective radii. We identify these wrong fits by comparing the flux of the bulge component to the one from the zoom-in process described in \sct \ref{methodo_srx}, that is calculated as the excess flux in the center of the galaxy isophotal print (it may correspond to some star-forming regions in bulgeless galaxies). In the 3 bands $g$, $r$, and $i$, we discard these wrong fits and their erroneous $B/T$ values by replacing them with the zoom-in estimate of $B/T$ for all galaxies with $B/T>f(B/T_{zoom})$ and morphological types Sc and later (the $f$ threshold function is empirically defined).

Through the bulge and disk SED model-fitting described in \sct \ref{sed_bulge_disk}, one can convert fluxes to stellar masses for either the bulge or the disk of each galaxy. Right panel of \fg\ref{BT} shows the distribution of the $B/T$ stellar mass ratio in the color-mass diagram and consequently looks very similar to the left panel of \fg\ref{BT}, because of the strong anticorrelation between $M_r$ and $M_\ast$ (see \fg\ref{mass_mag}): the fraction of mass in the bulge of a galaxy increases along the color-mass sequence from the Blue Cloud, through the Green Plain, and into the Red Sequence, hence from irregulars and late spirals to lenticulars and ellipticals.

To examine with more details the changes in the galaxy properties across the Green Plain, we zoom-in the region $\log(M_\ast/M_\odot) \in [10.5, 12.0]$, which also includes both the Red Sequence and the massive part of the Blue Cloud. In this series of graphs, we bin the bulge stellar mass values shown in \fg\ref{BT} by intervals of 0.5 dex in $\log(M_\ast/M_\odot)$, and the $NUV-r$ color as [2.5, 4], [4, 5.5], and [5.5, 7] intervals, corresponding to the Blue Cloud, the Green Plain, and the Red Sequence, respectively. Within these cells defined by color and mass interval, we calculate the mean value of the $B/T$ (left panel) and the bulge stellar mass (right panel). We also indicate the size of the sample $N$ in each bin and estimate the associated error on the mean as the root-mean-square deviation within each cell divided by $\sqrt{N}$. We compared these errors with those derived from the quadratic mean of the errors on the individual points, and found that the latter range from similar to a factor of 10 lower than former, hence their use in the graphs. 

One can see in the resulting graphs of \fg\ref{GV_bulge_ratio} that the fraction of flux (left) and stellar mass (right) comprised in the bulge increases significantly through the Green Plain: it is doubled to tripled between the [2.5, 4.0] and the [5.5, 7.0] $NUV-r$ ranges (depending on the mass interval), and already almost doubled from [2.5, 4.0] to [4.0, 5.5], thus quantifying the fact that early-type spirals have a more prominent bulge than their late-type counterparts. We therefore confirm results by \citet{2018MNRAS.476...12B} that the bulge is already significant in the Green Plain, but contrary to their study, we show that bulge growth occurs all across this entire region (and does not precede the star formation decline). The use of dust corrected $u-r$ color rather than $NUV-r$, and the fact that the Galaxy And Mass Assembly (GAMA) survey is at more distant redshifts ($z<0.2$), hence based on galaxies that are less resolved angularly than in EFIGI, might explain this discrepancy. One can further notice in both panels of \fg\ref{GV_bulge_ratio} that there is no total mass effect on the fraction of luminosity and mass in the bulge in the Green Plain, similar to the morphological trend in and around the Green Plain, which is purely a trend in color (see \fg\ref{mean-NUV-r-face-on} and \sct \ref{res_color_morpho}). There is nevertheless a moderate increase in both panels of \fg\ref{GV_bulge_ratio} with the total stellar mass across the massive end of the Blue Cloud, where mass growth is linked to morphological type change with a large dispersion (Sb to Scd), and across the Red Sequence due to the type transition from lenticulars to ellipticals. The fraction of light or mass in the bulge is therefore related to the morphology rather than to the galaxy stellar mass.

To characterize the growth of the bulge mass $M_{B,\ast}$, the left panel of \fg\ref{bulge_mass} shows its variation across the color-mass diagram: one can see a straightforward increase in the bulge stellar mass with the total galaxy stellar mass as an horizontal trend. As the bulge-to-total ratio of stellar mass also increases along the Hubble sequence, hence when going from the Blue Cloud through the Green Plain to the Red Sequence, and because there is on average a steady increase in total stellar mass from the irregulars to the ellipticals. One also expects an horizontal increase in the bulge mass across the massive part of the Blue Cloud, the Green Plain and the Red Sequence. The effect is indeed visible in the zoom-in graph of the right panel of \fg\ref{GV_bulge_ratio}; it is barely visible in the left panel of \fg\ref{bulge_mass} due to the color-map encompassing several orders of magnitude to accommodate all of the galaxy populations.
Indeed, the right panel of \fg\ref{bulge_mass} confirms that in addition to the increase in bulge mass for more massive galaxies (horizontal gradient) there is an increase in bulge mass for redder galaxies, even at fixed total mass (vertical gradient).

\begin{figure}
    \includegraphics[width=\linewidth]{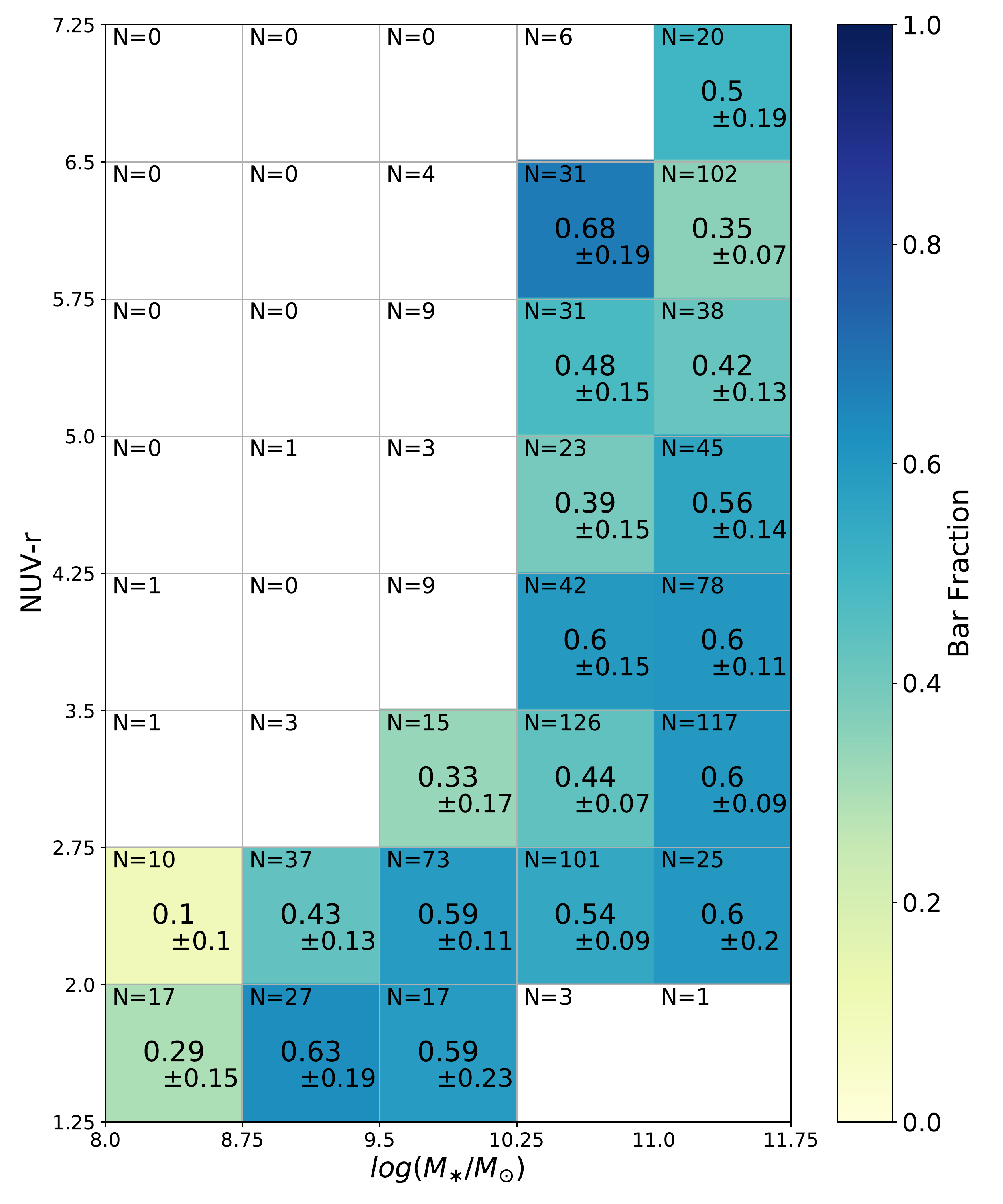}
    \caption{Fraction of EFIGI $\cap$ GALEX galaxies with {\tt Incl-Elong} $\leq 2$ and types S0$^-$ to Im having an attribute value {\tt Bar Length} $>0$, in color-mass cells of 0.75 dex in stellar mass and 0.75 in $NUV-r$ color, all the way from the Blue Cloud to the Red sequence. Apart from the low-mass irregulars having fractions below $30\%$, the fraction of EFIGI visually barred galaxies is large as it varies between $33\%$ to $68\%$ with a global mean of $52\%$, with no significant trend in either color or mass.}
    \label{bar_frequency}
\end{figure}

\begin{figure}
    \includegraphics[width=\linewidth]{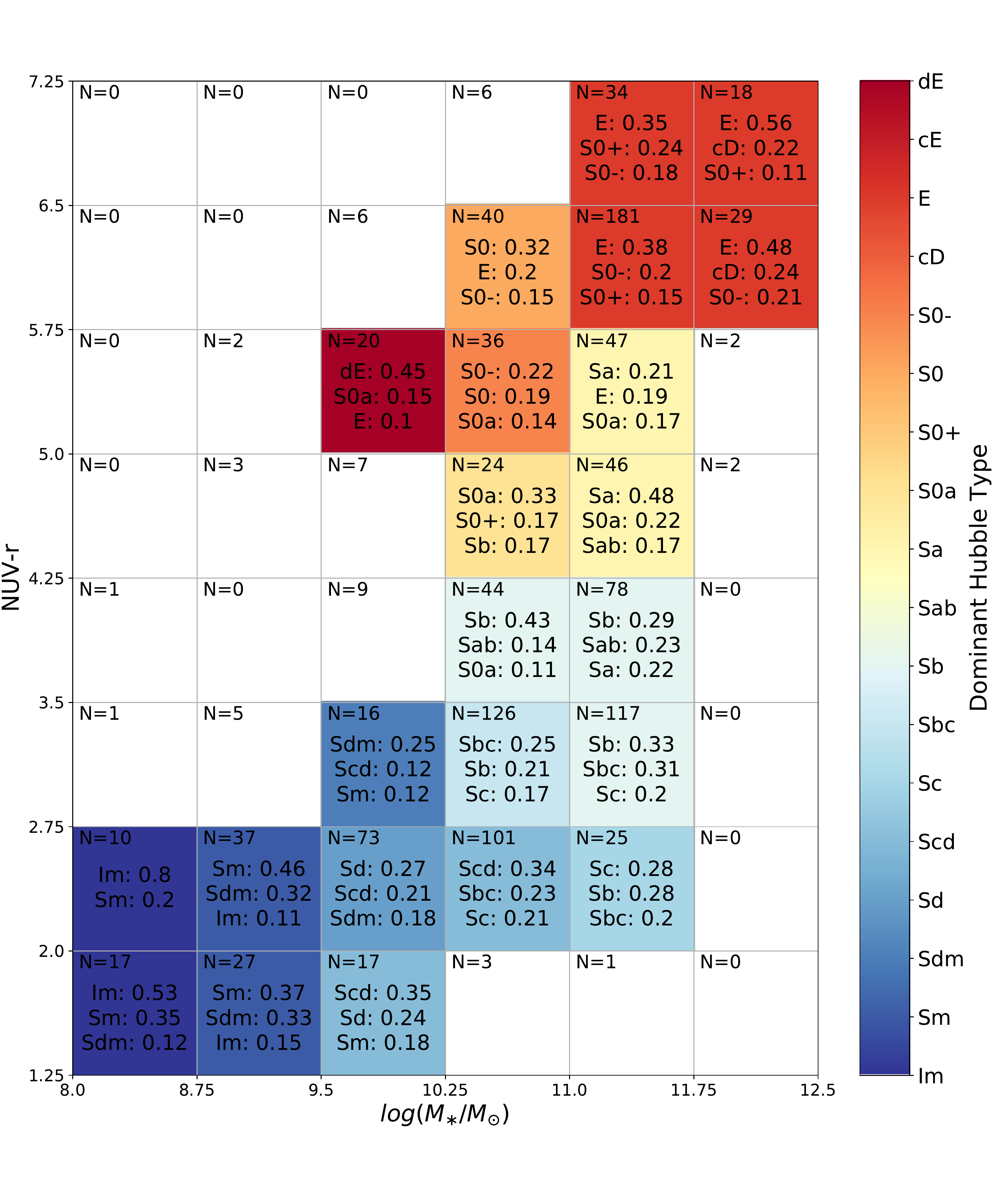}
    \caption{Fractions of morphological types for EFIGI $\cap$ GALEX galaxies with {\tt Incl-Elong} $\leq 2$: the three most represented Hubble types are listed in color-mass cells of 0.75 dex in both color and mass. The associated error on the mean, and the number of galaxies in each cell are indicated. The color of the cells represents the most represented morphological type. 
    Similarly to the effect seen in \fg\ref{color_mass}, the dominant Hubble type varies continuously along the color-mass diagram from the Im bluest and lowest mass galaxies to the blue and more massive spirals, through the massive and redder early-type spirals, and up to the red lenticulars and even more massive ellipticals.}
    \label{dom_types}
\end{figure}

\begin{figure*}
\includegraphics[width=1.06\columnwidth]{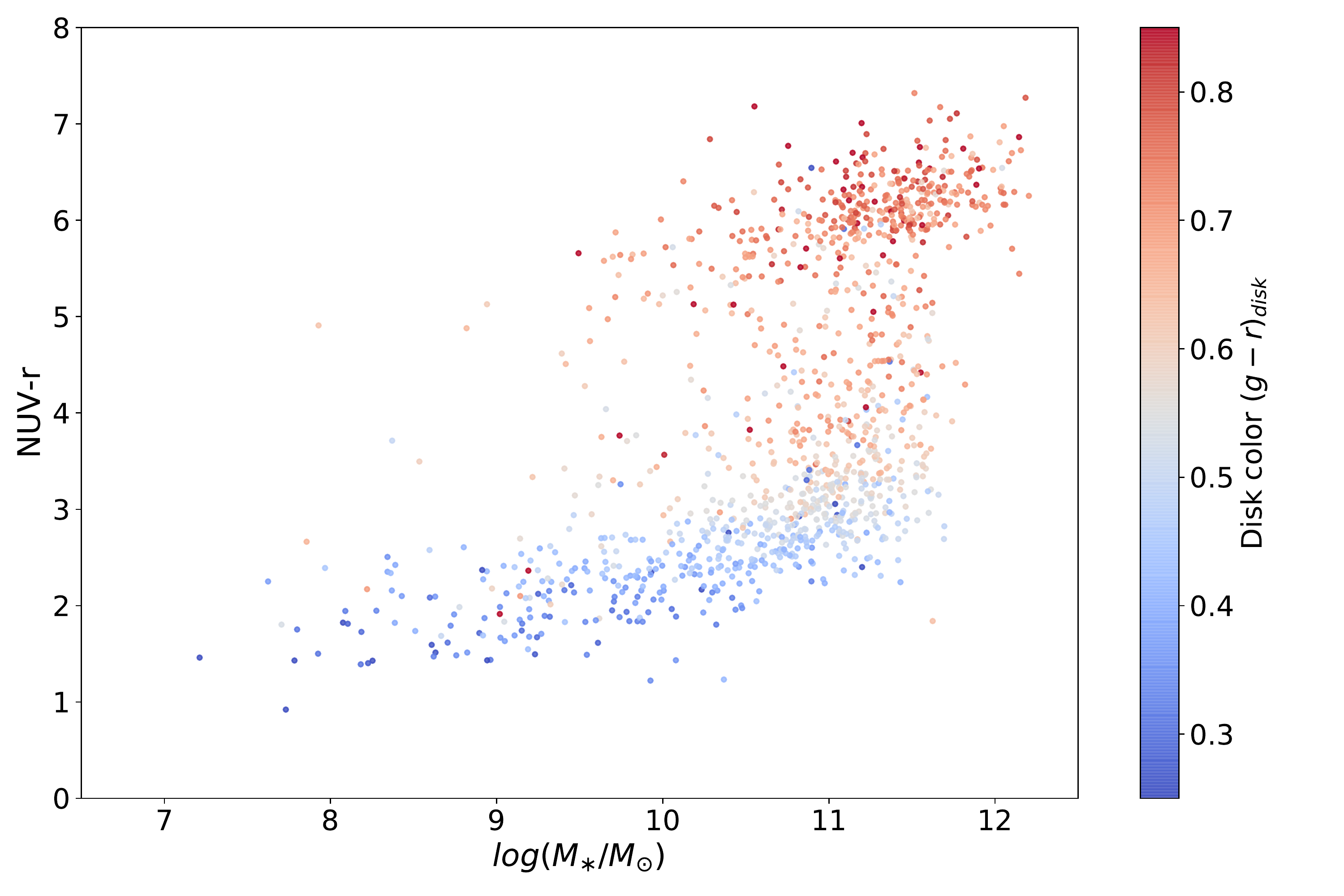}
\includegraphics[width=0.94\columnwidth]{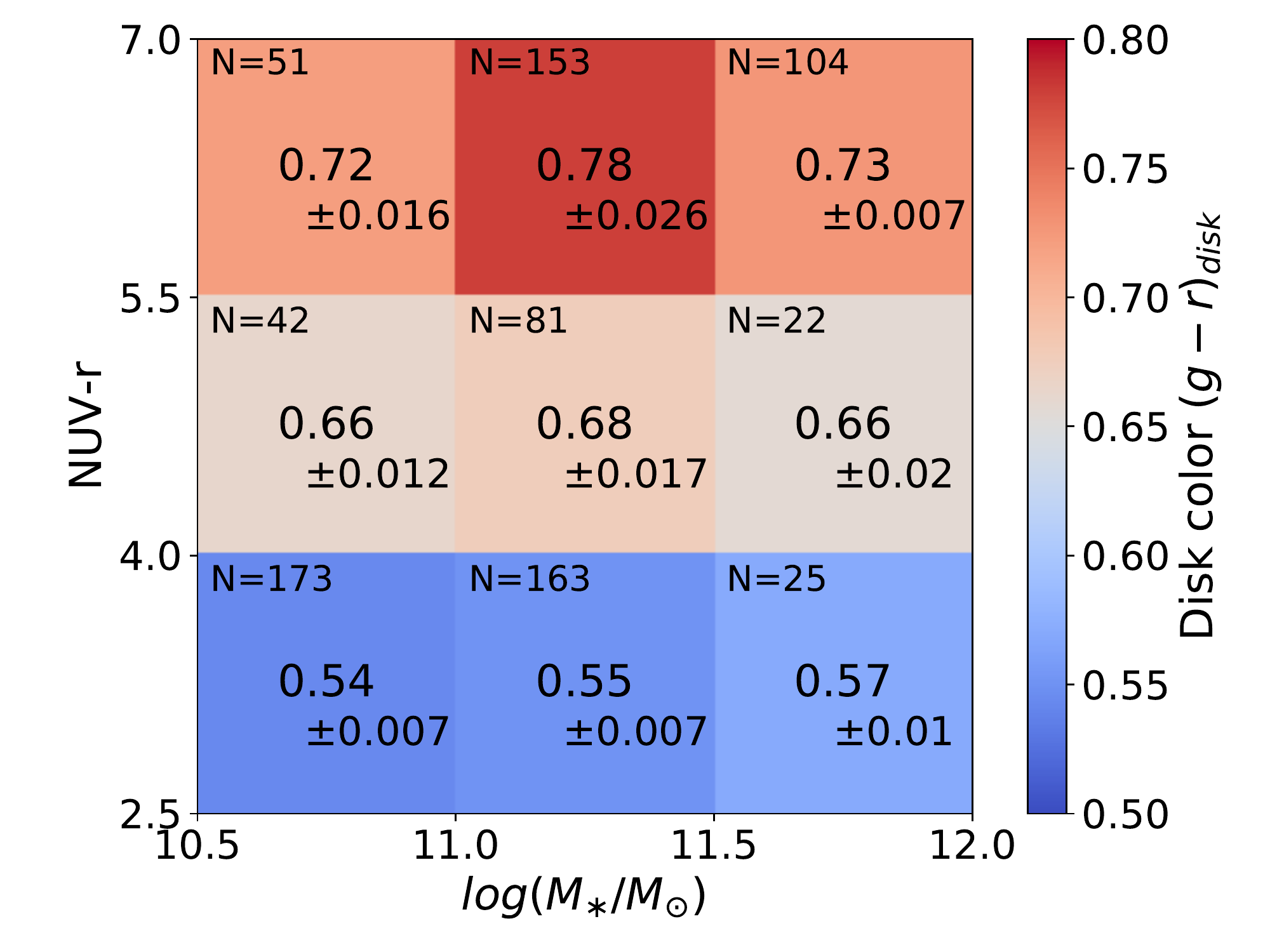}
\caption{Disk $g-r$ absolute color evolution in the galaxy color-mass diagram. \textbf{Left:} Color-mass diagram for all EFIGI $\cap$ GALEX galaxies with {\tt Incl-Elong} $\leq 2$, in which the color of the points represent the disk $g-r$ absolute color inferred from bulge and disk model-fitting. There is a systematic reddening of the disk along the Hubble sequence from irregulars to ellipticals. \textbf{Right:} Mean value of the disk color $(g-r)_{disk}$ for EFIGI $\cap$ GALEX galaxies in color-mass cells of 0.5 dex in total stellar mass and 1.5 in $NUV-r$ color. Only the bright part of the Blue Cloud, the Green Plain and the Red Sequence are shown. The associated error on the mean and the number of galaxies in each cell are indicated. There is a significant reddening of the disk for redder $NUV-r$ colors, as well as for earlier and earlier types along the color-mass sequence.}
\label{disk_color}
\end{figure*}

\subsubsection{Frequent bars in all disk galaxies   \label{bars}}

Bars are frequent and observed in all types of disk galaxies, that is both in spirals and lenticulars, as well as in Irregulars: a total of 52\% of EFIGI galaxies from types S0$^-$ to Im have the {\tt Bar Length} attribute $>0$. They are an important dynamical feature to investigate when studying bulges, as they contribute to their formation by driving both stellar migration \citep{2011A&A...534A..75B, 2011A&A...527A.147M, 2013A&A...553A.102D} and gas migration \citep{2004ARA&A..42..603K} toward the centers of galaxies. \fg\ref{bar_frequency} shows the fraction of EFIGI barred galaxies in color-mass cells of the color-mass diagram after excluding E, cE, cD and dE types (as these types are not expected to host a bar due to the absence of a disk). The error in the plotted fractions is estimated as Poissonian, using $f_{err} = f*\sqrt{1/N_{barred} + 1/N_{total}}$ where $f = N_{barred}/N_{total}$. 

In order to locate the dominant Hubble types within \fg\ref{bar_frequency}, \fg\ref{dom_types} shows the three highest Hubble types fractions in cells of 0.75 dex in color and 0.75 dex in mass as well, the color of the bin corresponding to the dominant Hubble type. This graph illustrates the continuous spanning of the Hubble sequence seen in \fg\ref{mean-NUV-r-face-on}, and can help to further read the subsequent color-mass diagrams (using the same binning) to relate a variation in a parameter to the variation in morphological mix.

Comparison of \fg\ref{bar_frequency} with \fg\ref{dom_types} shows that except for the low-mass end of the Blue Cloud dominated by Irregulars, the junction pixel between the Green Plain and the Red sequence with a $0.35\pm0.07$ fraction of bars, in which lie many bar-less S0 or S0$^-$, and the crook of the knee with a $0.33\pm0.17$ fraction of bars for only $N=15$ objects, there is a high fraction of galaxies with a bar, from $0.39\pm0.15$ to $0.68\pm0.19$ (with no significant trend) across the entire Blue Cloud and Green Plain all the way to the lenticular galaxies of the Red Sequence. We therefore suggest that bars may play a role in the marked growth of $B/T$ across the full color-mass diagram and in particular the Green Plain, as seen in \fgs\ref{BT} and \ref{GV_bulge_ratio}. Additional mechanisms may nevertheless be needed to contribute to both the high bar fraction and the bulge growth. We suggest in \sct \ref{discussion-mergers} that mergers may play this role.

\subsubsection{Disk reddening    \label{disk_reddening}}

By performing bulge and disk decomposition of galaxies from the GAMA survey, \cite{2018MNRAS.476...12B} highlight that the change in color from the Blue Cloud to the Red Sequence is driven by disk color.
Examination of EFIGI disk parameters may also help us to understand the morphological transformations undergone by galaxies through the Green Plain. \fg\ref{disk_color} shows the distribution of the absolute color $g-r$ of the SourceXtractor++ disk component in the color-mass diagram (individual points on the left, and values in cells on the right). Ideally, one should examine the $NUV-r$ color of the disk in order to best trace star formation in the disk, but the GALEX $NUV$ magnitude is only available for the whole galaxy. Left panel of \fg\ref{disk_color} displays, similarly to \fgs\ref{BT} and \ref{bulge_mass}, a trend related to the Hubble types: the disks of EFIGI galaxies are systematically redder for larger total $NUV-r$, and the continuous variation in Hubble types along the color-mass sequence implies that the reddening of the disk also takes place along the Hubble sequence, from late to early spirals, and all the way to lenticulars. Right panel of \fg\ref{disk_color} also shows the mean values and associated \rms deviation of the $g-r$ disk color in the Green Plain and around it, indicating a significant and $\sim0.2$ magnitude $g-r$ reddening of the disk as the total galaxy $NUV-r$ color increases (reddens). The frequent bars detected in all types of disk galaxies (see \fg\ref{bar_frequency}, \sct \ref{bars}) may play a role in the reddening of the disks of Blue Cloud and Green Plain galaxies.

The progressive disk reddening across the Blue Cloud can also be seen in right panel of \fg\ref{comparing_EFIGI_and_ZPEG_bulge_disk}: it shows that the $gri$ photometry of EFIGI disks are on average best fitted by earlier and earlier PEGASE.2 templates from EFIGI galaxies with Hubble types from Im to Sa. In contrast, right panel of \fg\ref{comparing_EFIGI_and_ZPEG_bulge_disk} shows that disks of lenticulars have red colors similar to those of ellipticals: indeed, the disks of S0$^-$, S0, S0$^+$ and E Hubble types are best fit by ZPEG with either the E or the Sab template, whereas the Sa and S0 templates rarely provide a good fit. While the necessity of a disk component in the bulge and disk profile modeling of a true morphological E galaxy may be due to the fact that a S\'ersic profile is inadequate (and nevertheless yields nearly identical colors for both components), the fact that disks of lenticulars have colors preferentially described by the E and Sab templates as for the E galaxies indicates a remarkable color stability across the total stellar mass range covered by the Red Sequence, that is also seen in both panels of \fg\ref{disk_color}. 

The right panel of \fg\ref{disk_color} also shows a stable mean $g-r$ disk color for Green Plain galaxies, in the central $NUV-r$ row, whatever the total stellar mass is. A similar independence to total mass is seen in \fg\ref{GV_bulge_ratio} showing the increasing fraction of luminosity or mass enclosed in the bulge from the tip of the Blue Cloud to the Red Sequence (except a weak increase with total mass in the massive end of the Blue Cloud). This independence of both the disk color and bulge ratio with total stellar mass, at constant $NUV-r$ color, results from the fact that the morphological type transitions in the Green Plain are characterized by strong color changes rather than stellar mass (see \fg\ref{mean-NUV-r-face-on}), to the contrary of the Blue Could and Red Sequence, in which type changes are dominated by total stellar mass variations.

\begin{figure}
  \includegraphics[width=\linewidth]{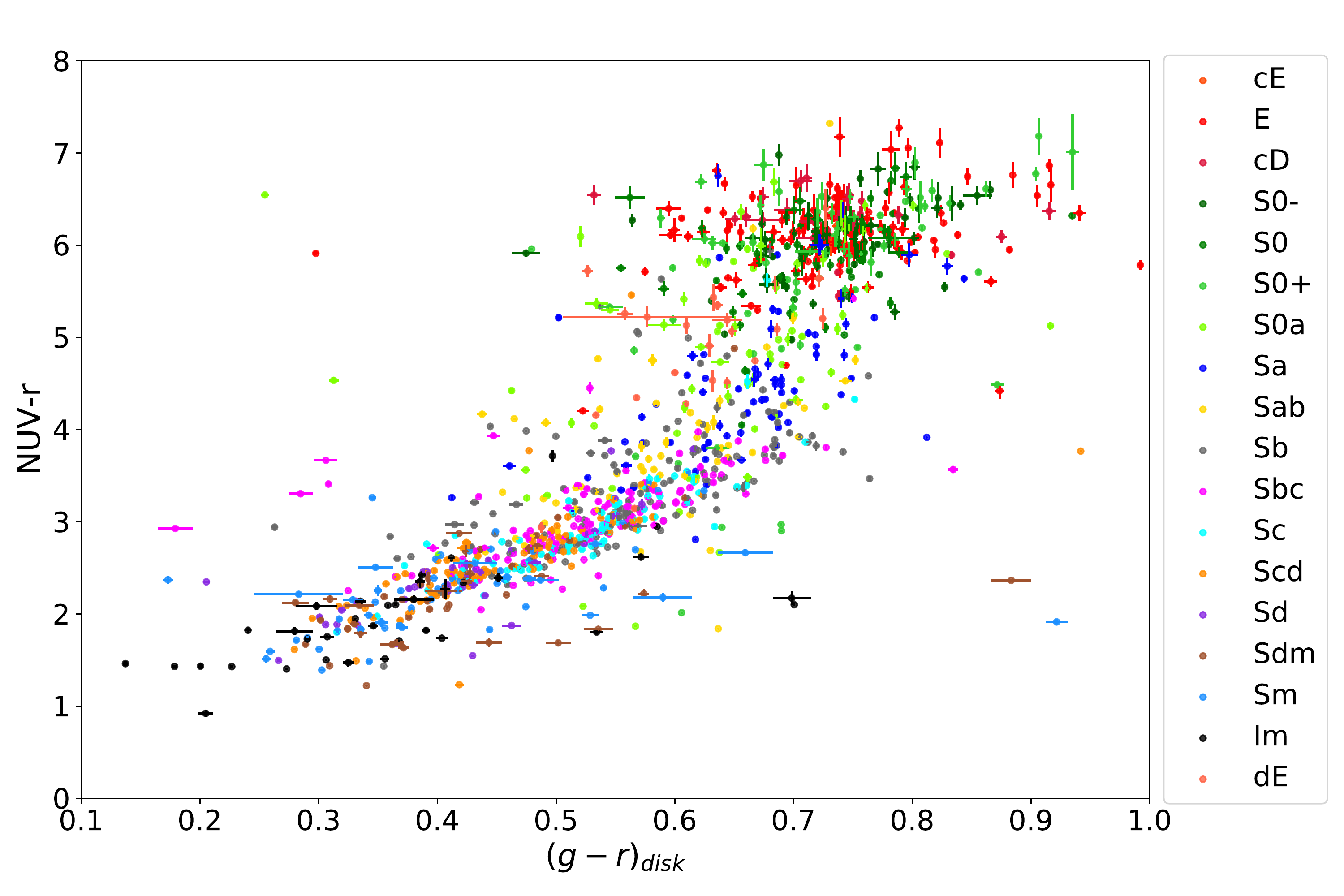}
    \caption{Absolute $NUV-r$ color of the galaxy versus absolute $g-r$ color of the disk for EFIGI $\cap$ GALEX galaxies with {\tt Incl-Elong} $\leq 2$. The color of the points indicates the Hubble Type. The strong correlation between total galaxy $NUV-r$ color and disk $g-r$ color for types Im to Sb, having no bulge, or less than a 10\% bulge-to-total ratio, suggests that their disk reddening translates into the fading of star formation.}
    \label{color_galaxy_color_disk}
\end{figure}

\begin{figure*}
    \includegraphics[width=0.33\linewidth]{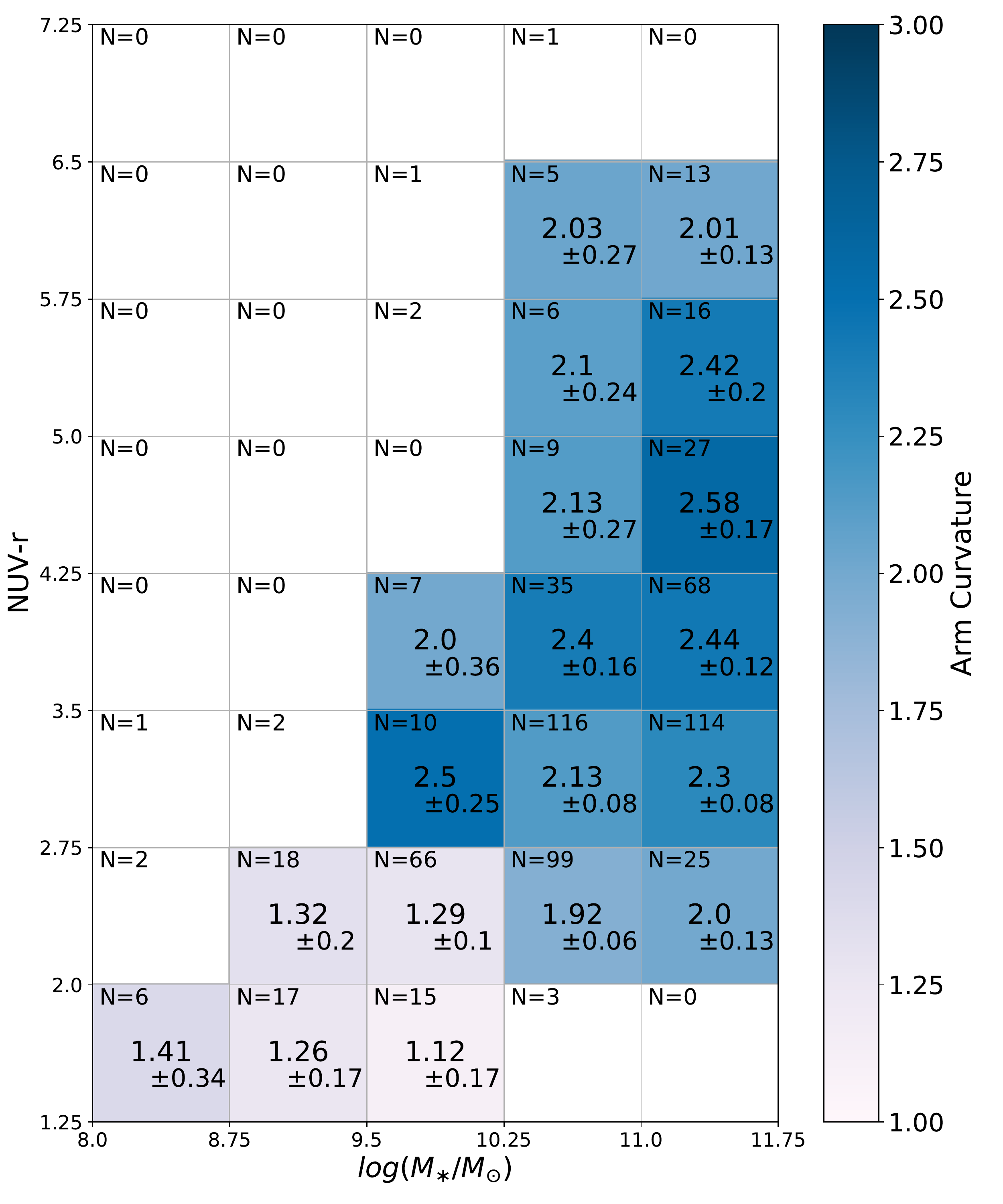}
    \includegraphics[width=0.33\linewidth]{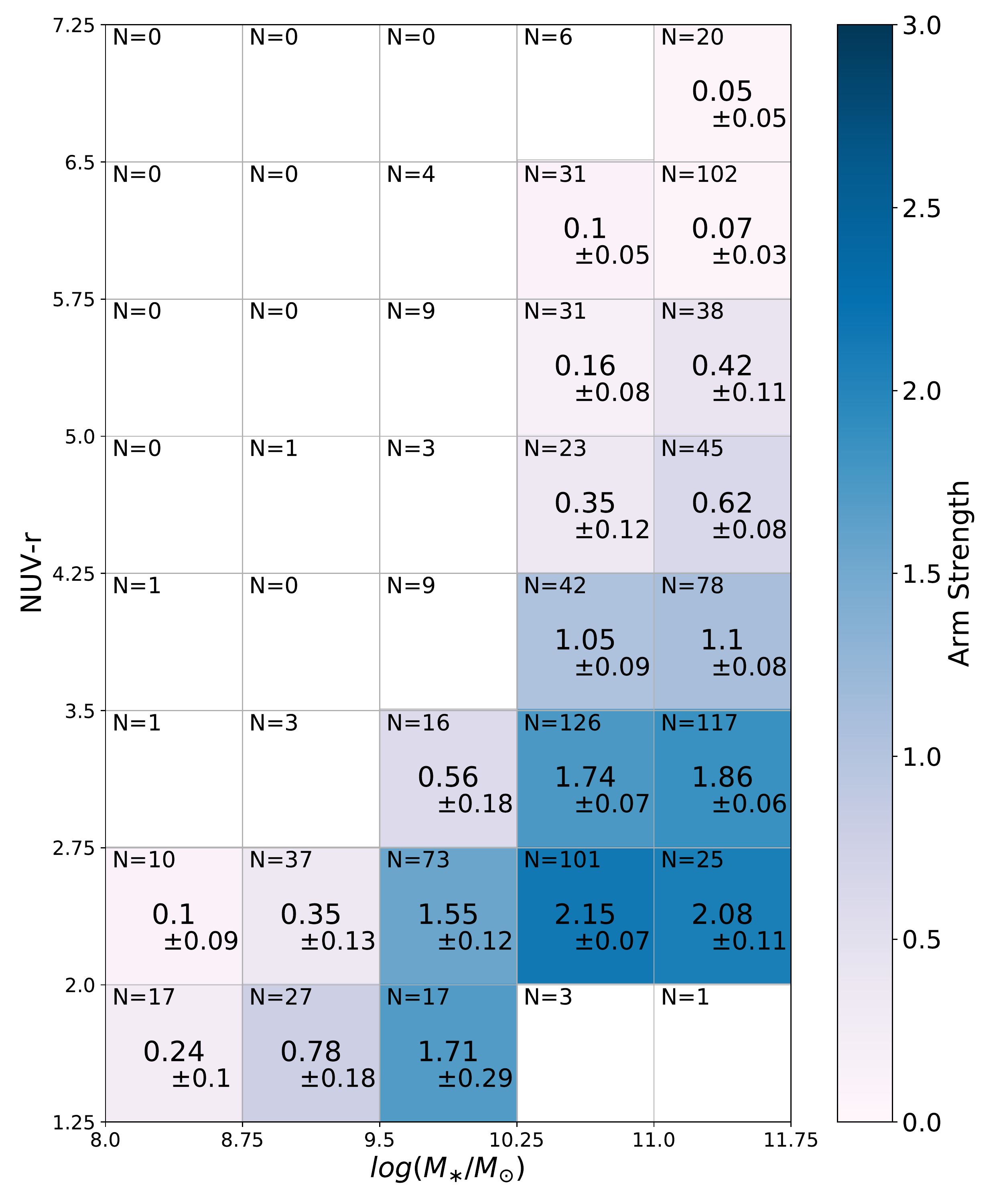}
    \includegraphics[width=0.33\linewidth]{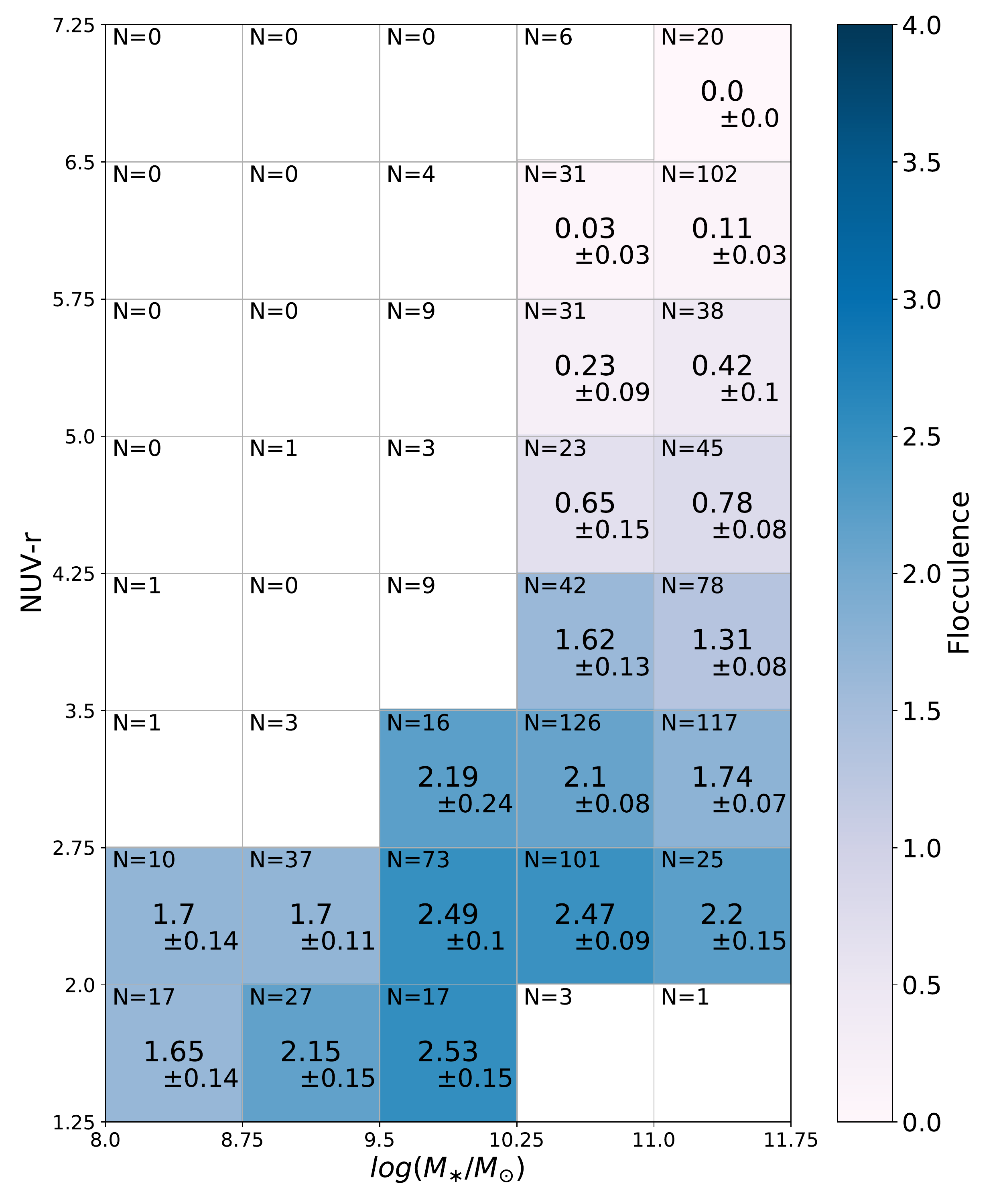}
    \caption{Evolution of EFIGI attributes related to disk galaxies.\textbf{Left:} Mean values of the EFIGI {\tt Arm Curvature} attribute for all EFIGI $\cap$ GALEX galaxies with {\tt Incl-Elong} $\leq 2$, in color-mass cells of 0.75 dex in stellar mass and 0.75 in $NUV-r$ color all the way from the Blue Cloud to the Red Sequence. {\tt Arm Curvature} describes the intrinsic curvature of the spiral arms, when they are visible (hence the smaller samples compared to the central and right panels). It increases from late-type to earlier type spirals, in agreement with the definition of these types in the Hubble-de Vaucouleurs sequence. \textbf{Center:} Same as the left panel but for the mean values of the EFIGI {\tt Arm Strength} attribute, which describes the strength of the spiral arms in terms of flux fraction relative to the whole galaxy. \textbf{Right:} Same as the left panel but for the mean values of the EFIGI {\tt Flocculence} attribute, which measures the relative importance of flocculent features, due to scattered HII regions. The joint fading of {\tt Flocculence} and {\tt Arm Strength} across the Green Plain hints at a restriction in neutral gas supply.}
    \label{disk_features}
\end{figure*}

Despite the major disk reddening detected for EFIGI galaxies across the color-magnitude sequence, left panel of \fg\ref{comparing_EFIGI_and_ZPEG_bulge_disk} shows that it operates at a constant bulge color: all SourceXtractor++ bulges are best fit by PEGASE.2 E template (but see Quilley \& de Lapparent, \textit{in prep.}, for evidence of subtle variations in bulge colors with morphological type). There is also an exception for the latest types Sdm, Sm and Im; in these galaxies, the bulge is very faint or does not exist, and a bright HII region near the galaxy center may be fitted by the bulge component of the model (the resulting SEDs are then best fit by the Im or starburst template, see \sct \ref{sed_bulge_disk}). The stability of bulge color, as long as the bulge is prominent enough to be fitted, confirms the results of \cite{2016ApJS..225....6K}, also based on bulge and disk decompositions of SDSS images, who find a constant bulge color at all $B/T$ (except for $B/T\le0.1$, where uncertainties dominate).

We now show that the $g-r$ disk reddening can be interpreted in terms of star formation fading of the disk. Because we do not have the disk $NUV$ photometry, we cannot relate directly the $g-r$ disk optical reddening to a fading of the sSFR. However, \fg\ref{color_galaxy_color_disk} shows the relation between the absolute $NUV-r$ galaxy color and the absolute $g-r$ disk color. There is a tight correlation all across the Blue Cloud ($NUV-r\lesssim 3.75$), that is from types Im to Sb, with both the disk and the full galaxies undergoing reddening simultaneously. Because the Blue Cloud morphological types (Im to Sb) are disk dominated ($B/T\lesssim0.1$, see \fg\ref{BT}), we can infer that the $NUV-r$ reddening of the entire galaxies must also be present in their disk component. As this $NUV-r$ reddening corresponds to a fading of star formation (and decrease in the sSFR; see \fgs\ref{color_galaxy_color_disk} and \ref{ssfr_color}), we can infer that the $g-r$ reddening of disk marks their star formation fading. 

\fg\ref{color_galaxy_color_disk} also shows that the reddening of earlier type disks (Sab, Sa and S0a) from the knee and across the Green Plain occurs within a wide $g-r$ disk color interval ($\sim[0.6, 0.8]$ that is $\sim0.2$ magnitude) while it also displays the stretching of $NUV-r$ due to the large range of sSFR across the Green Plain (discussed in \sct \ref{result-fading}). This graph illustrates the fact that the $NUV-r$ reddening and star formation fading across the Green Plain can also be detected in the optical via the disk $g-r$ fading. But as the early spiral types have significant bulge-to-total light and mass ratio of $\sim0.4$ (see \fg\ref{GV_bulge_ratio}), the total optical colors fail in highlighting the Green Plain, as shown for total $u-r$ colors in \fg\ref{u-r_density} in \sct \ref{NUV_detect_GV}). Note also that in the Red Sequence of \fg\ref{color_galaxy_color_disk}, where galaxies gets redder in $NUV-r$, so does their disk in $g-r$ whether real (in S0), or only an artifact (in E, that could be due to either an inadequate profile fit for the bulge, or to a visual misclassification of an S0 into an E).

\subsubsection{Spiral arms and flocculence as markers of star formation and gas supply 
\label{results-floc-spiral}}

The fading of star formation in disks along the Blue Cloud and across the Green Plain (described in the previous subsection) leads us to further examine the morphological features within the disks of galaxies: spiral arms and flocculence. The RC3 classification \citep{1991rc3..book.....D} rests on how tightly wound are the spirals arms of spiral galaxies, whereas lenticulars have arm-less disks. A progressive decrease in spiral arms curvature is therefore expected for later and later spiral types, and is confirmed by the mean cell values of the {\tt Arm Curvature} EFIGI attribute that is shown in the left panel of \fg\ref{disk_features}: {\tt Arm Curvature} increases systematically from the Blue Cloud (late spirals) to the Green Plain (early spirals and lenticulars) \citep[see also][]{2011A&A...532A..75D}. We note that if one only considers bins with more than $50$ galaxies, the abrupt step from $1.29$ to $1.92$ and higher to the left of the knee between the Blue Cloud and the Green Plain (in the second row) appears to correspond to the presence of Sc types and earlier, as shown in \fg\ref{dom_types}. Sc galaxies are called ``grand design spirals'' as their spiral arms are better defined (less broken by wide-scale flocculence) than those of the earlier and later types. 

The central panel of \fg\ref{disk_features} shows the EFIGI {\tt Arm Strength} attribute, that estimates the light fraction in spiral arms compared to the total galaxy flux. There is a clear peak at the knee between the Blue Cloud and Green Plain, that can be defined by the 2 rightmost cells of the second row, and exhibits 2 dark blue pixels with {\tt Arm Strength}$=2.15\pm0.07$ and {\tt Arm Strength}$=2.08\pm0.11$, dominated by Sb, Sbc, Sc and Scd intermediate spiral types in comparable proportions (see \fg\ref{dom_types}). From these 2 pixels, {\tt Arm Strength} decreases systematically across both the Blue Cloud and the Green Plain, at constant $NUV-r$ color when the stellar mass decreases, and at constant stellar mass across both the Blue Cloud and the Green Plain when $NUV-r$ increases (becomes redder), that is when the sSFR decreases. At $NUV-r>5.75$, {\tt Arm Strength} becomes negligible, as the low mass part of the Red Sequence is composed mostly of lenticulars, hence with no visible spiral arms.

The right panel of \fg\ref{disk_features} shows the EFIGI  {\tt Flocculence} attribute, that measures the flocculent aspect of galaxies due to scattered HII regions relative to the galaxy disk profile. As for the {\tt Arm Strength} attribute, {\tt Flocculence} shows a systematic decrease for redder $NUV-r$ color at all spiral stellar masses, except in the two lowest mass cells dominated by irregulars (see \fg\ref{dom_types}). The {\tt Flocculence} attribute corresponds to inhomogeneities relative to the smooth profiles of all dynamical features: disk, bar, spirals arms, rings etc. Given that the Green Plain is the region where star formation fades upward, the concomitant decrease in both {\tt Arm Strength} and {\tt Flocculence} (see central and right panel of \fg\ref{disk_features}) suggests a decreasing gas content of early spirals along the Green Plain. In contrast, Blue Sequence galaxies maintain a significant {\tt Flocculence} despite the vanishing {\tt Arm Strength} in the latest spiral types (Sdm, Sm, Im). If spiral arms trigger star-formation by compressing the interstellar gas and dust on their edges, they are not indispensable for star formation to occur: Im galaxies have no spiral arms and strong sSFR (see \fg\ref{ssfr_color}). The EFIGI {\tt Arm Strength} and {\tt Flocculence} attributes therefore bring additional evidence that the reddening of Green Plain galaxy disks may be due to some restriction in the neutral gas supply available for stellar formation (called quenching, see \sct \ref{discussion-quenching}).

\section{Discussion                               \label{discussion}}

The present analysis of EFIGI data shows systematic and major morphological variations in the color-mass diagram (see \fg\ref{color_mass}), encompassing all galaxy types from the Main Sequence of star-forming galaxies (Blue Cloud) to quiescent ones (Red Sequence), through an intermediate region which we rename Green Plain (see \scts \ref{from_efigi_to_morcat} and \ref{result-fading}). We characterize these changes not only in terms of Hubble type, but quantitatively in terms of bulge growth (see \fg\ref{BT}) and disk reddening (see \fg\ref{disk_color}),  and show below that altogether, they strongly advocate for the Green Plain being a slow transition region between star-forming and quiescence.

We also discuss below the results of our analysis and show that these results obtained from low redshift galaxies confirm what was suggested from high redshift observations in the Hubble Deep Field \citep{1996MNRAS.279L..47A}: that irregulars and the very late spiral types are the building blocks of the more massive early spiral type galaxies, and eventually of the lenticular and elliptical galaxies (see \sct \ref{discussion-mergers}). Altogether, the mass distributions and limits for the bulges, disks, and more generally the various morphological types suggest that changes may take place on long timescales, as they may be triggered by flybys or mergers, and are tightly intertwined.

\subsection{Quenching   \label{discussion-quenching}}

Transition of galaxies from the Blue Cloud (star-forming) to the Red Sequence (quiescent) is obviously related to the gas cycle and resulting star formation in galaxies. Numerical simulations suggest that galaxies are gas fed from the cosmic streams \citep{2006MNRAS.368....2D}, the circumgalactic medium (\citealt{2004MNRAS.355..694M}, \citealt{2008MNRAS.383..119D}, \citealt{2009ApJ...700L...1K}), as well as during mergers. The decrease in star formation of galaxies with time \citep{2014ARA&A..52..415M} has been attributed to quenching that would limit the cold gas supply. Quenching is theorized to either prevent gas from forming stars, or to prevent the building up of the gas reservoir, or even to expel gas from the galaxy. AGN feedback has been considered as playing a crucial role, by thermal feedback - causing gas heating \citep{2006MNRAS.365...11C}, or mechanical feedback - expelling the gas \citep{1998A&A...331L...1S}. There is however a significant scale ratio between the parsec scale of the central engine of an AGN, namely the supermassive black hole, and the more than kiloparsec scale of a galaxy disk and halo, questioning the ability of the AGN to affect the cold gas content of entire disks. 

We emphasize that there is often confusion in articles between the decrease in star formation (the symptom) and the quenching (the cause), by invoking the latter when only observing the former. Here we make the distinction and, when appropriate, specifically refer to the symptom as \textbf{star formation fading}.

We note that the 4 order of magnitude range in stellar mass identified here for irregulars to spirals along the Blue Cloud (see \fg\ref{color_mass}) may have implications in terms of quenching. If the gas accretion from the cosmic streams \citep{2006MNRAS.368....2D} is independent of a galaxy mass but rather on its location in the cosmic web, this accretion may correspond to a smaller fraction of the gas reservoir in more massive galaxies. It may therefore not be sufficient to replenish the gas reservoir consumed by the on-going star formation above some mass threshold. This would lead to a quenching of star formation, hence redden the most massive spirals into the Green Plain. Combined with the mass limit described in \sct \ref{results_ssfr_m}, this could design the knee shape of the Green Plain.

\subsection{Major mergers along the Blue Cloud      \label{discussion-mergers}}

The 4 order of magnitude extent in $M_\ast$ of the Blue Cloud (see \fg\ref{color_mass}) can be partly explained by galaxies consuming their large HI reservoir \citep{2016A&A...595A.118V} to form stars, which is made possible by their high sSFR values in the range of $10^{-9.7} yr^{-1}$ for irregulars to $10^{-10.5} yr^{-1}$ for Sc types (see \fg\ref{ssfr_color}). However, these galaxies would need several to almost 10 Gyr years to double their mass, so the scenario of an irregular or spiral galaxy evolving passively across the entire Blue Cloud can be ruled out. This indicates that mergers are also needed to explain the stellar mass increase for late-type spirals, and that these mergers should be major in order to allow the building-up of the massive spirals at the tip of the Blue Cloud.

\begin{figure}
    \includegraphics[width=\columnwidth]{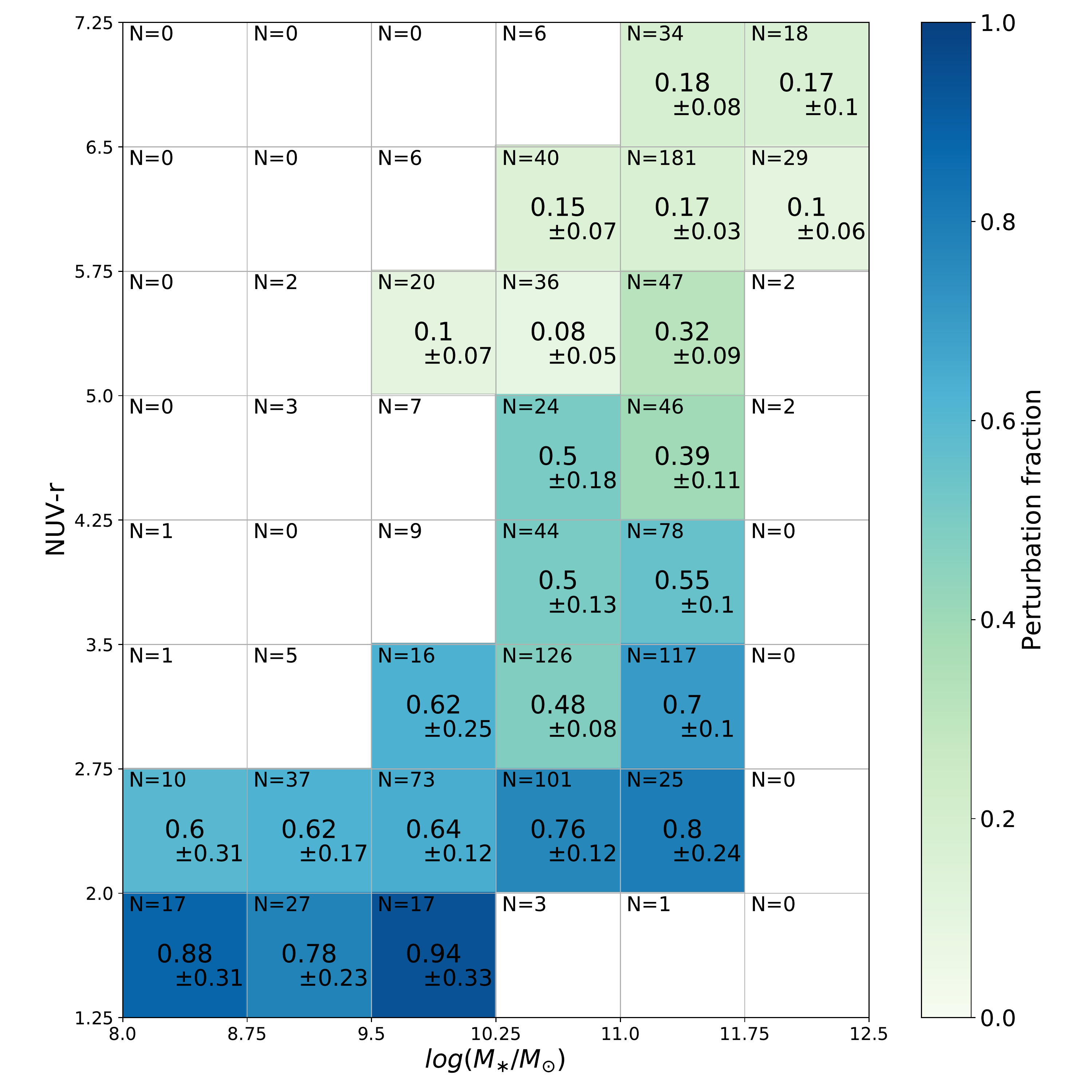}
    \caption{Fraction of EFIGI $\cap$ GALEX galaxies with {\tt Incl-Elong} $\leq 2$ having an attribute value of {\tt Perturbation} $>0$, all the way from the Blue Could to the Red Sequence. This attribute measures the distortion of the galaxy profile from rotational symmetry, with a nonzero value thus indicating an asymmetry. Galaxies are binned in cells of 0.75 dex in stellar mass and 0.75 in $NUV-r$ color, and cells with less than 10 galaxies are discarded. The sample size as well as the error on the mean are indicated in each cell. Red Sequence galaxies are mostly unperturbed, while many of the Blue Cloud galaxies show various levels of {\tt Perturbation}, with a trend of bluer galaxies being more frequently perturbed.}
    \label{perturbation}
\end{figure}

Another piece of evidence for mergers along the Blue Cloud can be derived from the {\tt Perturbation} attribute available in the EFIGI catalog, which measures the amplitude of distortions in the galaxy luminosity profile from rotational symmetry. Galaxies that went through a recent merger event, or only a flyby are expected to undergo strong to weak gravitational tidal field effects and consequently to have a non null value of the {\tt Perturbation} attribute (see the study of peculiar galaxies to understand interactions, initiated by \citealt{1966ApJS...14....1A}). \fg\ref{perturbation} shows the fraction of galaxies having an EFIGI  {\tt Perturbation} attribute $>0$. One can see that the Blue Cloud ($NUV-r \leq 3.5$) is the region of the graph where there is the highest fraction of perturbed galaxies, with only $34\%$ showing no perturbation, against $62\%$ for the bins roughly corresponding to the Green Plain ($NUV-r \in [3.5,5.75]$) and $83\%$ in the Red Sequence ($NUV-r > 5.75$).
There is also a vertical trend over most of the Blue Cloud, in which the bluest galaxies are more frequently perturbed at a fixed mass (at the $\sim2\sigma$ level). We interpret both trends as evidence that merger events increase star-formation in already star-forming irregular and spiral galaxies.

Another piece of evidence for mergers along the Blue Cloud is the presence of very bright regions, which are frequently visible along the spiral arms, but with no preferred location in Im galaxies, and they are characterized by the EFIGI {\tt Hot Spots} attribute - on a scale of 0 to 4. It measures the strength of (a) region(s) with a very high surface brightness standing out in a galaxy and that correspond(s) to either a giant star-forming region, an active nuclei (very rarely), or a stellar nuclei (in the nucleated dE galaxies). The vast majority of the visually detected hot spots with low to intermediate values of the attribute (1-2) are strong HII regions, appearing as blue dots, contrary to the flocculence patches that are more diffuse and on larger spatial scales. In contrast, the highest values of the {\tt Hot Spots} attribute (3,4) correspond to (often a single) huge star-bursting region, and are rarer by a factor of $\sim 10$. 

\begin{figure}
    \includegraphics[width=\columnwidth]{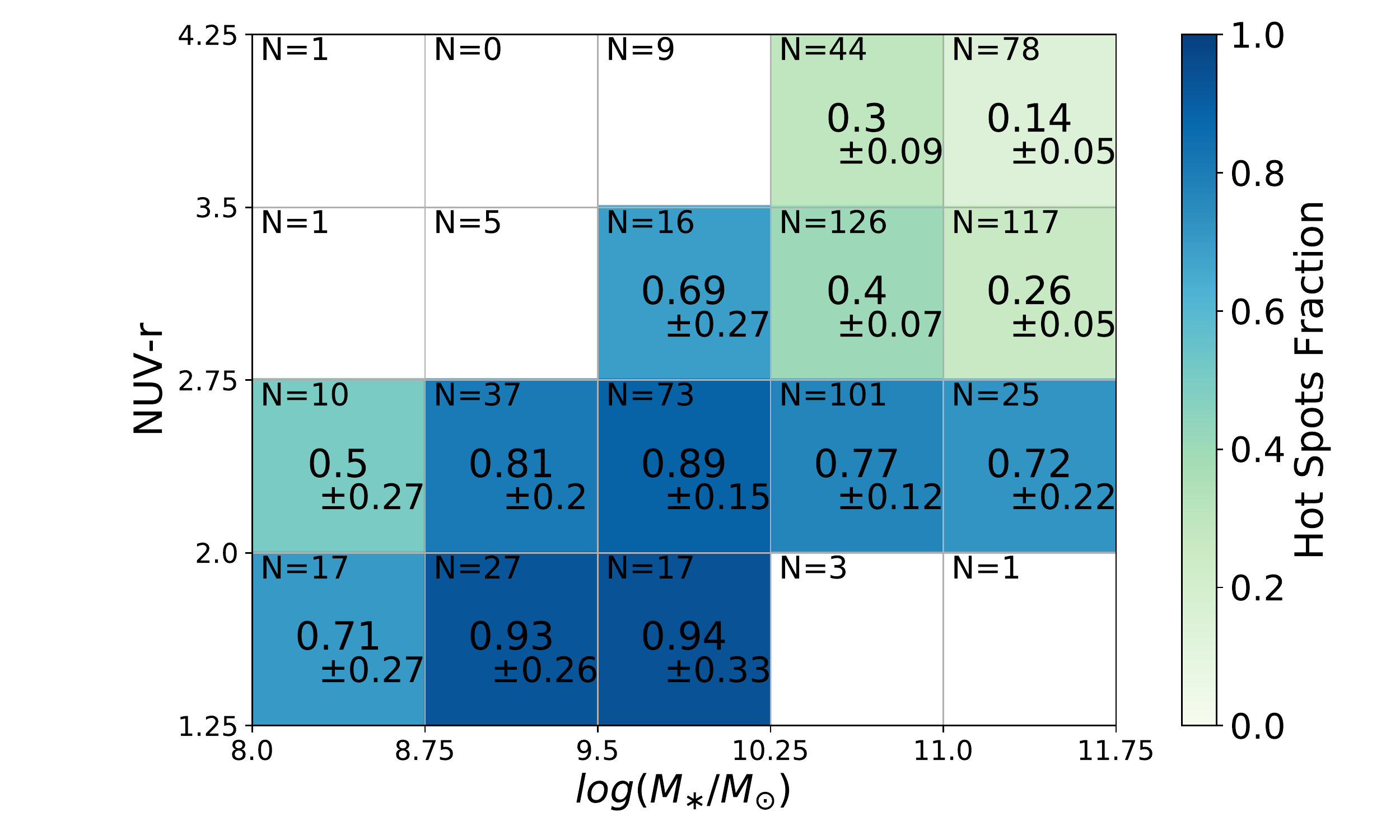}
    \caption{Fraction of EFIGI $\cap$ GALEX galaxies with {\tt Incl-Elong} $\leq 2$ having an attribute value of {\tt Hot Spots} $>0$. The attribute measures the relative intensity of regions with a very high surface brightness. Galaxies are binned in cells of 0.75 dex in stellar mass and 0.75 in $NUV-r$ color. The sample size as well as the error on the mean are indicated in each cell. Cells with less than 10 galaxies are discarded, and the graph is limited to $NUV-r\leq4.25$, as the fraction of {\tt Hot Spots}$>0$ is below $5\%$ for redder galaxies.}
    \label{hot_spots}
\end{figure}

\fg\ref{hot_spots} shows the fraction of galaxies having some weak to strong evidence for hot spots within the Blue Cloud, in color-mass cells. There is, similarly to the {\tt Perturbation} attribute a color trend with the bluest galaxies showing the highest frequencies of hot spot presence, at any given mass. Comparison with \fg\ref{dom_types} shows that the vast majority of late-type spirals (Sd to Sm) and Im show evidence for some hot spot(s), as galaxies with $\log(M_\ast/M_\odot) \in [8.75,10.25]$ and $NUV-r < 2.75$ show a fraction $>0.81$ (the slight decrease in the Irregular dominated bin with a $0.5 \pm 0.27$ fraction seen in \fg\ref{dom_types} is not significant given the quoted error). As for the {\tt Perturbation} attribute, there is a vertical trend over most of the Blue Cloud, as well as the shown portion of the Green Plain, in which the bluest galaxies have more frequent visible hot spots in the form of bright HII regions. This attribute value distribution is also compatible with gravitational interactions and mergers driving the stellar mass build-up in the Blue Cloud: a minor encounter or a cosmic stream connecting to a specific region of a disk could lead to locally increased star formation; a tidal interaction on larger scales (due to a flyby) may lead to a general enhancement of the star formation in the disk and the rise of several bright HII regions.

In contrast, \fg\ref{perturbation} shows that in the Red Sequence, there is a rather low fraction of galaxies showing perturbed isophotes ($< 20\%$), at all masses, so it is valid for both lenticulars and ellipticals. There are two ways to interpret this lower fraction, either mergers are not as common for red galaxies, or the distortion of the light profile is often too weak or too short-lived to be visible.
\citet{1995ApJ...438L..75M} suggests that major mergers of spiral galaxies into ellipticals at $z\sim0.4-1$ erase signs of distorted isophotes in only 0.2 Gyr, due to violent relaxation. This could explain the very low fraction of perturbed EFIGI galaxies in the massive part ($\log(M_{\ast}/M_{\odot})\gtrsim11.75$) of the Red Sequence, where ellipticals dominate. The similarly low frequency of perturbed galaxies in the low mass portion of the Red Sequence, in which S0 dominate, may indicate that the transformation from Green Plain spirals to relaxed lenticulars occurs on a comparable time-scale, despite the different nature and frequency of these mergers: minor and frequent mergers. We actually provide in \sct and \ref{discussion-bulge} and \sct \ref{discussion-masslimit} arguments in favors of mergers forming galaxies in the Red Sequence, that are not contradicted by the {\tt Perturbation} attribute distribution.

\subsection{Main sequence of star-forming galaxies         \label{main-seq}}

The Blue Cloud is often referred to as the Main Sequence of star-forming galaxies, which was first discovered as a linear relation between the star formation rate and the stellar mass (in log-log), onto which an overwhelming majority of star-forming galaxies align \citep{2004MNRAS.351.1151B, 2007A&A...468...33E, 2007ApJ...660L..43N}. The rather small scatter of this relation, on the order of $0.3\, \mathrm{dex}$, raises the question of the star-forming mechanisms that could explain this dispersion.

In the present analysis, we show that the variation along the Main Sequence can be described by the continuous change of morphological type from Im to Sbc (see \fg\ref{mean-NUV-r-face-on}). Moreover, the Main Sequence is curved at its massive end when reaching Sb and earlier types, corresponding to the knee of the Green Plain.
As far as the scatter around the Main Sequence is concerned, we have brought to light several EFIGI morphological attributes that follow an increasing trend transverse to the Blue Cloud (hence the Main Sequence) when moving toward bluer $NUV-r$ colors: these attributes are spiral {\tt Arm Strength} (\fg\ref{bar_frequency}), {\tt Flocculence} (\fg\ref{disk_features}), {\tt Perturbation} (\fg\ref{perturbation}) and {\tt Hot Spots} (\fg\ref{hot_spots}), which are all linked to the stellar formation processes in galaxies, either as a direct measure (flocculence, bright HII regions), or its triggering (strength of spiral arms, amplitude of distorted isophotes).
We thus confirm that the scatter around the Main Sequence corresponds to variations in stellar formation, and identify dynamical processes that could cause them.

\subsection{Transition through the Green Plain       \label{discussion-transition}}

In the following, we further examine and discuss the properties of the Green Plain, and suggest that it is a transitional region through which galaxies may pass when evolving from the Blue Cloud to the Red Sequence.

\subsubsection{Bulge growth requires mergers    \label{discussion-bulge}}

We propose here that the strong increase in bulge mass fraction across the Green Plain excludes a secular and passive reddening of galaxies from the Blue Cloud to the Red Sequence. Indeed, one can make the distinction between two kinds of bulges according to their formation processes: classical bulges, that were formed through violent events such as mergers, and pseudo-bulges, that were formed by secular evolution, with inward stellar migration from the disk to the bulge, over longer timescales \citep{2004ARA&A..42..603K}. These different formation scenarios lead to different properties.

\begin{figure}
  \includegraphics[width=\linewidth]{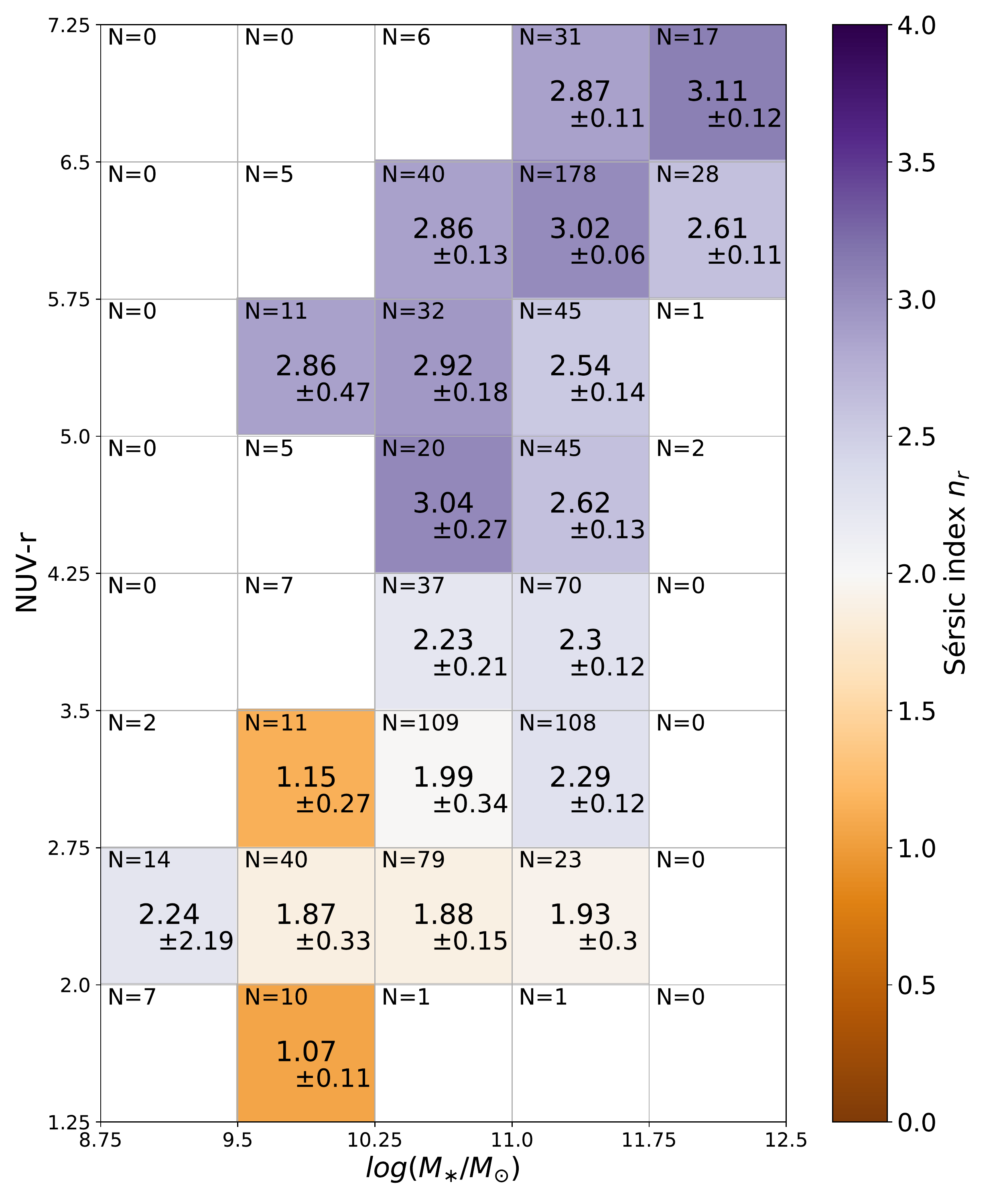}
    \caption{Mean value of the S\'ersic index of the bulge profile in the $r$ band for galaxies in cells of 0.75 dex in stellar mass and 0.75 in $NUV-r$ color. Cells with less than ten galaxies are discarded, and the sample size as well as the maximum of the error on the mean and the quadratic mean of the errors on the individual points. Pseudo-bulges populate the Blue Cloud, whereas the Green Plain and the Red Sequence are dominated by classical bulges.}
    \label{sersic}
\end{figure}

The S\'ersic index of a bulge light profile (obtained through the bulge and disk modeling, see \sct \ref{methodo_srx}) has been proposed as a criterion to differentiate pseudo-bulges from classical bulges, with $n \leq 2$ and $n \geq 2$ respectively (\citealt{2008AJ....136..773F}, \citealt{2009MNRAS.393.1531G}). In order to further examine the pseudo-bulge and classical bulge distributions, we look in \fg\ref{sersic} at the variations of the bulge profile S\'ersic index as measured in the $r$ band, averaged in cells across the color-mass diagram. Compared to previous parameters, the quadratic mean of errors for the S\'ersic index can be higher than the error on the mean we defined as $\sigma_{RMS}/\sqrt(N)$ ($N$ being the number of galaxies in the bin) because of large error bars on this index. The plotted error in \fg\ref{sersic} is therefore the maximum of the two estimates. We precise that the value of $2.191$ for the leftmost bin is caused by the large uncertainties for individual galaxies with very low bulge fluxes (most of these bulges are probably not real, and due to bright HII regions).

\fg\ref{sersic} shows that there is a systematic increase in the S\'ersic index of the bulge profile when $NUV-r$ increases in the color-mass diagram, that is from late spiral types in the Blue Cloud to early types in the Red Sequence, meaning galaxies have more concentrated bulge light profiles when moving from the bottom-left to the top-right of the color-mass diagram. In the Blue Cloud and all the way to the knee of the Green Plain (for $NUV-r<3.5$), galaxies have a mean S\'ersic index of $n \sim 1.07-2.3$, thus indicating a mix of pseudo-bulges and classical bulges, or composite bulges which may have both a pseudo and a classical component. In contrast, the Green Plain galaxies have an average $n \sim2.2-2.9$, and the Red Sequence an average $n \sim2.6-3.1$, making all these galaxies dominated by classical bulges. Therefore, there is not only a growth of the bulge across the Green Plain (as seen in \sct\ref{bulge}) but also a change in the nature of these bulges, from pseudo-bulges to classical ones, as evidenced here by the sharpness of their light profile. These results are in agreement with \cite{2016ApJS..225....6K}, who showed a positive correlation between $B/T$ and bulge S\'ersic index for SDSS galaxies in a similar redshift range as for EFIGI and MorCat ($0.005 < z < 0.05$).

\begin{figure}
  \includegraphics[width=\linewidth]{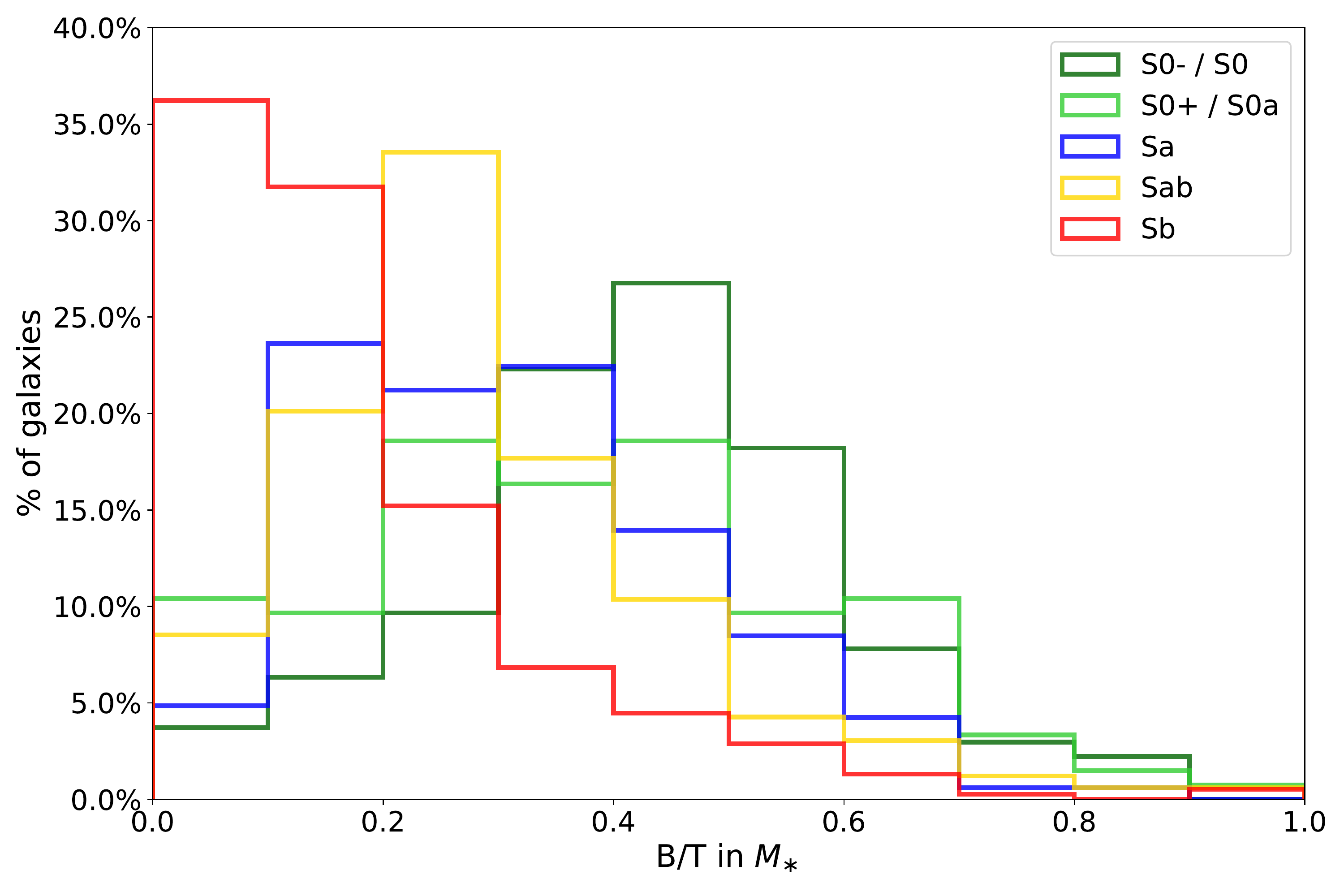}
    \caption{Relative distributions of the bulge-over-total stellar mass ratio for early spirals and lenticulars. Each color represents a specific Hubble Type (or a combination of subsequent Hubble types). Despite a large dispersion, $B/T$  systematically increases with morphological type. In particular, transformation of Sb types into S0$^-$-S0 or S0$^+$-S0a often requires bulge growth.}
    \label{BT_hist}
\end{figure}

In \fg\ref{BT_hist}, we plot the relative distributions of the bulge-over-total stellar mass ratio for early spirals and groups of lenticular types, each histogram being normalized to 100\%. Moving along the EFIGI Hubble sequence from Sb to S0, one can see that the peak of the distribution corresponds to increasing values of $B/T$, thus indicating a growth of the bulge prominence in galaxies of these types: $B/T$ peaks at $\sim0.45$, $\sim0.35$, $\sim0.25$, $\sim0.25$ and $\sim0.1$ for S0$^-$-S0, S0$^+$-S0, Sa, Sab and Sb types respectively, and to lower values for later spiral types; and median values are 0.42, 0.37, 0.30, 0.26, 0.14 for S0$^-$-S0, S0$^+$-S0, Sa, Sab and Sb types respectively. Therefore, the dominant fractions of Sb types and later in the Blue Cloud (see \fg\ref{dom_types}) imply that pseudo-bulges correspond to the $B/T$ values found in the Blue Cloud, whereas the $B/T$ of bulges of S0$^{-}$, S0, S0$^+$ lenticulars located in the Red Sequence correspond to the values for classical bulges.
\citet{2017ApJ...840...79S} show from HST and SDSS data that, from $z=1$ to $z=0$, both pseudo-bulges and classical bulges grow significantly, with $B/T$ in optical bands increasing from $21\%$ to $52\%$ for classical bulges, and pseudo-bulges from $10\%$ to $26\%$. If we compare these values at $z=0$ to the peak values of \fg\ref{BT_hist}, we find that Sab and later types have pseudo-bulges, lenticulars have classical bulges, while the picture is a bit more mixed for Sa galaxies due to their wide distribution of $B/T$. This analysis of the $B/T$ distributions per morphological type further confirms the shift from pseudo-bulges in the Blue Cloud to classical bulges in the Red Sequence. \cite{2017ApJ...840...79S} also conclude that minor mergers are needed to obtain sufficient mass growth to produce classical bulges. This translates in our analysis into a need for minor mergers to explain the main trend of bulge growth seen across the Green Plain, and to reach the $B/T$ observed for the lenticular galaxies.

Moreover, \fg\ref{BT_hist} shows that  there is a wide distribution of $B/T$ for each group of morphological types. We calculate that only $17.8\%$ of Sb have a $B/T \geq 0.3$, and $19.7\%$ of S0$^-$-S0 have a $B/T \leq 0.3$. More generally, there are no bins in $B/T$ that include significant fractions of both S0 and Sb galaxies, suggesting that only a minority of Sb galaxies could evolve from spirals to lenticulars without seeing a significant change in their $B/T$. Therefore, secular evolution cannot be the main evolutionary pathway across the Green Plain, because processes that increase significantly the $B/T$ are required for galaxies to transition across this region.

We emphasize that this picture of galaxy evolution in which the fading of star formation is linked to the growth of the bulge (through mergers), is compatible with the concept of ``morphological quenching'' proposed by \cite{2009ApJ...707..250M}. The growth of a stellar bulge stabilizes the gas disk and quenches star formation, leading to a red quiescent galaxy without the need to deplete its gas supply.

\subsubsection{Slow transition through the Green Plain        \label{discussion-noquick}}

If there is a mass change in the color-mass diagram, the process that drives the mass growth is expected to impact the star forming state of the galaxy, hence its color. Gas-rich mergers initially create a burst of star formation which makes temporarily the galaxy bluer, before it proceeds with its secular fading. 

The fact that the Green Plain is nearly vertical in the color-mass diagram, with a mass range of $\log(M_\ast/M_\odot)\sim 10.5-11.7$, that is a factor of $\sim10$, leaves room for major mergers to play a role in the process of type transformation across this region. However, as the rate of major mergers decreases sharply from $z\sim1$ to $z\sim0$ \citep{2009A&A...501..505L, 2009MNRAS.394L..51B, 2010ApJ...709.1067B}, they can only play a minor role in recent galaxy evolution. The building-up of the bulge across the Green Plain is expected to take several Gyr, if we consider the timescales quoted in \cite{2017ApJ...840...79S}, as a growth of $B/T$ by a factor $\sim2.5$ occurs between $z=1$ and $z=0$.

The coupling of star formation fading and morphological transformation detected across the Green Plain suggests that the timescales of these two phenomena may be of the same order. Indeed, \cite{2022MNRAS.509..567S} derive from the analysis of 257 SDSS clusters that the morphological transition happens prior to the fading, or quenching, of star formation, with respective timescales of $\SI{1}{Gyr}$ and $\SI{3}{Gyr}$. Note however that the authors define morphological transformations as when the galaxy converts from Sa to S0, hence the short timescale of 1 Gyr. Nevertheless, we observe in their \fg 9 that the full morphological transformation from type Sbc (Blue Cloud) to S0, along with the decrease in sSFR, follows the same trend over the 5 Gyr that these changes take altogether. The dense environments of galaxy clusters analyzed by \cite{2022MNRAS.509..567S}, in which ram pressure effects are efficient at removing gas from disks differ from the EFIGI and MorCat catalogs analyzed here that sample the large-scale galaxy distribution, with some clusters and many groups \citep{2011A&A...532A..74B}. Therefore even longer timescales than 1 and 5 Gyr may be needed to convert Sa and Sbc types to S0 respectively in the more representative environment of the general galaxy distribution provided by EFIGI and MorCat. Indeed, \cite{2019MNRAS.485.5559P} derived from SED fitting the star formation histories of galaxies in the GAMA survey (at redshifts  $0.1<z<0.2$), assuming an exponential decay of the SFR: their conclusions are that Green Plain galaxies are experiencing a star formation decrease, due to the gradual shortage of their gas reservoir, over a timescale of 2 to 4 Gyrs.

\subsection{Mass limits and formation of massive disks and ellipticals \label{discussion-masslimit}}

Another interesting result seen in the color-mass diagram of \fg\ref{color_mass}, is that the Green Plain is nearly vertically aligned with the low-mass end of the Red Sequence and the high-mass end of the Blue Cloud ($\log(M_\ast/M_\odot) \in [10.5,11.7]$; \fg\ref{color_mass}). Furthermore, there is a common mass limit of $\log(M_\ast) \sim 11.7 M_\odot$ for all early-type spirals from Sa to Sc. This probably corresponds to the exponential decrease in the galaxy luminosity function at bright luminosities \citep{2003A&A...408..845D}, which is successfully reproduced in numerical simulations when introducing AGN feedback \citep{2006MNRAS.370..645B, 2017MNRAS.467.4739K}. This mass limit also constitutes an almost upper mass limit for the lenticular galaxies within the Red Sequence, as shown in \fg\ref{ell_vs_len_mass}, whereas elliptical galaxies dominate the Red Sequence at high masses with $\log(M_\ast/M_\odot) \gtrsim 11.5$. This is consistent with kinematic studies which show that there exists a critical mass $\log(M_{crit}/M_\odot)\simeq 11.3$ separating slow and fast rotators \citep{Capellari2016}.

We now examine how these mass limits and mass segregation with galaxy morphology can provide useful clues on the evolution of the various morphological types. Secular evolution from the low-mass lenticular part of the Red Sequence into its massive part dominated by ellipticals is not possible because the sSFR of both types of galaxies is too low to significantly increase their stellar mass by secular evolution ($\mathrm{sSFR} < 10^{-11} {yr}^{-1}$ or $10^{-2} {Gyr}^{-1}$; the sSFR can be seen as the inverse of the time needed to produce the stellar mass of a galaxy at its current SFR). Mergers (either major or many minor) are definitely needed to explain the existence of Red Sequence ellipticals with stellar masses larger than $ log(M_\ast/M_\odot) \sim 11.5$.

This result is in good agreement with the results of \cite{2018MNRAS.480.2266M} who show from the cosmological hydrodynamical Horizon-AGN simulation, that mergers play a major part in driving morphological transformation, and especially disk-to-spheroid evolution, with this transformation being more drastic as the stellar mass of galaxies involved in the merger increases. Indeed, the difference in the dispersion of the stellar velocities between the most massive progenitor and the galaxy formed by the merger event is negative for a disk progenitor, indicating a shift toward a more dispersion-supported object, while the absolute value of the dispersion increases with stellar mass. The fact that the more massive spirals tend to form spheroids after a merger event is expected to deplete the color-mass diagram from early-spiral galaxies at the highest masses, as we obtain in \fg\ref{color_mass}.

In \cite{2018MNRAS.480.2266M}, the highest mass bin where the morphological transformation is strongest at $z=0$ is $\log(M_\ast/M_\odot) \in [11.5, 12.0]$, which corresponds to the mass limit of $\log(M_\ast/M_\odot)> 11.7$ detected in the present study. Moreover, the shift in morphology from disk to spheroid triggered by the merger event strongly depends on the mass ratio between the two merging galaxies, with an effect significantly higher for major mergers. Nevertheless, their low frequency rate, especially since $z \sim 1$ means that minor mergers are also needed to explain the formation of elliptical galaxies. The formation of either disk-dominated or bulge-dominated galaxies also cannot be solely explained by a difference in the merger history, and, still according to \cite{2018MNRAS.480.2266M}, it is expected to mostly depend on the gas fraction of the progenitors, which is key to determine the remnant morphology. 

Furthermore, our results are compatible with the samples analyzed in \cite{2020MNRAS.494.5568J} and \cite{2022MNRAS.511..607J} that focus on the most massive spirals, with $\log(M_\ast/M_\odot) \geq 11.4$, with the goal to understand why, if the most massive galaxies have a rich merger history, these successive mergers do not systematically lead to an elliptical morphology. The authors answer to the issue of the existence of massive spiral galaxies by showing that gas-rich minor mergers allow for the persistence of a star-forming disk. \cite{2018MNRAS.480.2266M} also mention gas-rich interactions, as well as cosmological accretion, to explain that disk survival in mergers is driven by the accretion of cold gas.

At higher redshifts, \citet{2015MNRAS.450.1268D, 2016MNRAS.461.4517D, 2017ApJ...834..109M, 2021ApJ...920...32C} observe in Red Sequence cluster galaxies a gradual disappearance of the disks, corresponding to the lenticular toward elliptical evolution reported here, and attribute it to disk fading and bar or bulge formation, without the need for mergers. So it seems that the morphological evolution around this mass limit could also happen through secular evolution, and mergers would then only be needed to explain the mass growth to form the most massive ellipticals.

\subsection{Constraining evolutionary pathways  \label{discussion-pathways}}

Numerical simulations propose scenarios of type transformations that we can confront with our observational results in order to check their compatibility.

\subsubsection{Evolution pathways of lenticulars in numerical simulations \label{discussion-S0}}

When discussing the growth of bulges and the transition across the Green Plain in \sct \ref{discussion-transition}, we extensively discussed the formation of lenticular galaxies. We here confront our observations with \cite{2021MNRAS.508..895D}
who study the formation pathways of lenticular galaxies in the Illustris numerical simulation. This analysis puts an emphasis on mergers (both minor and major) in order to explain the formation of lenticulars from spirals in $57\%$ of cases. Another major pathway to form $37\%$ of lenticulars in this simulation is through tidal stripping of spiral disks. This process is however at play only in clusters, which are rare in surveys of the large-scale galaxy distribution that are more extended than the typical diameter of the voids in the cosmic web \citep{1986ApJ...302L...1D}, which the EFIGI and MorCat samples are; such surveys are therefore dominated by group galaxies in which stripping mechanisms are less efficient. A third and less dominant pathway to form $5\%$ of lenticular galaxies identified by \citet{2021MNRAS.508..895D} is by passive evolution of spiral galaxies. We already discuss this scenario when describing the growth of bulges, and conclude that it is possible for only a minority of objects (see \sct\ref{discussion-bulge}), hence in agreement with \cite{2021MNRAS.508..895D}.

\subsubsection{Evolution pathways of ellipticals in numerical simulations \label{discussion-ell}}

Unfortunately, we do not find as much common ground as with \cite{2021MNRAS.508..895D} (discussed in the previous subsection), in other studies based on numerical simulations. For instance, \cite{2021arXiv211207679P} also used Illustris TNG50 cosmological hydrodynamical simulations of galaxy formation, and show that there are three main pathways leading to massive quiescent galaxies with $\log(M_\ast/M_\odot) \in [10.5,11.5]$. Even though such galaxies are mainly lenticulars in the EFIGI sample (as we show in \fgs\ref{mass_mag}, \ref{color_mass} and \ref{ell_vs_len_mass}), our observations are not in agreement with the evolutionary trajectories seen in \cite{2021arXiv211207679P}. We emphasize that understanding their results in light of ours can be prejudiced by their choice in definition for the two splits describing galaxies : star-forming and quiescent as well as disk and elliptical. While the cut in sSFR shown in their \fg 1 is coherent with our measured values presented in \fg\ref{ssfr_color}, their disk-to-total ratio $D/T$ defined as the fractions of stars with circular orbits is different from the mass or luminosity ratios considered here, and a cut at 0.5 in stellar mass would not be adequate to split disks from ellipticals: \fg 26 shows that 14\%, 10\% and 6\% of early-spiral galaxies, of Sa, Sab and Sb morphological types respectively, have a $B/T > 0.5$ (hence $D/T<0.5$), as well as up to a third of lenticulars.

Nevertheless, there are three evolutionary pathways to form massive quiescent ellipticals that are identified in the TNG50 numerical simulation, and that one can examine in light of the color-mass diagram for the EFIGI galaxies. The first scenario is a slow decrease in star formation turning star-forming disks into quiescent disks, which could be compatible with our observations, if we consider transition from early-type spiral to lenticulars that have similar values of $B/T$ (see \fg\ref{BT_hist}).

The second scenario is in two phases: a rapid star formation fading caused by strong AGN feedback (gas ejection) turns star-forming disks into quiescent disks; it is then followed by a disk dispersion mechanism turning the galaxy into a rather low-mass elliptical. The lenticulars and the less massive ellipticals overlap in the color-mass diagram (see \fg\ref{ell_vs_len_mass}), so this second phase morphological change between the two is possible; however, our EFIGI results strongly advocate against a rapid star formation fading or quenching due to the significant bulge growth across the Green Plain (see \sct \ref{bulge}).

The third scenario involves (minor or major) mergers or misaligned gas accretion transforming a star-forming disk into a star-forming elliptical, that becomes quiescent through a delayed-then-rapid star formation fading or quenching process. This scenario seems incompatible with the EFIGI sample because it contains only 2 elliptical galaxies, out of the 158 in EFIGI $\cap$ GALEX with a non zero value of sSFR, and they are the lowest possible values (corresponding to the Sa SED template). These blue, hence star-forming, ellipticals seen in the simulation do not seem to exist in the nearby Universe: this class of object could be an artifact of the simulation. We therefore consider that the described pathway is at best a rare evolution pattern for the formation of massive quiescent galaxies.

At last, our analysis, by showing that Hubble types span the color-mass diagram at specific locations (see \fgs\ref{color_mass} and \ref{mean-NUV-r-face-on}), and that many structural parameters follow the same trend, seems to indicate that the fading of star formation and the morphological transformations are intertwined. Therefore, we reach an opposite conclusion to \cite{2021arXiv211207679P} who finds that ``quenching and morphological transformation are decoupled''. We can temper our conclusion by mentioning the dispersion seen for each Hubble Type, that allows a galaxy to transform between subsequent types while maintaining its star formation constant, or to undergo a small decrease in its sSFR with no change in its morphology. Nevertheless, only minor changes are compatible with the values of the dispersion in both color and mass.

\section{Conclusions}

In this article, we examine the distribution of Hubble types for the EFIGI catalog of nearby galaxies with detailed visual classification in the $NUV-r$ versus $r$ absolute color-magnitude and color-mass planes. The goal is to relate the morphology of galaxies to their evolution in terms of stellar mass and the star-forming state. We obtained precise apparent magnitudes and colors for all types of galaxies, as well as for their bulge and disk components by modeling the luminosity profiles of galaxy images from the SDSS using the Euclid SourceXtractor++ software, which allows one to perform simultaneous multiband model-fitting with priors. We then used ZPEG software to adjust SED based on templates from the PEGASE.2 scenarios of spectrophotometric galaxy evolution, to the apparent magnitudes of EFIGI galaxies in GALEX $NUV$ and our SourceXtractor++ photometry in the $gri$ bands from the SDSS images. From the ZPEG fits, we derived absolute magnitudes, stellar masses ($M_\ast$) and sSFRs for each galaxy. We repeated this process to bulge and disk apparent magnitudes in the $gri$ bands to derive separate parameters for each component.

The distribution of $NUV-r$ absolute colors versus $r$ absolute magnitudes for EFIGI galaxies recovers the well-known bimodality in the galaxy distribution between star-forming and quiescent galaxies (\fgs\ref{NUV-r-face-on} and \ref{density_NUV}). We provide new information by showing that the different Hubble types occupy specific locations in the color-magnitude diagram (\fg\ref{mean-NUV-r-face-on}). The Hubble sequence draws an ``S shape'' in this plane, corresponding to three galaxy groupings: the \textbf{Red Sequence} composed of elliptical and lenticular galaxies, the \textbf{Green Plain} (formerly Green Valley, see below) populated by intermediate type S0a and early spirals Sa, and the \textbf{Blue Cloud} encompassing all spirals from the Sab types to the irregular galaxies.

We also confirm that the use of the optical $u-r$ photometry overlays the Green Plain and the Red Sequence and prevents one from studying these regions (\fg\ref{u-r_density}). Complementing optical data with a UV band is pivotal to this study and we show that there is a strong correlation between sSFR and $NUV-r$ color (\fg\ref{ssfr_color}). There is also a tight correlation between the stellar mass $M_\ast$ and absolute magnitude $M_r$ (\fg\ref{mass_mag}), which allowed us to convert the color-magnitude diagram into a color-mass diagram (\fg\ref{color_mass}), and interpret it in terms of sSFR.

The Blue Cloud and Red Sequence are confirmed to correspond to star-forming and quiescent galaxies respectively, while the Green Plain is a region of intermediate sSFR. Ellipticals dominate the massive part of the Red Sequence, while lenticulars do so in the lower mass part, and the full sequence spans 2 order of magnitudes in stellar mass (\fg\ref{ell_vs_len_mass}). In contrast, the Blue Cloud extends over 4 orders of magnitudes in stellar mass. $NUV-r$ varies only by $\lesssim1$ and $\sim2$ magnitudes across the Red Sequence and the Blue Cloud respectively, despite the large mass intervals they span. In contrast, the intermediate region encompassing the Green Plain is characterized by as wide a color range as the Blue Cloud ($\sim2$ magnitudes in $NUV-r$), but over a much smaller stellar mass interval of $\sim1$ dex. From irregulars to ellipticals, the steps in $NUV-r$ color between types are small ($\sim 0.1$) in the Red Sequence and the Blue Cloud, but much larger in the Green Plain (0.78 for the Sab-Sa, and 0.9 for the S0a-S0$^+$ types), thus indicating a wide stretching in $NUV-r$ in the intermediate region, which justifies it being renamed from ``Green Valley''.

We use the complete magnitude-limited MorCat catalog to show that the bimodality of the Red Sequence and Blue Cloud, as well as the under-dense Green Plain connecting them are not specific to the EFIGI catalog, nor caused by its selection effects aimed at densely populating all morphological types. We also argue, using MorCat, that the wider range of $NUV-r$ colors (hence sSFR) across the Green Plain, and for morphological types S0a and Sa, causes the lower galaxy density in this region of the color-magnitude and color-mass diagram, making it an under-density. In this picture, the Green Plain is neither due to an intrinsic lower frequency of the S0a and Sa galaxy types that populate it, nor to a shorter lifetime of these types compared to those in the Red Sequence and Blue Cloud: to our knowledge, there is no evidence that these types are more ephemeral than any other disk type (\ie , lenticular or spiral type) along the Hubble sequence.

Our analysis of the EFIGI sample allows us to quantify the morphological transformations that galaxies undergo when evolving from the Blue Cloud to the Red Sequence. To this end, we use their morphological types and detailed morphological features, their bulge-to-total flux ratios ($B/T$), their derived total and separate bulge and disk masses, as well as their total and separate bulge and disk colors, all based on our luminosity profiles and SED fittings. We quantify the morphological changes through the Green Plain as follows:
\begin{itemize}
    \item \textbf{Bulge growth:} From intermediate spirals in the knee (in which the Blue Cloud breaks into the Green Plain) to S0 types in the Red Sequence, $B/T$ varies from $\sim 0.15$ to $\sim 0.5$ (\fg\ref{GV_bulge_ratio}), and the bulge stellar masses more than double at all stellar masses (\fg\ref{bulge_mass}). There is also an increase in the S\'ersic index of bulges across the Green Plain, with redder galaxies having more concentrated bulges (\fg\ref{sersic}), marking a transition from pseudo-bulges to classical ones.
    \item \textbf{Disk reddening:} The bulge and disk decompositions of EFIGI galaxies show a systematic reddening in $g-r$ of all disks along the Blue Cloud (starting from the blue colors of irregular Magellanic types) through the Green Plain, which we demonstrate to indicate a fading of the disk sSFR (\fgs\ref{disk_color} and \ref{color_galaxy_color_disk}). In contrast, the bulges of all spiral types display a remarkable stability in color, which is identical to those of lenticulars and ellipticals (\fg\ref{comparing_EFIGI_and_ZPEG_bulge_disk}), as well as for pseudo-bulges. The decrease in the EFIGI spiral {\tt Arm Strength} and {\tt Flocculence} morphological attributes across the Green Plain (\fg\ref{disk_features}) suggest that the star formation fading of galaxy disks may be due to some restriction in the neutral gas supply available for stellar formation.
\end{itemize}

This analysis of the EFIGI bimodal color-mass diagram provides new clues on the evolution pathways of galaxies in the present and recent past of the Universe. In the Blue Cloud, the 4 orders of magnitude in stellar mass encompassed by late-type spirals, and the high fractions of galaxies with perturbed isophotes ($60\%$ to $90\%$, see \fg\ref{perturbation}) and bright HII regions ($70\%$ to $90\%$, see \fg\ref{hot_spots}), advocate for a mass build-up through numerous mergers. 

The fact that the bulge light and mass ratios of Green Plain galaxies are intermediate between those for the Blue Cloud and Red Sequence, together with the systematic and progressive bulge growth of galaxies from the former to the latter, provides evidence that the Green Plain is a \textbf{transition} region of galaxies in their evolution pathway from star-forming to quiescence. The characteristics of Green Plain galaxies moreover supports the idea that this transition operates through mergers, as stellar inward migration from the disk cannot transfer enough mass to grow a classical bulge. This is in agreement with the results from the numerical simulations that are reviewed in the discussion (\sct \ref{discussion-pathways}), and that suggest the predominance of minor mergers in the formation of the massive spirals and lenticulars. In addition, the common mass limit of $\log(M_\ast)/log(M_\odot)=11.7$ that we measured for all galaxies from Sc to S0 types in the Green Plain matches the mass at which ellipticals become dominant in the Red Sequence (\fg\ref{ell_vs_len_mass}). This suggests that some of the most massive ellipticals were formed by major mergers of the most massive early-type spirals (after the eventual fading of the starbursts that may be produced).

We emphasize that the known high frequency of flybys and mergers in the lifetime of a galaxy, as well as the fact that all parts of the color-mass diagram are populated with standard types from the full Hubble sequence are not compatible with the idea that the Green Plain is the locus of a rapid star formation fading (hence rapid quenching) that would happen while galaxies transit from the Blue Cloud to the Red Sequence. The high frequency of bars (\fg\ref{bar_frequency}) for all spirals as well as the stronger spiral arms and flocculence in the knee of the Green Plain (\fg\ref{disk_features}), suggest that internal dynamics, likely triggered by flybys or mergers, as well as the gas content and physics of star formation may be the key to the bulge growth, disk reddening and aging of galaxy disks from the Blue Cloud to the Red Sequence. Because EFIGI and MorCat are catalogs of nearby galaxies, our analysis is focused on the present and recent evolution of the Universe. Nevertheless, it provides useful insights into the transformations undergone by galaxies of similar morphological types at higher redshifts, whenever these types existed with similar masses.

Interestingly, our analysis suggests that the Hubble sequence can be considered as a reverse evolutionary sequence, with monotonous changes in the color-mass diagram characterized by bulge growth and disk reddening, and with the limitation that galaxies do not go linearly or systematically through all individual types. Indeed, despite a systematic decrease in the $B/T$ bulge-to-total ratio along the sequence (from E to Im), there is a significant dispersion in the measured values of $B/T$ with morphological type (\fg\ref{BT_hist}). However, a major mass build-up in the history of a galaxy requires many mergers that cause a significant $B/T$ increase, and therefore a shift to earlier morphological types. Forward changes in morphology along the Hubble sequence can nevertheless occur as a result of individual mergers that are likely to temporarily boost the star formation rate, hence making the disk bluer. Nevertheless, as the Universe expands, the cosmic streams get thinner with time which may reduce the gas accretion from the cosmic web to galaxies, and the numerous mergers decrease the number density of galaxies, in particular of low mass, so galaxies will inexorably all age toward the Red Sequence. 

Although we have reviewed some results from numerical simulations and assessed their compatibility with our observations (see \sct\ref{discussion-pathways}), a direct comparison by deriving from cosmological volumes the color-mass diagram complemented by $B/T$ and disk color would be a simple and critical test of the simulations. Ideally, use of the same approach based on producing projected image of the simulated galaxies and performing bulge and disk modeling with SourceXtractor++ would allow one to monitor the possible biases due to observational effects. To bring new insights into the characteristics of nearby galaxies as a function of their morphological types and location within the evolutive sequence displayed by the $NUV-r$ color-mass plane, we will report in Quilley \& de Lapparent (\textit{in prep.}) on other properties of the bulge and disks of EFIGI and MorCat galaxies, such as their distributions of bulge-to-total ratios, effective radii and S\'ersic indexes, as well as the scaling relations.

\section{Acknowledgements}
We warmly thank Matthew Lehnert for enriching discussions, thorough
reading, critical comments on this manuscript, and for suggesting the
``Green Plain'' renaming. We are very grateful to Emmanuel Bertin for
initiating and leading the EFIGI project, and for maintaining the
computer cluster on which all calculations were performed. We also thank Damien Le Borgne for his great help in using the ZPEG software. We are
indebted to Aline Chu for her significant internship work that led to
the selection of the MorCat catalog from HyperLeda galaxies, and for her
making of the MorCat sky mask. We also thank the anonymous referee for
their careful reading and important suggestions and comments.

This research made use of {\tt SourceXtractor++}\footnote{\url{https://github.com/astrorama/SourceXtractorPlusPlus}}, an open source software package developed for the Euclid satellite project. This research made use of the VizieR catalog access tool, CDS, Strasbourg, France (DOI: 10.26093/cds/vizier). The original description of the VizieR service was published in A$\&$AS 143, 23. This research also made use of NASA's Astrophysics Data System.

This work is based on observations made with the NASA Galaxy Evolution Explorer. GALEX is operated for NASA by the California Institute of Technology under NASA contract NAS5-98034. Funding for the SDSS and SDSS-II has been provided by the Alfred P. Sloan Foundation, the Participating Institutions, the National Science Foundation, the U.S. Department of Energy, the National Aeronautics and Space Administration, the Japanese Monbukagakusho, the Max Planck Society, and the Higher Education Funding Council for England. The SDSS Web Site is http://www.sdss.org/.
The SDSS is managed by the Astrophysical Research Consortium for the Participating Institutions. The Participating Institutions are the American Museum of Natural History, Astrophysical Institute Potsdam, University of Basel, University of Cambridge, Case Western Reserve University, University of Chicago, Drexel University, Fermilab, the Institute for Advanced Study, the Japan Participation Group, Johns Hopkins University, the Joint Institute for Nuclear Astrophysics, the Kavli Institute for Particle Astrophysics and Cosmology, the Korean Scientist Group, the Chinese Academy of Sciences (LAMOST), Los Alamos National Laboratory, the Max-Planck-Institute for Astronomy (MPIA), the Max-Planck-Institute for Astrophysics (MPA), New Mexico State University, Ohio State University, University of Pittsburgh, University of Portsmouth, Princeton University, the United States Naval Observatory, and the University of Washington.

\bibliography{biblio.bib}

\end{document}